\documentclass[11pt]{article}

\usepackage{amsfonts,amssymb,chet,dsfont,graphicx,mathrsfs,pgfplots,tikz}
\usetikzlibrary{decorations.markings}
\usetikzlibrary{decorations.pathreplacing}
\allowdisplaybreaks

\newcommand{\1}{\mathds{1}}
\newcommand{\zb}{\bar{z}}
\newcommand{\vp}[2]{\varphi_{#1}(z_{#2})}
\newcommand{\vpb}[2]{\varphi_{#1}(z_{#2},\zb_{#2})}
\newcommand{\vpbb}[2]{\bar{\varphi}_{#1}(z_{#2},\zb_{#2})}
\newcommand{\hb}{\bar{h}}
\newcommand{\cOPE}[2]{c_{#1}^{\phantom{#1}#2}}
\newcommand{\D}{\mathcal{D}}
\newcommand{\DOPE}[4]{\D_{#1}^{\phantom{#1}#2}(z_{#3},z_{#4})}
\newcommand{\DOPEb}[4]{\D_{#1}^{\phantom{#1}#2}(z_{#3},\zb_{#3},z_{#4},\zb_{#4})}
\newcommand{\vev}[1]{\langle{#1}\rangle}

\newcommand{\etab}{\bar{\eta}}

\title{All Global One- and Two-Dimensional\\Higher-Point Conformal Blocks}

\author{Jean-Fran\c{c}ois Fortin$^{\ast,}$\email{jean-francois.fortin@phy.ulaval.ca}, Wen-Jie Ma$^{\ast,}$\email{wenjie.ma.1@ulaval.ca} and Witold Skiba$^{\dagger,}$\email{witold.skiba@yale.edu}}

\affiliation{
$^\ast$D\'epartement de Physique, de G\'enie Physique et d'Optique\\Universit\'e Laval, Qu\'ebec, QC G1V 0A6, Canada\\
$^\dagger$Department of Physics, Yale University, New Haven, CT 06520, USA
}

\abstract{We introduce a full set of rules to directly express all $M$-point conformal blocks in one- and two-dimensional conformal field theories, irrespective of the topology.  The $M$-point conformal blocks are power series expansion in some carefully-chosen conformal cross-ratios.  We then prove the rules for any topology constructively with the help of the known position space operator product expansion.  To this end, we first compute the action of the position space operator product expansion on the most general function of position space coordinates relevant to conformal field theory.  These results provide the complete knowledge of all $M$-point conformal blocks with arbitrary external and internal quasi-primary operators (including arbitrary spins in two dimensions) in any topology.}

\date{September 2020} 

\begin{document}

\maketitle



\section{Introduction}\label{SecIntro}

Conformal field theories (CFTs) are special quantum field theories (QFTs) with extended spacetime symmetry groups.  CFTs are important in QFTs as fixed points of the renormalization group flow and in condensed matter theory as descriptions of second-order phase transitions.  Their extra symmetries lead to a separation of operators into quasi-primaries and descendants, that then imply a very powerful operator product expansion (OPE).  Indeed, in CFTs the OPE re-expresses the product of two quasi-primaries at different points into an infinite sum of quasi-primaries.  To generate the descendants, the latter quasi-primaries are acted upon by differential operators that are completely fixed by conformal invariance, up to the OPE coefficients which encode the different CFTs.  Hence, once the OPE is determined, it is straightforward to generate arbitrary correlation functions in terms of infinite sums of products of the OPE coefficients and the so-called conformal blocks that are generated by the multiple action of the OPE.

For arbitrary correlation functions, conformal blocks are functions of sets of conformal cross-ratios (ratios of position space distances that are invariant under the conformal group) that are completely fixed by conformal covariance from, \textit{e.g.}\, the conformal covariance of the OPE.  Conformal blocks are however exceptionally hard to compute in general.  Nevertheless, when conformal blocks are determined, it is possible to constrain the allowed values of the OPE coefficients by relying on the associativity of the correlation functions, the celebrated conformal bootstrap approach \cite{Ferrara:1973yt,Polyakov:1974gs}.  Generally, the constraints originating from the four-point conformal bootstrap are all there is---higher-point conformal bootstrap being redundant, although they could help by considering external quasi-primary operators in scalar representations only.

Hence, much of the work done has been towards the computation of four-point conformal blocks following different approaches.  See for example the Casimir equations \cite{Dolan:2003hv,Dolan:2011dv,Kravchuk:2017dzd}, the shadow formalism \cite{Ferrara:1972xe,Ferrara:1972uq,SimmonsDuffin:2012uy}, the weight-shifting formalism \cite{Karateev:2017jgd,Costa:2018mcg}, integrability \cite{Isachenkov:2016gim,Schomerus:2016epl,Schomerus:2017eny,Isachenkov:2017qgn,Buric:2019dfk}, AdS/CFT \cite{Hijano:2015zsa,Nishida:2016vds,Castro:2017hpx,Dyer:2017zef,Chen:2017yia,Sleight:2017fpc}, and the OPE \cite{Ferrara:1971vh,Ferrara:1971zy,Ferrara:1972cq,Ferrara:1973eg,Ferrara:1973vz,Ferrara:1974nf,Dolan:2000ut,Fortin:2016lmf,Fortin:2016dlj,Comeau:2019xco,Fortin:2019fvx,Fortin:2019dnq,Fortin:2019xyr,Fortin:2019pep,Fortin:2019gck,Fortin:2020ncr}.

With respect to higher-point correlation functions, most of the work is fairly new, with scalar $M$-point blocks in the comb topology\footnote{Sometimes topologies are referred to as channels in the literature. Here, we use channels to distinguish different external field assignments in a given topology, as is customary for four-point functions.} in one- and two-dimensional CFTs as well as scalar five-point blocks in any spacetime dimensions first computed in \cite{Alkalaev:2015fbw,Rosenhaus:2018zqn,Goncalves:2019znr,Parikh:2019ygo,Jepsen:2019svc,Alkalaev:2020kxz}.  Higher-point conformal blocks in the comb topology for arbitrary spacetime dimensions have been obtained in \cite{Parikh:2019dvm,Fortin:2019zkm}.  Scalar six-point conformal blocks in the first non-comb topology, the so-called snowflake topology, have been presented in \cite{Fortin:2020yjz}.\footnote{See also \cite{Anous:2020vtw} for specific snowflake conformal blocks in two-dimensional CFTs.}  Scalar seven-point conformal blocks in the extended snowflake topology, scalar higher-point conformal blocks in the OPE topology, as well as plausible rules for scalar higher-point conformal blocks in higher-dimensional CFTs were introduced in \cite{Fortin:2020bfq,Hoback:2020pgj}.

In \cite{Fortin:2019zkm,Fortin:2020yjz,Fortin:2020bfq}, the authors relied on the embedding space OPE developed in \cite{Fortin:2019fvx,Fortin:2019dnq} to recursively compute higher-point conformal blocks.  It is well-known that CFTs in one and two spacetime dimensions are much simpler than higher-dimensional CFTs.  For example, they do not need to use the machinery of the embedding space.  Moreover, the possible irreducible representations of the Lorentz group are much simpler, leading to all four-point conformal blocks \cite{Bowcock:1990ku,Osborn:2012vt}.  Strangely enough, although the position space OPE has been known in one- and two-dimensional CFTs for a very long time (see for example \cite{Ferrara:1974nf,Ferrara:1974ny,Belavin:1984vu,Dolan:2000ut}), it has not been used to compute arbitrary conformal blocks.\footnote{CFTs in two spacetime dimensions are not only invariant under the standard conformal group, dubbed the global conformal group, they are also invariant under the local conformal group, leading to the Virasoro algebra.  This distinction leads to a separation between the quasi-primaries and the primaries.  There has been a lot of very important work on local conformal invariance and their associated Virasoro blocks, \textit{e.g.}\ \cite{Belavin:1984vu,Zamolodchikov:1985ie,Ginsparg:1988ui,DiFrancesco:1997nk,Perlmutter:2015iya}.  In this work, we are only concerned about the global conformal blocks.}

In this paper, we use the known position space OPE in one- and two-dimensional CFTs to compute any conformal partial wave for arbitrary internal and external quasi-primary operators, including spinning quasi-primary operators in $2d$ CFTs, irrespective of the topology.  We introduce a set of rules to decompose higher-point correlation functions in sums of higher-point conformal blocks (depending on carefully-chosen conformal cross-ratios), with the proper leg factors and OPE coefficient functions.  We then determine the action of the position space OPE on products of powers of position space distances and use it recursively to prove the rules.  With these results, all quantities in arbitrary one- and two-dimensional correlation functions that are prescribed by global conformal invariance can be determined.  Moreover, since there exists $\max\{1,T_0(M)-1\}$ independent conformal bootstrap equations for $M$-point correlation functions \cite{Fortin:2020yjz}, we present the $M$-point conformal bootstrap equations for four-, five-, six-, seven-, and eight-point correlation functions.  Here $T_0(M)$ is the number of inequivalent $M$-point topologies, or the number of unrooted binary trees with $M$ unlabeled leaves.\footnote{$T_0(M)$ does not have an analytic expression.  Starting at $M=2$, the first few numbers in the sequence are $(1,1,1,1,2,2,4,6,11,\ldots)$.  See \textit{The On-line Encyclopedia of Integer Sequences} at \href{https://oeis.org/A000672}{https://oeis.org/A000672} and \href{https://oeis.org/A129860}{https://oeis.org/A129860} for more details.}

This paper is organized as follows: Section \ref{SecOPE} discusses the simplifications occurring in low-dimensional CFTs, reviews the position space OPE, and determines its action on products of powers of position space distances.  In Section \ref{SecCF}, we first review $(M<4)$-point correlation functions and then introduce our notation for $(M\geq4)$-point correlation functions.  For the latter, we decompose correlation functions in sums of OPE coefficient functions times conformal partial waves dependent on the chosen topology.  We also write conformal partial waves in terms of leg factors times conformal blocks which are functions of the conformal cross-ratios.  We finally introduce the rules by defining different OPE vertices according to their number of internal legs.  The rules determine the OPE coefficient factors, the leg factors, the conformal blocks, and the conformal cross-ratios.  We then present several examples in Section \ref{SecExample}, giving the complete set of conformal bootstrap equations for four-, five-, six-, seven-, and eight-point correlation functions.  Finally, we conclude in Section \ref{SecConc} with a discussion of the generalization to higher-dimensional CFTs while Appendix \ref{SAppProof} presents the proof of the rules using the position space OPE recursively.


\section{Operator Product Expansion}\label{SecOPE}

After describing the simplifications occurring in one- and two-dimensional CFTs, this section reviews the position space OPE and presents its action on the most general function of position space coordinates relevant to CFTs in one and two spacetime dimensions.


\subsection{Simplifications in Low Dimensions}

Before discussing the position space OPE, we first survey the simplifications occurring in global CFTs in one and two spacetime dimensions.

First, global CFTs in one and two spacetime dimensions are much simpler than in higher spacetime dimensions due to the allowed irreducible Lorentz group representations.  Indeed, the possible irreducible representations of $1d$ and $2d$ CFTs are all trivial.

In $d=1$, all quasi-primary operators $\vp{}{}$ with conformal dimensions $h$ are in the trivial irreducible representation.  Thus, for a triplet of $1d$ quasi-primary operators there is only one (trivial) OPE tensor structure.  In other words, there exists only one OPE coefficient per triplet of quasi-primary operators.

In $d=2$, all irreducible Lorentz representations have at most two independent components.  As such, a $2d$ quasi-primary operator in a non-trivial irreducible representation with two components can be split into two quasi-primary operators (its holomorphic and anti-holomorphic parts), each effectively in the trivial irreducible representation.  They are denoted by $\vpb{}{}$ [labeled by $(h,\hb)$] and $\vpbb{}{}$ [labeled by $(\hb,h)$], respectively, and their conformal dimension and spin are $\Delta=h+\hb$ and $s=h-\hb$ (with $2s\in\mathbb{Z}$).  Considering the latter quasi-primary operators instead of the former, the OPE for any triplet of quasi-primary operators in the complete set $\{\vpb{i}{},\vpbb{i}{}\}$ (all in the trivial irreducible representation) has only one trivial OPE tensor structure.  Once again, there exists only one OPE coefficient per triplet of quasi-primary operators.

Hence, CFTs in one and two spacetime dimensions are not plagued by the intricacies originating from non-trivial irreducible representations that are ubiquitous in higher spacetime dimensions.  They can be fully investigated by considering only quasi-primary operators in the trivial irreducible representation.

Second, the number of independent conformal cross-ratios, which are ratios of position space coordinates $z_{ij}=z_i-z_j$ of the type
\eqn{\eta_{ij;kl}=\frac{z_{ij}z_{kl}}{z_{il}z_{kj}},}[Eqeta]
and are invariant under conformal transformations, is much smaller in one- and two-dimensional CFTs.

Indeed, for $(M\geq4)$-point correlation functions, there are $M-3$ conformal cross-ratios in $d=1$.  Hence only the analog of the $u_a$ conformal cross-ratios in higher-dimensional CFTs exist in one-dimensional CFTs, all the higher-dimensional $v_{ab}$ conformal cross-ratios are redundant.

For two-dimensional $(M\geq4)$-point correlation functions, the number of conformal cross-ratios is $2(M-3)$, twice as much as in $d=1$.  This fact can be deduced from the factorization of the $2d$ OPE discussed below.  The factorization property also implies that the extra conformal cross-ratios of the $v$-type appearing in two-dimensional CFTs are easily deduced from the $1d$ conformal cross-ratios as
\eqn{\etab_{ij;kl}=\frac{\zb_{ij}\zb_{kl}}{\zb_{il}\zb_{kj}}.}[Eqetab]

As a consequence, control over the conformal cross-ratios is much simpler in one- and two-dimensional CFTs when compared to CFTs in higher spacetime dimensions.  For one, the action of the OPE differential operator does not involve as many re-summations.  Moreover, the observation that only the $u$-type conformal cross-ratios exist in one- and two-dimensional CFTs directly leads to a proof of the higher-point correlation function rules presented in this paper.


\subsection{Action of the Operator Product Expansion}

The position space OPE in one and two spacetime dimensions is well known \cite{Ferrara:1974nf,Ferrara:1974ny,Belavin:1984vu,Dolan:2000ut}.  In one-dimensional CFTs, it is given by
\eqn{
\begin{gathered}
\vp{i}{1}\vp{j}{2}=\sum_k\cOPE{ij}{k}\DOPE{ij}{k}{1}{2}\vp{k}{2},\\
\DOPE{ij}{k}{1}{2}=\frac{1}{z_{12}^{h_i+h_j-h_k}}{}_1F_1(h_i-h_j+h_k,2h_k;z_{12}\partial_2),
\end{gathered}
}[EqOPE1d]
while it is
\eqn{
\begin{gathered}
\vpb{i}{1}\vpb{j}{2}=\sum_k\cOPE{ij}{k}\DOPEb{ij}{k}{1}{2}\vpb{k}{2},\\
\DOPEb{ij}{k}{1}{2}=\frac{1}{z_{12}^{h_i+h_j-h_k}}{}_1F_1(h_i-h_j+h_k,2h_k;z_{12}\partial_2)\\
\qquad\qquad\qquad\qquad\qquad\qquad\times\frac{1}{\zb_{12}^{\hb_i+\hb_j-\hb_k}}{}_1F_1(\hb_i-\hb_j+\hb_k,2\hb_k;\zb_{12}\bar{\partial}_2),
\end{gathered}
}[EqOPE2d]
in two-dimensional CFTs (with the extra requirement that $h_i-\hb_i+h_j-\hb_j+h_k-\hb_k\in\mathbb{Z}$ from spin statistics).  Here it is understood that the partial derivatives in the expansion of the Kummer confluent hypergeometric function act first, \textit{i.e.}\
\eqn{{}_1F_1(a,b;z_{12}\partial_2)\equiv\sum_{n\geq0}\frac{(a)_n}{(b)_n}\frac{z_{12}^n\partial_2^n}{n!}.}[Eq1F1]
Clearly, the $2d$ OPE factorizes into two $1d$ OPEs---the holomorphic and anti-holomorphic OPEs.  As a consequence, two-dimensional higher-point conformal blocks factorize into their one-dimensional holomorphic (functions of $\eta_a$) and anti-holomorphic (functions of $\etab_a$) factors.  Thus, without loss of generality, we can focus solely on one-dimensional CFTs from now on.

The most general function of position space coordinates that can appear in a CFT is made out of products of powers of $z_{ij}$.  Since
\eqn{\partial_j^n\prod_{a\neq i,j}\frac{1}{z_{ja}^{p_a}}=(-1)^nn!\sum_{\substack{\{m_a\}\geq0\\\sum_am_a=n}}\prod_{a\neq i,j}\frac{(p_a)_{m_a}}{m_a!z_{ja}^{p_a+m_a}},}
from the multinomial theorem, we have
\eqn{\DOPE{12}{3}{i}{j}\prod_{a\neq i,j}\frac{1}{z_{ja}^{p_a}}=\sum_{\{m_a\}\geq0}\frac{(-1)^{\bar{m}}(h_1-h_2+h_3)_{\bar{m}}}{(2h_3)_{\bar{m}}z_{ij}^{h_1+h_2-h_3-\bar{m}}}\prod_{a\neq i,j}\frac{(p_a)_{m_a}}{m_a!z_{ja}^{p_a+m_a}},}[EqIb]
where $\bar{m}=\sum_{a\neq i,j}m_a$.  Equation \eqref{EqIb} is the analog of the $\bar{I}$-function of \cite{Fortin:2019dnq}.  Its knowledge will allow us to construct and prove the rules for building $M$-point correlation functions, to which we now turn.


\section{Higher-Point Correlation Functions}\label{SecCF}

This section relies on the position space OPE in one spacetime dimension \eqref{EqOPE1d} and its action on products of powers of position space coordinates $z_{ij}$ \eqref{EqIb} to generate all correlation functions in any topology, the generalization to two spacetime dimensions is straightforward.  After reviewing the one-, two-, and three-point correlation functions, we present a complete set of rules to explicitly write any $M$-point correlation function.  The proof of the rules and detailed computations are left for the appendix.


\subsection{\texorpdfstring{$M<4$}{M<4}-Point Correlation Functions}

In a CFT, the only non-trivial one-point correlation function involves the identity operator $\1$, which is invariant under conformal transformations with $h_\1=0$ (and $\hb_\1=0$ in two spacetime dimensions).  The identity operator is defined such that $\vev{\1}=1$.

From the OPE \eqref{EqOPE1d} and the one-point correlation function $\vev{\1}$, non-vanishing two-point correlation functions are given by
\eqn{\vev{\vp{i}{1}\vp{j}{2}}=\cOPE{ij}{\1}\DOPE{ij}{\1}{1}{2}\vev{\1}=\frac{\cOPE{ij}{\1}}{z_{12}^{2h}},}[Eq2ptCF]
with $h_i=h_j=h$.  As expected, two-point correlation functions vanish unless both quasi-primary operators have the same conformal dimension.

By applying the OPE \eqref{EqOPE1d} on the two-point correlation functions \eqref{Eq2ptCF}, it is straightforward to compute three-point correlation functions as
\eqn{\vev{\vp{i}{1}\vp{j}{2}\vp{k}{3}}=\sum_{k'}\cOPE{ij}{k'}\DOPE{ij}{k'}{1}{2}\vev{\vp{k'}{2}\vp{k}{3}}=\frac{c_{ijk}}{z_{12}^{h_i+h_j-h_k}z_{23}^{h_j+h_k-h_i}z_{13}^{h_k+h_i-h_j}},}[Eq3ptCF]
with the help of \eqref{EqIb}.  We note that in \eqref{Eq3ptCF} we defined three-point coefficients $c_{ijk}$ from OPE coefficients $\cOPE{ij}{k}$ as $c_{ijk}=\sum_{k'}\cOPE{ij}{k'}\cOPE{k'k}{\1}$.

The well-known results \eqref{Eq2ptCF} and \eqref{Eq3ptCF}, and their straightforward generalizations to $2d$ CFTs, show that the one-, two-, and three-point correlation functions are completely fixed by global conformal invariance up to some overall constants, as expected.\footnote{As is common knowledge, conformal invariance fixes the form of the $(M<4)$-point correlation functions.  From the familiar two- and three-point correlation functions \eqref{Eq2ptCF} and \eqref{Eq3ptCF}, it is then straightforward to obtain the $1d$ OPE \eqref{EqOPE1d} [as well as \eqref{EqOPE2d} in two spacetime dimensions].}


\subsection{\texorpdfstring{$M\geq4$}{M>=4}-Point Correlation Functions}

Due to the presence of conformal cross-ratios, $(M\geq4)$-point correlation functions are not completely fixed by conformal invariance.  Hence, they are technically more difficult to determine.  Nevertheless, higher-point correlation functions can be separated through the OPE into their fundamental constituents, the conformal blocks, which are completely fixed by conformal invariance.  However, traditionally conformal blocks have been technically challenging to compute.  We introduce here a complete set of rules to explicitly write down $(M\geq4)$-point conformal blocks in any topology.  Before proceeding, we first discuss our notation.

Any $M$-point correlation function can be expanded through the OPE in several different ways.  By consistency, these different ways must lead to the same answer, an observation at the core of the conformal bootstrap \cite{Ferrara:1973yt,Polyakov:1974gs}.  When the number of quasi-primary operators is larger than five, the OPE leads to decompositions with different topologies. By choosing one specific OPE decomposition, $(M\geq4)$-point correlation functions can be divided into conformal partial waves as
\eqn{\vev{\vp{i_1}{1}\cdots\vp{i_M}{M}}=\sum_{\{k_a\}}f_{M(k_1,\ldots,k_{M-3})}^{(i_1,\ldots,i_M)}\left.W_{M(h_{k_1},\ldots,h_{k_{M-3}})}^{(h_{i_1},\ldots,h_{i_M})}\right|_{\text{topology}}
,}[EqCPW]
where the summation is over the $M-3$ exchanged quasi-primary operators $\vp{k_1}{},\ldots,\vp{k_{M-3}}{}$ appearing in the OPE decomposition.  In \eqref{EqCPW}, the products of OPE coefficients (including the proper sign for fermion crossings in two spacetime dimensions) are denoted by $f_M$ and the conformal partial waves $W_M$ are expressible in terms of conformal blocks $G_M$ following\footnote{In two-dimensional CFTs, $M$-point correlation functions are given by
\eqn{\vev{\vpb{i_1}{1}\cdots\vpb{i_M}{M}}=\sum_{\{k_a\}}f_{M(k_1,\ldots,k_{M-3})}^{(i_1,\ldots,i_M)}\left.W_{M(h_{k_1},\ldots,h_{k_{M-3}})}^{(h_{i_1},\ldots,h_{i_M})}\bar{W}_{M(\hb_{k_1},\ldots,\hb_{k_{M-3}})}^{(\hb_{i_1},\ldots,\hb_{i_M})}\right|_{\text{topology}},}[EqCPW2d]
in terms of one-dimensional conformal partial waves.  Here the bar on top of the second conformal partial wave simply means that $z_{ab}\to\zb_{ab}$ and $\eta_a\to\etab_a$ as dictated by the factorization of the $2d$ OPE.}
\eqna{
\left.W_{M(h_{k_1},\ldots,h_{k_{M-3}})}^{(h_{i_1},\ldots,h_{i_M})}\right|_{\text{topology}}&=L_{M|\text{topology}}^{(h_{i_1},\ldots,h_{i_M})}\left[\prod_{1\leq a\leq M-3}(\eta_a^M)^{h_{k_a}}\right]G_{M|\text{topology}}^{(\boldsymbol{h})}(\boldsymbol{\eta}^M).
}[EqW]
In \eqref{EqW}, $L_M$ represents the leg which is made out of position space coordinates $z_{ab}$ and is responsible for the proper behavior of the conformal partial wave under scale transformations, while $\boldsymbol{\eta}^M$ is the vector of conformal cross-ratios.  Moreover, the conformal blocks are power series expansion in the conformal cross-ratios of the form
\eqn{G_{M|\text{topology}}^{(\boldsymbol{h})}(\boldsymbol{\eta}^M)=\sum_{\{n_a\}\geq0}C_{M|\text{topology}}^{(\boldsymbol{h})}(\boldsymbol{n})F_{M|\text{topology}}^{(\boldsymbol{h})}(\boldsymbol{n})\prod_{1\leq a\leq M-3}\frac{(\eta_a^M)^{n_a}}{n_a!},}[EqG]
with $\boldsymbol{n}$ the vector of indices of summation $\{n_a\}$, $C_M$ a summand of the hypergeometric type, and $F_M$ encoding extra sums.  Although in more dimensions it was useful to treat $C_M$ and $F_M$ separately, here we always provide the product $C_MF_M$ combined.  Our goal in this section is thus to provide rules for the determination of the leg $L_M$ and the explicit definitions of the conformal cross-ratios $\boldsymbol{\eta}^M$ in terms of the position space coordinates as well as the conformal block $G_M$ for an arbitrary topology.


\subsection{Rules for \texorpdfstring{$M\geq4$}{M>=4}-Point Correlation Functions}

To begin, we note that the OPE can be used recursively to increase the number of points in an arbitrary correlation function.  This technique depends on the OPE differential operator acting on the initial correlation function using \eqref{EqIb}, followed by re-summations to eliminate superfluous sums.  In principle, one can generate any correlation function following this prescription.  However, to keep a suitable handle on the conformal cross-ratios and the associated re-summations, it is necessary to build the conformal partial waves constructively.  Hence our strategy relies on applying the OPE following a fixed, ordered, procedure to reach the appropriate topology.

\begin{figure}[t]
\centering
\resizebox{16cm}{!}{%
\begin{tikzpicture}[thick]
\begin{scope}
\node at (0,2.5) {$1I$ OPE vertex};
\draw[-] (0,0)--+(90:1) node[above]{$\vp{i_\alpha}{\alpha}$};
\draw[-,postaction={decorate,decoration={markings,mark=at position 0.5 with {\arrow{<}}}}] (0,0)--+(180:1) node[left]{$\vp{i_\beta}{\beta}$};
\draw[-,dotted,postaction={decorate,decoration={markings,mark=at position 0.5 with {\arrow{>}}}}] (0,0)--+(0:1) node[right]{$\vp{k_{j_3}}{\gamma}$};
\node at (0,-1.5) {$\vp{i_\alpha}{\alpha}\vp{i_\beta}{\beta}\sim\vp{k_{j_3}}{\beta}$};
\node at (0,-3.0) {$(-1)^{-h_{k_{j_3}}}\cOPE{i_\alpha i_\beta}{k_{j_3}}$};
\node at (0,-4.5) {$z_{\beta\gamma;\alpha}^{h_{i_\alpha}}z_{\gamma\alpha;\beta}^{h_{i_\beta}}$};
\node at (0,-6.0) {$(h_{i_\alpha}-h_{i_\beta}+h_{k_{j_3}})_{n_{j_3}}$};
\end{scope}
\begin{scope}[xshift=6cm]
\node at (0,2.5) {$2I$ OPE vertex};
\draw[-] (0,0)--+(90:1) node[above]{$\vp{i_\alpha}{\alpha}$};
\draw[-,dotted,postaction={decorate,decoration={markings,mark=at position 0.5 with {\arrow{<}}}}] (0,0)--+(180:1) node[left]{$\vp{k_{j_2}}{\beta}$};
\draw[-,dotted,postaction={decorate,decoration={markings,mark=at position 0.5 with {\arrow{>}}}}] (0,0)--+(0:1) node[right]{$\vp{k_{j_3}}{\gamma}$};
\node at (0,-1.5) {$\substack{\vp{i_\alpha}{\alpha}\vp{k_{j_2}}{\beta}\sim\vp{k_{j_3}}{\beta}\\\vp{k_{j_3}}{\gamma}\vp{i_\alpha}{\alpha}\sim\vp{k_{j_2}}{\alpha}}$};
\node at (0,-3.0) {$(-1)^{-h_{k_{j_3}}}\cOPE{i_\alpha k_{j_2}}{k_{j_3}}$};
\node at (0,-4.5) {$z_{\beta\gamma;\alpha}^{h_{i_\alpha}}$};
\node at (0,-6.0) {$\substack{(h_{i_\alpha}-h_{k_{j_2}}-n_{j_2}+h_{k_{j_3}})_{n_{j_3}}\\\times(-h_{i_\alpha}+h_{k_{j_2}}+h_{k_{j_3}})_{n_{j_2}}}$};
\end{scope}
\begin{scope}[xshift=12cm]
\node at (0,2.5) {$3I$ OPE vertex};
\draw[-,dotted] (0,0)--+(90:1) node[above]{$\vp{k_{j_1}}{\alpha}$};
\draw[-,dotted,postaction={decorate,decoration={markings,mark=at position 0.5 with {\arrow{<}}}}] (0,0)--+(180:1) node[left]{$\vp{k_{j_2}}{\beta}$};
\draw[-,dotted,postaction={decorate,decoration={markings,mark=at position 0.5 with {\arrow{>}}}}] (0,0)--+(0:1) node[right]{$\vp{k_{j_3}}{\gamma}$};
\node at (0,-1.5) {$\substack{\vp{k_{j_1}}{\alpha}\vp{k_{j_2}}{\beta}\sim\vp{k_{j_3}}{\beta}\\\vp{k_{j_3}}{\gamma}\vp{k_{j_1}}{\alpha}\sim\vp{k_{j_2}}{\alpha}\\\vp{k_{j_3}}{\gamma}\vp{k_{j_2}}{\beta}\sim\vp{k_{j_1}}{\beta}}$};
\node at (0,-3.0) {$(-1)^{h_{k_{j_1}}-h_{k_{j_3}}}\cOPE{k_{j_1}k_{j_2}}{k_{j_3}}$};
\node at (0,-4.5) {---};
\node at (0,-6.0) {$\substack{(h_{k_{j_1}}-h_{k_{j_2}}-n_{j_2}+h_{k_{j_3}})_{n_{j_1}+n_{j_3}}\\\times(-h_{k_{j_1}}+h_{k_{j_2}}+h_{k_{j_3}})_{n_{j_2}}{}_3F_2}$};
\end{scope}
\end{tikzpicture}
}
\caption{$1I$, $2I$, and $3I$ OPE vertices with their associated OPE limits, OPE coefficient contributions, leg factors, and conformal block factors (from top to bottom).  Here, solid (dotted) lines represent external (internal, or exchanged) quasi-primary operators while the arrows depict the flow of position space coordinates, \textit{i.e.}\ the chosen OPE limits relevant for the gluing procedure representing the OPE action.  The hypergeometric function that appears in the conformal block factor for the $3I$ OPE vertex is given by ${}_3F_2\equiv{}_3F_2\left[\protect\begin{array}{c}-n_{j_1},-n_{j_2},1-2h_{k_{j_2}}-n_{j_2}\\h_{k_{j_1}}-h_{k_{j_2}}-n_{j_2}+h_{k_{j_3}},1+h_{k_{j_1}}-h_{k_{j_2}}-n_{j_2}-h_{k_{j_3}}\protect\end{array};1\right]$.  We note that the internal quasi-primary operator without an arrow in the $3I$ OPE vertex serves as an anchor point for an extra comb structure.}
\label{FignIOPE}
\end{figure}
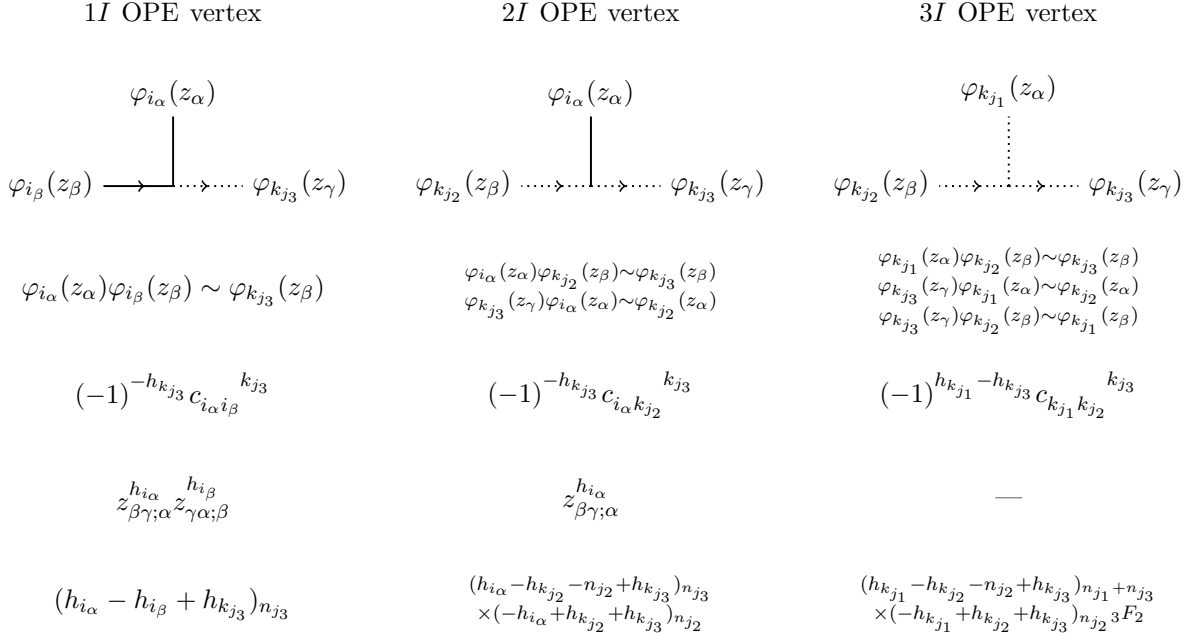
To this end, we divide the OPE into three different groups---$1I$, $2I$, and $3I$ OPEs---where an $nI$ OPE vertex in a given topology has $n$ internal lines (representing internal, or exchanged, quasi-primary operators) and $3-n$ external lines (representating external quasi-primary operators).\footnote{In this notation, $0I$ OPE vertices never appear in $(M\geq4)$-point correlation functions.  Obviously, an $0I$ OPE vertex always appears alone---it represents a three-point correlation function---and its associated set of rules is derived straightforwardly from \eqref{Eq3ptCF}.}  We also introduce an extra $1I$ OPE vertex which corresponds to the initial $1I$ OPE vertex from which the full topology will be constructed.  The $nI$ OPE vertices with their associated rules are shown in Figures \ref{FignIOPE} and \ref{FigInitialOPE}.

\begin{figure}[!t]
\centering
\resizebox{5cm}{!}{%
\begin{tikzpicture}[thick]
\begin{scope}
\node at (0,2.5) {Initial $1I$ OPE vertex};
\draw[-] (0,0)--+(90:1) node[above]{$\vp{i_\alpha}{\alpha}$};
\draw[-,dotted,postaction={decorate,decoration={markings,mark=at position 0.5 with {\arrow{<}}}}] (0,0)--+(180:1) node[left]{$\vp{k_{j_2}}{\beta}$};
\draw[-,postaction={decorate,decoration={markings,mark=at position 0.5 with {\arrow{>}}}}] (0,0)--+(0:1) node[right]{$\vp{i_\gamma}{z_\gamma}$};
\node at (0,-1.5) {$\vp{i_\gamma}{\gamma}\vp{i_\alpha}{\alpha}\sim\vp{k_{j_2}}{\alpha}$};
\node at (0,-3.0) {$c_{i_\alpha k_{j_2}i_\gamma}$};
\node at (0,-4.5) {$z_{\beta\gamma;\alpha}^{h_{i_\alpha}}z_{\alpha\beta;\gamma}^{h_{i_\gamma}}$};
\node at (0,-6.0) {$(h_{i_\gamma}-h_{i_\alpha}+h_{k_{j_2}})_{n_{j_2}}$};
\end{scope}
\end{tikzpicture}
}
\caption{Initial $1I$ OPE vertex with its OPE limit, OPE coefficient contribution, leg factor, and conformal block factor.  The notation matches the one of Figure \ref{FignIOPE}.}
\label{FigInitialOPE}
\end{figure}
For these figures, solid (dotted) lines represent external (internal, or exchanged) quasi-primary operators while arrows depict the flow of position space coordinates.  The latter fix the choice of OPE limits relevant when appending OPE vertices together following the gluing procedure.  The OPE limits determine fully the set of rules, with each $nI$ OPE vertex having a specific leg and conformal block factor, while two glued OPE vertices are necessary to obtain the conformal cross-ratios.

Hence specific rules for the OPE coefficient contributions (up to fermion crossings in two spacetime dimensions),\footnote{The overall minus signs in the OPE coefficient contributions originate from the choice of leg factors since $z_{ji}=-z_{ij}$, contrary to higher-dimensional CFTs.} the legs with the notation
\eqn{z_{\alpha\beta;\gamma}=\frac{z_{\alpha\beta}}{z_{\alpha\gamma}z_{\beta\gamma}},}[EqL]
(and its obvious generalization with $z_i\to\zb_i$ for $2d$ CFTs), and the conformal block factors are associated to each $nI$ OPE vertex while the conformal cross-ratios, defined in \eqref{Eqeta} and \eqref{Eqetab}, are not yet included in the rules of Figures \ref{FignIOPE} and \ref{FigInitialOPE} since they are built from the gluing of two OPE vertices.

Although the leg rules are included in Figures \ref{FignIOPE} and \ref{FigInitialOPE}, they require the knowledge of the position space coordinates of all quasi-primary operators, including the exchanged quasi-primary operators.  Consequently, it is also necessary to know how the different OPE vertices are combined together in an arbitrary topology to determine the proper leg factors, as for the conformal cross-ratios.

These observations lead us to the gluing procedure and the flow of position space coordinates depicted by the arrows in Figures \ref{FignIOPE} and \ref{FigInitialOPE}, or in other words the chosen OPE limits.  To elucidate the gluing procedure, we first note that any topology has at least two $1I$ OPE vertices, with the comb topology saturating the bound.  We now choose one $1I$ OPE vertex (any will do) that plays the role of the initial $1I$ OPE vertex of Figure \ref{FigInitialOPE}.  We then start gluing $2I$ and $3I$ OPE vertices in the proper order until we reach another $1I$ OPE vertex, where this procedure stops.  This procedure produces a comb-like topology, but some of the teeth of this comb correspond to internal lines that need to be glued further.

When $2I$ OPE vertices are included in this initial comb-like structure there is nothing further to do since the corresponding tooth represents an external operator.  This is not the case $3I$ OPE vertices.  From this initial comb topology, we select one of the $3I$ OPE vertices and repeat the procedure above by gluing $2I$ and $3I$ OPE vertices in the correct order corresponding to the associated OPE decomposition until we reach another $1I$ OPE vertex.  We note that this new comb-like structure needs another arrow type to differentiate its flow of position space coordinates.  To systematically construct the conformal partial wave of interest, we continue this procedure with each additional comb structure and their associated arrows until all the $3I$ OPE vertices have been completely glued, \textit{i.e.}\ until the number of $3I$ OPE vertices added in the final comb structure is zero.

\begin{figure}[!t]
\centering
\resizebox{12cm}{!}{%
\begin{tikzpicture}[thick]
\begin{scope}
\node[circle,draw,left,minimum size=24pt] at (0.5,0) {$\alpha_2$};
\draw[-,postaction={decorate,decoration={markings,mark=at position 0.6cm with {\arrow{>>}},mark=at position 1.6cm with {\arrow{>>}},mark=at position 2.6cm with {\arrow{>>}},mark=at position 3.6cm with {\arrow{>>}}}}] (0.5,0)--(4.5,0);
\draw[-] (1.5,0)--(1.5,1) node[circle,draw,above,minimum size=24pt] {$\beta_2$};
\fill (1.5,0) circle[radius=3pt];
\draw[-] (2.5,0)--(2.5,1) node[circle,draw,above,minimum size=24pt] {$\beta_1$};
\draw[-] (3.5,0)--(3.5,1) node[circle,draw,above,minimum size=24pt] {$\alpha_1$};
\node at (2,-0.5) {$k_{j_1}$};
\end{scope}
\begin{scope}[xshift=7cm]
\node[circle,draw,left,minimum size=24pt] at (0.5,0) {$\alpha_2$};
\draw[-,postaction={decorate,decoration={markings,mark=at position 0.6cm with {\arrow{>>}},mark=at position 1.6cm with {\arrow{>>}},mark=at position 2.6cm with {\arrow{>>}}}}] (0.5,0)--(3.5,0);
\draw[-] (1.5,0)--(1.5,1) node[circle,draw,above,minimum size=24pt] {$\beta_2$};
\fill (1.5,0) circle[radius=3pt];
\draw[-] (2.5,0)--(2.5,1) node[circle,draw,above,minimum size=24pt] {$\beta_1$};
\draw[-,postaction={decorate,decoration={markings,mark=at position 0.5 with {\arrow{<}}}}] (3.5,0)--+(60:1) node[circle,draw,above right,minimum size=24pt] {$\alpha_1$};
\draw[-,postaction={decorate,decoration={markings,mark=at position 0.5 with {\arrow{>}}}}] (3.5,0)--+(-60:1);
\node at (2,-0.5) {$k_{j_1}$};
\end{scope}
\begin{scope}[yshift=-4cm]
\node[circle,draw,left,minimum size=24pt] at (0.5,0) {$\alpha_2$};
\draw[-,postaction={decorate,decoration={markings,mark=at position 0.6cm with {\arrow{>>}},mark=at position 1.6cm with {\arrow{>>}}}}] (0.5,0)--(2.5,0);
\draw[-] (1.5,0)--(1.5,1) node[circle,draw,above,minimum size=24pt] {$\beta_2$};
\fill (1.5,0) circle[radius=3pt];
\draw[-,postaction={decorate,decoration={markings,mark=at position 0.5 with {\arrow{<}}}}] (2.5,0)--+(60:1) node[circle,draw,above right,minimum size=24pt] {$\beta_1$};
\draw[-,postaction={decorate,decoration={markings,mark=at position 0.6 with {\arrow{>}}}}] (2.5,0)--+(-60:1) node[circle,draw,below right,minimum size=24pt] {$\alpha_1$};
\node at (2,-0.5) {$k_{j_1}$};
\end{scope}
\begin{scope}[xshift=7cm,yshift=-4cm]
\node[circle,draw,left,minimum size=24pt] at (0.5,0) {$\alpha_2$};
\draw[-,postaction={decorate,decoration={markings,mark=at position 0.6cm with {\arrow{>}},mark=at position 1.6cm with {\arrow{>}},mark=at position 2.6cm with {\arrow{>}}}}] (0.5,0)--(3.5,0);
\draw[-] (1.5,0)--(1.5,1) node[circle,draw,above,minimum size=24pt] {$\beta_2$};
\fill (1.5,0) circle[radius=3pt];
\draw[-] (2.5,0)--(2.5,1) node[above] {$\beta_1$};
\fill ([xshift=-3pt,yshift=-3pt]2.5,0) rectangle ++(6pt,6pt);
\node[right] at (3.5,0) {$\alpha_1$};
\node at (2,-0.5) {$k_{j_1}$};
\end{scope}
\end{tikzpicture}
}
\caption{The conformal cross-ratio associated to the exchanged quasi-primary operator $\vp{k_{j_1}}{}$ is given by $\eta_{j_1}=\eta_{\alpha_2\beta_2;\beta_1\alpha_1}$ while the leg factor for the $1I$, $2I$, or $3I$ OPE vertex denoted by a dot is $z_{\alpha_2\beta_1;\beta_2}^{h_{i_{\beta_2}}}z_{\beta_1\beta_2;\alpha_2}^{h_{i_{\alpha_2}}}$ ($1I$ OPE vertex), $z_{\alpha_2\beta_1;\beta_2}^{h_{i_{\beta_2}}}$ ($2I$ OPE vertex), or $1$ ($3I$ OPE vertex).  Finally, the leg factor associated to the initial $1I$ OPE vertex, denoted by a square, is $z_{\alpha_2\alpha_1;\beta_1}^{h_{i_{\beta_1}}}z_{\beta_1\alpha_2;\alpha_1}^{h_{i_{\alpha_1}}}$.}
\label{FigGlue}
\end{figure}
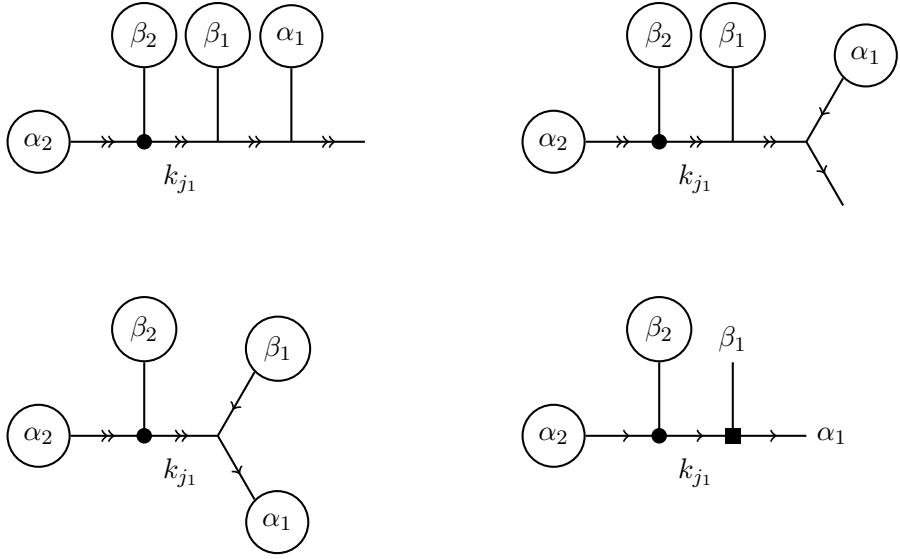

With our specific choice of OPE limits on the $nI$ OPE vertices, the gluing procedure leads to well-defined rules for the leg factors and the conformal cross-ratios appearing in the conformal partial wave of interest.  These rules are shown in Figure \ref{FigGlue}.

To simplify the notation, we draw all lines as solid ones.  Moreover, external quasi-primary operators are denoted only by their position space coordinates.  Hence $\alpha_1$ in Figure \ref{FigGlue} corresponds to $\vp{i_{\alpha_1}}{\alpha_1}$.  For internal quasi-primary operators, we include an index with a subscript on each internal line.  Thus, $k_{j_1}$ in Figure \ref{FigGlue} denotes the exchanged quasi-primary operator $\vp{k_{j_1}}{z}$.  Circles with outgoing arrows (implicit when not depicted) represent arbitrary contributions, with the numbers corresponding to the first external quasi-primary operators from which the arrows flow out.  Finally, circles with incoming arrows correspond to arbitrary contributions, but with the numbers standing for the first external quasi-primary operators appearing in the contributions.

The complete set of rules are thus given in Figures \ref{FignIOPE}, \ref{FigInitialOPE} and \ref{FigGlue} with the following recipe.  First, the product of OPE coefficients $f_M$ in \eqref{EqCPW} is computed by multiplying the OPE coefficient contributions for each vertex (up to an overall sign for fermion crossings in two spacetime dimensions).  In the same manner, the leg is the product of the leg factors.  Finally, the conformal block \eqref{EqG}, more precisely the product $C_MF_M$, is calculated from the product of the conformal block factors divided by $\prod_{1\leq a\leq M-3}(2h_{k_a})_{n_a}$, with each exchanged quasi-primary operator having its associated conformal cross-ratio.  Conveniently, in $1d$ there are as many exchange operators as conformal cross-ratios.

The complete set of rules are proven by induction in Appendix \ref{SAppProof}.  To demonstrate the rules better, we now turn to concrete examples.


\section{Example}\label{SecExample}

In this section, we use the rules of Section \ref{SecCF} to build conformal partial waves for arbitrary topologies.  The goal is to illuminate the procedure.  We note that according to our rules, there are several ways of writing the same conformal partial waves.  Indeed, the choice of the initial $1I$ OPE vertex, of the initial comb structure, and of the flow of position space coordinates, lead to different-looking answers that must be equal by consistency at the level of the conformal partial waves.  At the level of the correlation functions, \textit{i.e.}\ for the conformal bootstrap, different orderings (non-trivial re-orderings) of the external quasi-primary operators and/or different topologies must be equated.

We present here the conformal partial waves for the four-, five-, six-, seven-, and eight-point conformal bootstrap equations.  To encode the OPE order, we organize the quasi-primary operators in the initial comb structure as follows: $\vev{\cdots\vp{i_4}{4}\vp{i_3}{3}\vp{i_2}{2}|\vp{i_1}{1}}$ where the first OPE is $\vp{i_3}{3}\vp{i_2}{2}\sim\vp{k_1}{2}$, the second OPE is $\vp{i_4}{4}\vp{k_1}{2}\sim\vp{k_2}{2}$, and so on until the last OPE which is $\vp{k_n}{n}\vp{i_1}{1}\sim\1$.  Moreover, we delimit all extra comb structures by curly brackets, as for example $\{\cdots\vp{i_3}{3}\vp{i_2}{2}\vp{i_1}{1}\}$ with the same pattern for the OPEs, \textit{i.e.}\ first $\vp{i_2}{2}\vp{i_1}{1}\sim\vp{k_1}{1}$ followed by $\vp{i_3}{3}\vp{k_1}{1}\sim\vp{k_2}{1}$ and so forth.  Finally, fermion crossings occurring in two spacetime dimensions lead to overall sign factors of the form $(-1)^{F_{i_1i_2}}$ that are $-1$ when both $\vp{i_1}{1}$ and $\vp{i_2}{2}$ are fermions and $1$ otherwise.


\subsection{Four-Point Correlation Functions}

\begin{figure}[!t]
\centering
\resizebox{12cm}{!}{%
\begin{tikzpicture}[thick]
\begin{scope}
\node[left] at (0.5,0) {$1$};
\draw[-,postaction={decorate,decoration={markings,mark=at position 0.6cm with {\arrow{>}},mark=at position 1.6cm with {\arrow{>}},mark=at position 2.6cm with {\arrow{>}}}}] (0.5,0)--(3.5,0);
\draw[-] (1.5,0)--(1.5,1) node[above] {$2$};
\draw[-] (2.5,0)--(2.5,1) node[above] {$4$};
\node[right] at (3.5,0) {$3$};
\node at (2,-0.5) {$k_1$};
\node at (5.5,0) {$=$};
\end{scope}
\begin{scope}[xshift=7cm]
\node[left] at (0.5,0) {$1$};
\draw[-,postaction={decorate,decoration={markings,mark=at position 0.6cm with {\arrow{>}},mark=at position 1.6cm with {\arrow{>}},mark=at position 2.6cm with {\arrow{>}}}}] (0.5,0)--(3.5,0);
\draw[-] (1.5,0)--(1.5,1) node[above] {$4$};
\draw[-] (2.5,0)--(2.5,1) node[above] {$2$};
\node[right] at (3.5,0) {$3$};
\node at (2,-0.5) {$k_1$};
\end{scope}
\end{tikzpicture}
}
\caption{Four-point conformal bootstrap equations.}
\label{FigEx4pt}
\end{figure}
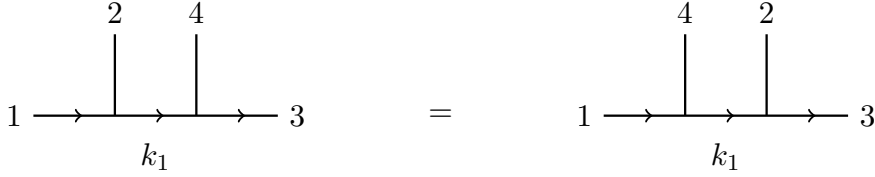
The conformal partial waves for four-point correlation functions of arbitrary quasi-primary operators were found in \cite{Bowcock:1990ku,Osborn:2012vt}.  The four-point conformal bootstrap equations are shown in Figure \ref{FigEx4pt} and correspond to
\eqn{\vev{\vp{i_4}{4}\vp{i_2}{2}\vp{i_1}{1}|\vp{i_3}{3}}=(-1)^{F_{i_2i_4}}\vev{\vp{i_2}{2}\vp{i_4}{4}\vp{i_1}{1}|\vp{i_3}{3}},}[EqBoot4]
where the overall minus sign appears in two-dimensional CFTs and comes from fermion crossings.  Demanding the equality \eqref{EqBoot4} for all external quasi-primary operators constitute the full set of four-point bootstrap equations (which is the complete set of bootstrap equations since higher-point bootstrap equations are redundant).

Following our rules for $\vev{\vp{i_4}{4}\vp{i_2}{2}\vp{i_1}{1}|\vp{i_3}{3}}$ (see the arrows for the left topology of Figure \ref{FigEx4pt}), we have
\eqn{
\begin{gathered}
f_4=(-1)^{-h_{k_1}}\cOPE{i_2i_1}{k_1}c_{i_4k_1i_3},\\
L_4=z_{14;2}^{h_{i_2}}z_{42;1}^{h_{i_1}}z_{13;4}^{h_{i_4}}z_{41;3}^{h_{i_3}},\\
\eta_1^4=\eta_{12;43},\\
C_4F_4=\frac{(h_{i_2}-h_{i_1}+h_{k_1})_{n_1}(h_{i_3}-h_{i_4}+h_{k_1})_{n_1}}{(2h_{k_1})_{n_1}},
\end{gathered}
}[Eq4pt1]
which is the usual result quoted in the literature.

For the right topology found in Figure \ref{FigEx4pt}, which is denoted by $\vev{\vp{i_2}{2}\vp{i_4}{4}\vp{i_1}{1}|\vp{i_3}{3}}$, we obtain instead
\eqn{
\begin{gathered}
f_4=(-1)^{-h_{k_1}}\cOPE{i_4i_1}{k_1}c_{i_2k_1i_3},\\
L_4=z_{12;4}^{h_{i_4}}z_{24;1}^{h_{i_1}}z_{13;2}^{h_{i_2}}z_{21;3}^{h_{i_3}},\\
\eta_1^4=\eta_{14;23}=1-\eta_{12;43},\\
C_4F_4=\frac{(h_{i_4}-h_{i_1}+h_{k_1})_{n_1}(h_{i_3}-h_{i_2}+h_{k_1})_{n_1}}{(2h_{k_1})_{n_1}}.
\end{gathered}
}[Eq4pt2]

We note that although the exchanged quasi-primary operators are denoted by $\vp{k_1}{}$ in both \eqref{Eq4pt1} and \eqref{Eq4pt2}, they do not necessarily represent the same sets.  Using \eqref{EqCPW}, demanding that \eqref{EqBoot4} is satisfied for all external quasi-primary operators leads to the full conformal bootstrap.


\subsection{Five-Point Correlation Functions}

\begin{figure}[!t]
\centering
\resizebox{12cm}{!}{%
\begin{tikzpicture}[thick]
\begin{scope}
\node[left] at (0.5,0) {$1$};
\draw[-,postaction={decorate,decoration={markings,mark=at position 0.6cm with {\arrow{>}},mark=at position 1.6cm with {\arrow{>}},mark=at position 2.6cm with {\arrow{>}},mark=at position 3.6cm with {\arrow{>}}}}] (0.5,0)--(4.5,0);
\draw[-] (1.5,0)--(1.5,1) node[above] {$2$};
\draw[-] (2.5,0)--(2.5,1) node[above] {$3$};
\draw[-] (3.5,0)--(3.5,1) node[above] {$4$};
\node[right] at (4.5,0) {$5$};
\node at (2,-0.5) {$k_1$};
\node at (3,-0.5) {$k_2$};
\node at (6,0) {$=$};
\end{scope}
\begin{scope}[xshift=7cm]
\node[left] at (0.5,0) {$2$};
\draw[-,postaction={decorate,decoration={markings,mark=at position 0.6cm with {\arrow{>}},mark=at position 1.6cm with {\arrow{>}},mark=at position 2.6cm with {\arrow{>}},mark=at position 3.6cm with {\arrow{>}}}}] (0.5,0)--(4.5,0);
\draw[-] (1.5,0)--(1.5,1) node[above] {$3$};
\draw[-] (2.5,0)--(2.5,1) node[above] {$4$};
\draw[-] (3.5,0)--(3.5,1) node[above] {$1$};
\node[right] at (4.5,0) {$5$};
\node at (2,-0.5) {$k_1$};
\node at (3,-0.5) {$k_2$};
\end{scope}
\end{tikzpicture}
}
\caption{Five-point conformal bootstrap equations.}
\label{FigEx5pt}
\end{figure}
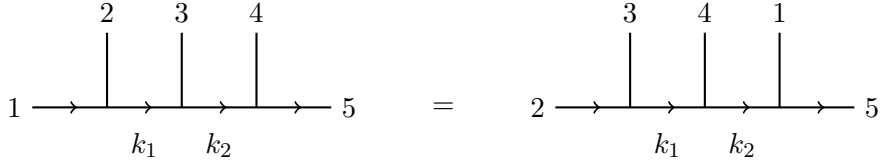
Five-point correlation functions are reminiscent of four-point correlation functions: they also have only one topology, the so-called comb topology \cite{Alkalaev:2015fbw,Rosenhaus:2018zqn}; and there exists only one set of conformal bootstrap equations, depicted in Figure \ref{FigEx5pt}.  Any other bootstrap equation is satisfied automatically due to the symmetries of the comb topology~\cite{Fortin:2020yjz}.  Figure \ref{FigEx5pt} leads to
\eqna{\vev{\vp{i_4}{4}\vp{i_3}{3}\vp{i_2}{2}\vp{i_1}{1}|\vp{i_5}{5}}=(-1)^{F_{i_1i_2}+F_{i_1i_3}+F_{i_1i_4}}\vev{\vp{i_1}{1}\vp{i_4}{4}\vp{i_3}{3}\vp{i_2}{2}|\vp{i_5}{5}},}[EqBoot5]
where again the overall minus sign exists only in two-dimensional CFTs and comes from fermion crossings.

From the rules of Section \ref{SecCF} applied to $\vev{\vp{i_4}{4}\vp{i_3}{3}\vp{i_2}{2}\vp{i_1}{1}|\vp{i_5}{5}}$ (the left topology of Figure \ref{FigEx5pt}), we can write
\eqn{
\begin{gathered}
f_5=(-1)^{-h_{k_1}-h_{k_2}}\cOPE{i_2i_1}{k_1}\cOPE{i_3k_1}{k_2}c_{i_4k_2i_5},\\
L_5=z_{13;2}^{h_{i_2}}z_{32;1}^{h_{i_1}}z_{14;3}^{h_{i_3}}z_{15;4}^{h_{i_4}}z_{41;5}^{h_{i_5}},\\
\eta_1^5=\eta_{12;34},\qquad\eta_2^5=\eta_{13;45},\\
C_5F_5=\frac{(h_{i_2}-h_{i_1}+h_{k_1})_{n_1}(h_{i_3}-h_{k_1}-n_1+h_{k_2})_{n_2}(-h_{i_3}+h_{k_1}+h_{k_2})_{n_1}(h_{i_5}-h_{i_4}+h_{k_2})_{n_2}}{(2h_{k_1})_{n_1}(2h_{k_2})_{n_2}},
\end{gathered}
}[Eq5pt1]
which matches the result found in \cite{Rosenhaus:2018zqn} after trivial manipulations.

Equivalently, for $\vev{\vp{i_1}{1}\vp{i_4}{4}\vp{i_3}{3}\vp{i_2}{2}|\vp{i_5}{5}}$, we reach
\eqn{
\begin{gathered}
f_5=(-1)^{-h_{k_1}-h_{k_2}}\cOPE{i_3i_2}{k_1}\cOPE{i_4k_1}{k_2}c_{i_1k_2i_5},\\
L_5=z_{24;3}^{h_{i_3}}z_{43;2}^{h_{i_2}}z_{21;4}^{h_{i_4}}z_{25;1}^{h_{i_1}}z_{12;5}^{h_{i_5}},\\
\eta_1^5=\eta_{23;41},\qquad\eta_2^5=\eta_{24;15},\\
C_5F_5=\frac{(h_{i_3}-h_{i_2}+h_{k_1})_{n_1}(h_{i_4}-h_{k_1}-n_1+h_{k_2})_{n_2}(-h_{i_4}+h_{k_1}+h_{k_2})_{n_1}(h_{i_5}-h_{i_1}+h_{k_2})_{n_2}}{(2h_{k_1})_{n_1}(2h_{k_2})_{n_2}},
\end{gathered}
}[Eq5pt2]
a simple rewriting of \eqref{Eq5pt1}.

From \eqref{EqCPW} and the conformal partial waves \eqref{Eq5pt1} and \eqref{Eq5pt2}, \eqref{EqBoot5} implements the five-point conformal bootstrap.


\subsection{Six-Point Correlation Functions}

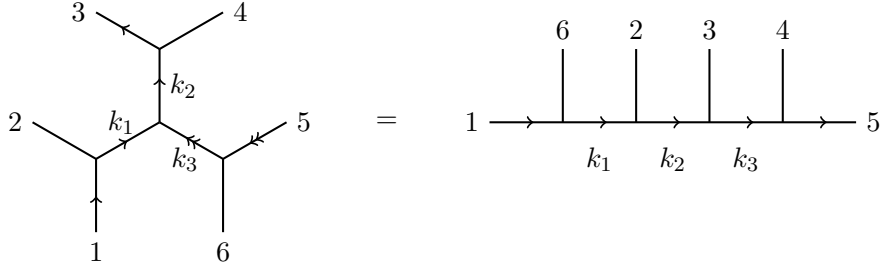
\begin{figure}[!t]
\centering
\resizebox{12cm}{!}{%
\begin{tikzpicture}[thick]
\begin{scope}
\draw[-,postaction={decorate,decoration={markings,mark=at position 0.6 with {\arrow{<}}}}] (0,0)--+(-150:1) node[pos=0.6,above]{$k_1$};
\draw[-,postaction={decorate,decoration={markings,mark=at position 0.6 with {\arrow{<}}}}] (0,0)++(-150:1)--+(-90:1) node[below]{$1$};
\draw[-] (0,0)++(-150:1)--+(150:1) node[left]{$2$};
\draw[-,postaction={decorate,decoration={markings,mark=at position 0.6 with {\arrow{>}}}}] (0,0)--+(90:1) node[pos=0.5,right]{$k_2$};
\draw[-,postaction={decorate,decoration={markings,mark=at position 0.6 with {\arrow{>}}}}] (0,0)++(90:1)--+(150:1) node[left]{$3$};
\draw[-] (0,0)++(90:1)--+(30:1) node[right]{$4$};
\draw[-,postaction={decorate,decoration={markings,mark=at position 0.6 with {\arrow{<<}}}}] (0,0)--+(-30:1) node[pos=0.4,below]{$k_3$};
\draw[-,postaction={decorate,decoration={markings,mark=at position 0.6 with {\arrow{<<}}}}] (0,0)++(-30:1)--+(30:1) node[right]{$5$};
\draw[-] (0,0)++(-30:1)--+(-90:1) node[below]{$6$};
\node at (3.1,0) {$=$};
\end{scope}
\begin{scope}[xshift=4cm]
\node[left] at (0.5,0) {$1$};
\draw[-,postaction={decorate,decoration={markings,mark=at position 0.6cm with {\arrow{>}},mark=at position 1.6cm with {\arrow{>}},mark=at position 2.6cm with {\arrow{>}},mark=at position 3.6cm with {\arrow{>}},mark=at position 4.6cm with {\arrow{>}}}}] (0.5,0)--(5.5,0);
\draw[-] (1.5,0)--(1.5,1) node[above] {$6$};
\draw[-] (2.5,0)--(2.5,1) node[above] {$2$};
\draw[-] (3.5,0)--(3.5,1) node[above] {$3$};
\draw[-] (4.5,0)--(4.5,1) node[above] {$4$};
\node[right] at (5.5,0) {$5$};
\node at (2,-0.5) {$k_1$};
\node at (3,-0.5) {$k_2$};
\node at (4,-0.5) {$k_3$};
\end{scope}
\end{tikzpicture}
}
\caption{Six-point conformal bootstrap equations.}
\label{FigEx6pt}
\end{figure}
Six-point correlation functions are interesting due to the appearance of a new topology, the so-called snowflake topology \cite{Fortin:2020yjz}.  Equating the snowflake and the comb as in Figure \ref{FigEx6pt} leads to the only independent set of six-point conformal bootstrap equations given by
\eqna{
&\vev{\vp{i_4}{4}\{\vp{i_6}{6}\vp{i_5}{5}\}\vp{i_2}{2}\vp{i_1}{1}|\vp{i_3}{3}}\\
&\qquad=(-1)^{F_{i_2i_3}+F_{i_3i_6}+F_{i_1i_3}+F_{i_3i_5}+F_{i_1i_5}+F_{i_2i_5}+F_{i_2i_6}}\vev{\vp{i_4}{4}\vp{i_3}{3}\vp{i_2}{2}\vp{i_6}{6}\vp{i_1}{1}|\vp{i_5}{5}},
}[EqBoot6]
where once again fermion crossings imply the overall minus signs of two-dimensional CFTs.

Applying the rules of Section \ref{SecCF} to the snowflake $\vev{\vp{i_4}{4}\{\vp{i_6}{6}\vp{i_5}{5}\}\vp{i_2}{2}\vp{i_1}{1}|\vp{i_3}{3}}$ implies that the conformal partial waves are
\eqn{
\begin{gathered}
f_6=(-1)^{-h_{k_1}-h_{k_2}}\cOPE{i_2i_1}{k_1}\cOPE{i_6i_5}{k_3}\cOPE{k_3k_1}{k_2}c_{i_4k_2i_3},\\
L_6=z_{15;2}^{h_{i_2}}z_{52;1}^{h_{i_1}}z_{41;3}^{h_{i_3}}z_{13;4}^{h_{i_4}}z_{51;6}^{h_{i_6}}z_{16;5}^{h_{i_5}},\\
\eta_1^6=\eta_{12;54},\qquad\eta_2^6=\eta_{15;43},\qquad\eta_3^6=\eta_{56;14},
\end{gathered}
}[Eq6pt1]
with
\eqna{
C_6F_6&=\frac{(h_{i_2}-h_{i_1}+h_{k_1})_{n_1}(h_{i_3}-h_{i_4}+h_{k_2})_{n_2}(h_{i_6}-h_{i_5}+h_{k_3})_{n_3}}{(2h_{k_1})_{n_1}(2h_{k_2})_{n_2}(2h_{k_3})_{n_3}}\\
&\phantom{=}\qquad\times(h_{k_3}-h_{k_1}-n_1+h_{k_2})_{n_3+n_2}(-h_{k_3}+h_{k_1}+h_{k_2})_{n_1}\\
&\phantom{=}\qquad\times{}_3F_2\left[\begin{array}{c}-n_3,-n_1,1-2h_{k_1}-n_1\\h_{k_3}-h_{k_1}-n_1+h_{k_2},1+h_{k_3}-h_{k_1}-n_1-h_{k_2}\end{array};1\right].
}

For the comb $\vev{\vp{i_4}{4}\vp{i_3}{3}\vp{i_2}{2}\vp{i_6}{6}\vp{i_1}{1}|\vp{i_5}{5}}$, we obtain instead
\eqn{
\begin{gathered}
f_6=(-1)^{-h_{k_1}-h_{k_2}-h_{k_3}}\cOPE{i_6i_1}{k_1}\cOPE{i_2k_1}{k_2}\cOPE{i_3k_2}{k_3}c_{i_4k_3i_5},\\
L_6=z_{12;6}^{h_{i_6}}z_{26;1}^{h_{i_1}}z_{13;2}^{h_{i_2}}z_{14;3}^{h_{i_3}}z_{15;4}^{h_{i_4}}z_{41;5}^{h_{i_5}},\\
\eta_1^6=\eta_{16;23},\qquad\eta_2^6=\eta_{12;34},\qquad\eta_3^6=\eta_{13;45},
\end{gathered}
}[Eq6pt2]
with
\eqna{
C_6F_6&=\frac{(h_{i_6}-h_{i_1}+h_{k_1})_{n_1}(h_{i_5}-h_{i_4}+h_{k_3})_{n_3}}{(2h_{k_1})_{n_1}(2h_{k_2})_{n_2}(2h_{k_3})_{n_3}}\\
&\phantom{=}\qquad\times(h_{i_2}-h_{k_1}-n_1+h_{k_2})_{n_2}(-h_{i_2}+h_{k_1}+h_{k_2})_{n_1}\\
&\phantom{=}\qquad\times(h_{i_3}-h_{k_2}-n_2+h_{k_3})_{n_3}(-h_{i_3}+h_{k_2}+h_{k_3})_{n_2},
}
in agreement with \cite{Rosenhaus:2018zqn}.

Starting from the six-point conformal bootstrap equations \eqref{EqBoot6} for all external quasi-primary operators, using the conformal partial wave decomposition \eqref{EqCPW} with the results \eqref{Eq6pt1} and \eqref{Eq6pt2}, generates the full six-point conformal bootstrap.


\subsection{Seven-Point Correlation Functions}

\begin{figure}[!t]
\centering
\resizebox{14cm}{!}{%
\begin{tikzpicture}[thick]
\begin{scope}
\draw[-,postaction={decorate,decoration={markings,mark=at position 0.6 with {\arrow{<}}}}] (0,0)--+(-150:1) node[pos=0.6,above]{$k_1$};
\draw[-,postaction={decorate,decoration={markings,mark=at position 0.6 with {\arrow{<}}}}] (0,0)++(-150:1)--+(-90:1) node[below]{$1$};
\draw[-] (0,0)++(-150:1)--+(150:1) node[left]{$2$};
\draw[-,postaction={decorate,decoration={markings,mark=at position 0.6 with {\arrow{<<}}}}] (0,0)--+(90:1) node[pos=0.5,right]{$k_3$};
\draw[-] (0,0)++(90:1)--+(150:1) node[left]{$3$};
\draw[-,postaction={decorate,decoration={markings,mark=at position 0.7 with {\arrow{<<}}}}] (0,0)++(90:1)--+(30:1) node[pos=0.5,above]{$k_4$};
\draw[-,postaction={decorate,decoration={markings,mark=at position 0.6 with {\arrow{<<}}}}] (0,0)++(90:1)++(30:1)--+(90:1) node[above]{$4$};
\draw[-] (0,0)++(90:1)++(30:1)--+(-30:1) node[right]{$5$};
\draw[-,postaction={decorate,decoration={markings,mark=at position 0.6 with {\arrow{>}}}}] (0,0)--+(-30:1) node[pos=0.4,below]{$k_2$};
\draw[-,postaction={decorate,decoration={markings,mark=at position 0.6 with {\arrow{>}}}}] (0,0)++(-30:1)--+(30:1) node[right]{$6$};
\draw[-] (0,0)++(-30:1)--+(-90:1) node[below]{$7$};
\node at (3.1,0.5) {$=$};
\end{scope}
\begin{scope}[xshift=4cm]
\node[left] at (0.5,0) {$3$};
\draw[-,postaction={decorate,decoration={markings,mark=at position 0.6cm with {\arrow{>}},mark=at position 1.6cm with {\arrow{>}},mark=at position 2.6cm with {\arrow{>}},mark=at position 3.6cm with {\arrow{>}},mark=at position 4.6cm with {\arrow{>}},mark=at position 5.6cm with {\arrow{>}}}}] (0.5,0)--(6.5,0);
\draw[-] (1.5,0)--(1.5,1) node[above] {$4$};
\draw[-] (2.5,0)--(2.5,1) node[above] {$2$};
\draw[-] (3.5,0)--(3.5,1) node[above] {$1$};
\draw[-] (4.5,0)--(4.5,1) node[above] {$7$};
\draw[-] (5.5,0)--(5.5,1) node[above] {$5$};
\node[right] at (6.5,0) {$6$};
\node at (2,-0.5) {$k_1$};
\node at (3,-0.5) {$k_2$};
\node at (4,-0.5) {$k_3$};
\node at (5,-0.5) {$k_4$};
\end{scope}
\end{tikzpicture}
}
\caption{Seven-point conformal bootstrap equations.}
\label{FigEx7pt}
\end{figure}
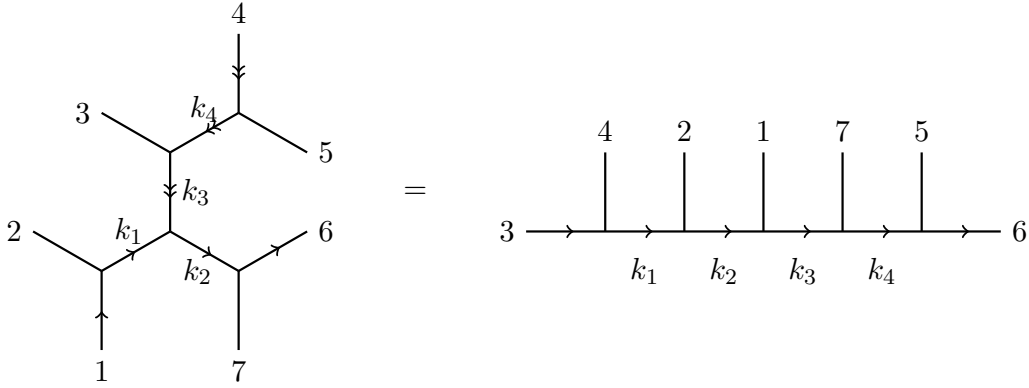
Seven-point correlation functions can be decomposed in conformal partial waves following two topologies: the comb and the extended snowflake topologies \cite{Fortin:2020bfq,Hoback:2020pgj}.  They are depicted in Figure \ref{FigEx7pt} with a given choice of OPE limits.  The equality shown in Figure \ref{FigEx7pt} translates into
\eqna{
&\vev{\vp{i_7}{7}\{\vp{i_3}{3}\vp{i_5}{5}\vp{i_4}{4}\}\vp{i_2}{2}\vp{i_1}{1}|\vp{i_6}{6}}\\
&\qquad=(-1)^{F_{i_1i_2}+F_{i_1i_3}+F_{i_1i_4}+F_{i_2i_3}+F_{i_2i_4}+F_{i_3i_4}+F_{i_3i_5}+F_{i_5i_7}}\\
&\qquad\phantom{=}\times\vev{\vp{i_5}{5}\vp{i_7}{7}\vp{i_1}{1}\vp{i_2}{2}\vp{i_4}{4}\vp{i_3}{3}|\vp{i_6}{6}},
}[EqBoot7]
and represents the sole set of seven-point conformal bootstrap equations, when considering all external quasi-primary operators.  In \eqref{EqBoot7}, the minus sign takes into account fermion crossings that are possible in two-dimensional CFTs only.

Looking at the extended snowflake topology with the choice of OPE limits seen in Figure \ref{FigEx7pt}, \textit{i.e.}\ the seven-point correlation functions $\vev{\vp{i_7}{7}\{\vp{i_3}{3}\vp{i_5}{5}\vp{i_4}{4}\}\vp{i_2}{2}\vp{i_1}{1}|\vp{i_6}{6}}$, the conformal partial waves are
\eqn{
\begin{gathered}
f_7=(-1)^{-h_{k_1}-h_{k_2}-h_{k_4}}\cOPE{i_2i_1}{k_1}\cOPE{k_3k_1}{k_2}\cOPE{i_3k_4}{k_3}\cOPE{i_5i_4}{k_4}c_{i_7k_2i_6},\\
L_7=z_{14;2}^{h_{i_2}}z_{42;1}^{h_{i_1}}z_{41;3}^{h_{i_3}}z_{43;5}^{h_{i_5}}z_{35;4}^{h_{i_4}}z_{16;7}^{h_{i_7}}z_{71;6}^{h_{i_6}},\\
\eta_1^7=\eta_{12;47},\qquad\eta_2^7=\eta_{14;76},\qquad\eta_3^7=\eta_{43;17},\qquad\eta_4^7=\eta_{45;31},
\end{gathered}
}[Eq7pt1]
with
\eqna{
C_7F_7&=\frac{(h_{i_2}-h_{i_1}+h_{k_1})_{n_1}(h_{i_6}-h_{i_7}+h_{k_2})_{n_2}(h_{i_5}-h_{i_4}+h_{k_4})_{n_4}}{(2h_{k_1})_{n_1}(2h_{k_2})_{n_2}(2h_{k_3})_{n_3}(2h_{k_4})_{n_4}}\\
&\phantom{=}\qquad\times(h_{i_3}-h_{k_4}-n_4+h_{k_3})_{n_3}(-h_{i_3}+h_{k_4}+h_{k_3})_{n_4}\\
&\phantom{=}\qquad\times(h_{k_3}-h_{k_1}-n_1+h_{k_2})_{n_3+n_2}(-h_{k_3}+h_{k_1}+h_{k_2})_{n_1}\\
&\phantom{=}\qquad\times{}_3F_2\left[\begin{array}{c}-n_3,-n_1,1-2h_{k_1}-n_1\\h_{k_3}-h_{k_1}-n_1+h_{k_2},1+h_{k_3}-h_{k_1}-n_1-h_{k_2}\end{array};1\right].
}

Focusing on $\vev{\vp{i_5}{5}\vp{i_7}{7}\vp{i_1}{1}\vp{i_2}{2}\vp{i_4}{4}\vp{i_3}{3}|\vp{i_6}{6}}$ instead, which corresponds to the comb topology of Figure \ref{FigEx7pt}, we get
\eqn{
\begin{gathered}
f_7=(-1)^{-h_{k_1}-h_{k_2}-h_{k_3}-h_{k_4}}\cOPE{i_4i_3}{k_1}\cOPE{i_2k_1}{k_2}\cOPE{i_1k_2}{k_3}\cOPE{i_7k_3}{k_4}c_{i_5k_4i_6},\\
L_7=z_{32;4}^{h_{i_4}}z_{24;3}^{h_{i_3}}z_{31;2}^{h_{i_2}}z_{37;1}^{h_{i_1}}z_{35;7}^{h_{i_7}}z_{36;5}^{h_{i_5}}z_{53;6}^{h_{i_6}},\\
\eta_1^7=\eta_{34;21},\qquad\eta_2^7=\eta_{32;17},\qquad\eta_3^7=\eta_{31;75},\qquad\eta_4^7=\eta_{37;56},
\end{gathered}
}[Eq7pt2]
with
\eqna{
C_7F_7&=\frac{(h_{i_4}-h_{i_3}+h_{k_1})_{n_1}(h_{i_6}-h_{i_5}+h_{k_4})_{n_4}}{(2h_{k_1})_{n_1}(2h_{k_2})_{n_2}(2h_{k_3})_{n_3}(2h_{k_4})_{n_4}}\\
&\phantom{=}\qquad\times(h_{i_2}-h_{k_1}-n_1+h_{k_2})_{n_2}(-h_{i_2}+h_{k_1}+h_{k_2})_{n_1}\\
&\phantom{=}\qquad\times(h_{i_1}-h_{k_2}-n_2+h_{k_3})_{n_3}(-h_{i_1}+h_{k_2}+h_{k_3})_{n_2}\\
&\phantom{=}\qquad\times(h_{i_7}-h_{k_3}-n_3+h_{k_4})_{n_4}(-h_{i_7}+h_{k_3}+h_{k_4})_{n_3},
}
as expected from the literature \cite{Rosenhaus:2018zqn}.

As usual, comparing the conformal partial wave decomposition \eqref{EqCPW} of the seven-point correlation functions appearing in \eqref{EqBoot7} that are given by \eqref{Eq7pt1} and \eqref{Eq7pt2} generate the seven-point conformal bootstrap.


\subsection{Eight-Point Correlation Functions}

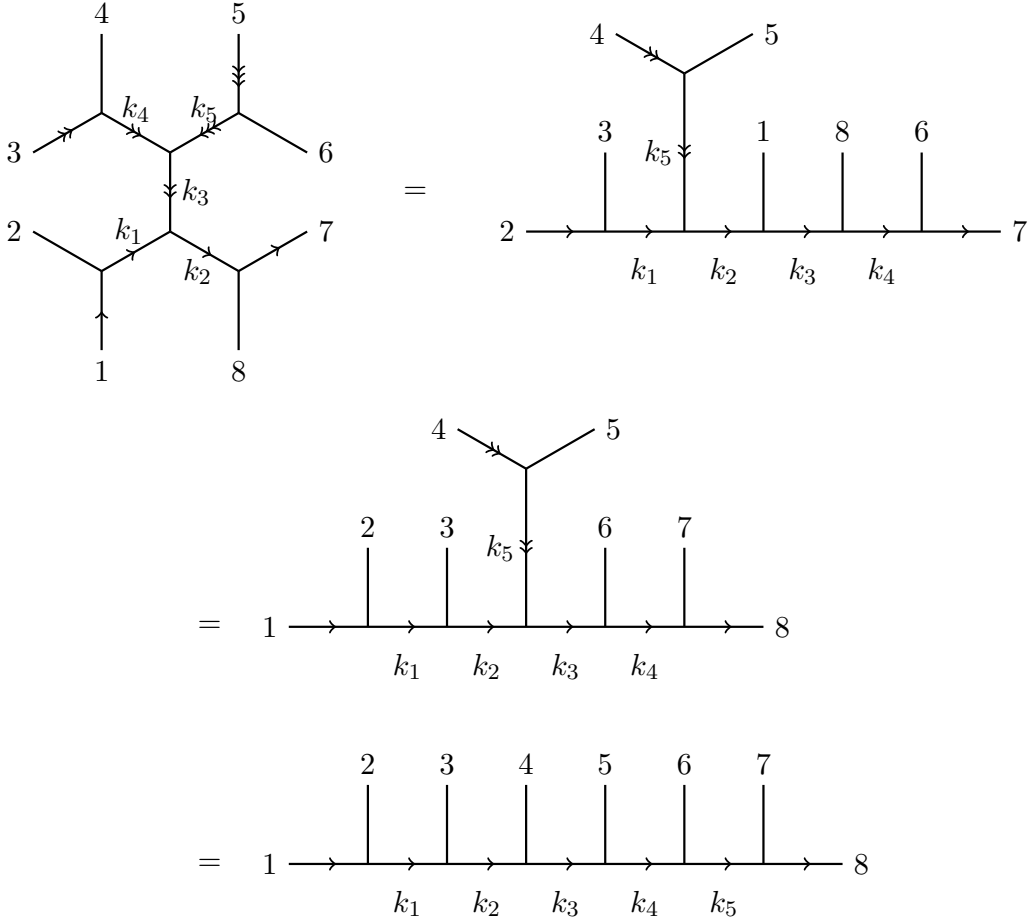
\begin{figure}[!t]
\centering
\resizebox{14cm}{!}{%
\begin{tikzpicture}[thick]
\begin{scope}
\draw[-,postaction={decorate,decoration={markings,mark=at position 0.6 with {\arrow{<}}}}] (0,0)--+(-150:1) node[pos=0.6,above]{$k_1$};
\draw[-,postaction={decorate,decoration={markings,mark=at position 0.6 with {\arrow{<}}}}] (0,0)++(-150:1)--+(-90:1) node[below]{$1$};
\draw[-] (0,0)++(-150:1)--+(150:1) node[left]{$2$};
\draw[-,postaction={decorate,decoration={markings,mark=at position 0.6 with {\arrow{<<}}}}] (0,0)--+(90:1) node[pos=0.5,right]{$k_3$};
\draw[-,postaction={decorate,decoration={markings,mark=at position 0.6 with {\arrow{<<}}}}] (0,0)++(90:1)--+(150:1) node[pos=0.5,above]{$k_4$};
\draw[-,postaction={decorate,decoration={markings,mark=at position 0.6 with {\arrow{<<}}}}] (0,0)++(90:1)++(150:1)--+(210:1) node[left]{$3$};
\draw[-] (0,0)++(90:1)++(150:1)--+(90:1) node[above]{$4$};
\draw[-,postaction={decorate,decoration={markings,mark=at position 0.7 with {\arrow{<<<}}}}] (0,0)++(90:1)--+(30:1) node[pos=0.5,above]{$k_5$};
\draw[-,postaction={decorate,decoration={markings,mark=at position 0.6 with {\arrow{<<<}}}}] (0,0)++(90:1)++(30:1)--+(90:1) node[above]{$5$};
\draw[-] (0,0)++(90:1)++(30:1)--+(-30:1) node[right]{$6$};
\draw[-,postaction={decorate,decoration={markings,mark=at position 0.6 with {\arrow{>}}}}] (0,0)--+(-30:1) node[pos=0.4,below]{$k_2$};
\draw[-,postaction={decorate,decoration={markings,mark=at position 0.6 with {\arrow{>}}}}] (0,0)++(-30:1)--+(30:1) node[right]{$7$};
\draw[-] (0,0)++(-30:1)--+(-90:1) node[below]{$8$};
\node at (3.1,0.5) {$=$};
\end{scope}
\begin{scope}[xshift=4cm]
\node[left] at (0.5,0) {$2$};
\draw[-,postaction={decorate,decoration={markings,mark=at position 0.6cm with {\arrow{>}},mark=at position 1.6cm with {\arrow{>}},mark=at position 2.6cm with {\arrow{>}},mark=at position 3.6cm with {\arrow{>}},mark=at position 4.6cm with {\arrow{>}},mark=at position 5.6cm with {\arrow{>}}}}] (0.5,0)--(6.5,0);
\draw[-] (1.5,0)--(1.5,1) node[above] {$3$};
\draw[-,postaction={decorate,decoration={markings,mark=at position 0.55 with {\arrow{<<}}}}] (2.5,0)--(2.5,2) node[pos=0.5,left] {$k_5$};
\draw[-,postaction={decorate,decoration={markings,mark=at position 0.55 with {\arrow{<<}}}}] (2.5,2)--+(150:1) node[left] {$4$};
\draw[-] (2.5,2)--+(30:1) node[right] {$5$};
\draw[-] (3.5,0)--(3.5,1) node[above] {$1$};
\draw[-] (4.5,0)--(4.5,1) node[above] {$8$};
\draw[-] (5.5,0)--(5.5,1) node[above] {$6$};
\node[right] at (6.5,0) {$7$};
\node at (2,-0.5) {$k_1$};
\node at (3,-0.5) {$k_2$};
\node at (4,-0.5) {$k_3$};
\node at (5,-0.5) {$k_4$};
\end{scope}
\begin{scope}[xshift=1cm,yshift=-5cm]
\node at (-0.5,0) {$=$};
\node[left] at (0.5,0) {$1$};
\draw[-,postaction={decorate,decoration={markings,mark=at position 0.6cm with {\arrow{>}},mark=at position 1.6cm with {\arrow{>}},mark=at position 2.6cm with {\arrow{>}},mark=at position 3.6cm with {\arrow{>}},mark=at position 4.6cm with {\arrow{>}},mark=at position 5.6cm with {\arrow{>}}}}] (0.5,0)--(6.5,0);
\draw[-] (1.5,0)--(1.5,1) node[above] {$2$};
\draw[-] (2.5,0)--(2.5,1) node[above] {$3$};
\draw[-,postaction={decorate,decoration={markings,mark=at position 0.55 with {\arrow{<<}}}}] (3.5,0)--(3.5,2) node[pos=0.5,left] {$k_5$};
\draw[-,postaction={decorate,decoration={markings,mark=at position 0.55 with {\arrow{<<}}}}] (3.5,2)--+(150:1) node[left] {$4$};
\draw[-] (3.5,2)--+(30:1) node[right] {$5$};
\draw[-] (4.5,0)--(4.5,1) node[above] {$6$};
\draw[-] (5.5,0)--(5.5,1) node[above] {$7$};
\node[right] at (6.5,0) {$8$};
\node at (2,-0.5) {$k_1$};
\node at (3,-0.5) {$k_2$};
\node at (4,-0.5) {$k_3$};
\node at (5,-0.5) {$k_4$};
\end{scope}
\begin{scope}[xshift=1cm,yshift=-8cm]
\node at (-0.5,0) {$=$};
\node[left] at (0.5,0) {$1$};
\draw[-,postaction={decorate,decoration={markings,mark=at position 0.6cm with {\arrow{>}},mark=at position 1.6cm with {\arrow{>}},mark=at position 2.6cm with {\arrow{>}},mark=at position 3.6cm with {\arrow{>}},mark=at position 4.6cm with {\arrow{>}},mark=at position 5.6cm with {\arrow{>}},mark=at position 6.6cm with {\arrow{>}}}}] (0.5,0)--(7.5,0);
\draw[-] (1.5,0)--(1.5,1) node[above] {$2$};
\draw[-] (2.5,0)--(2.5,1) node[above] {$3$};
\draw[-] (3.5,0)--(3.5,1) node[above] {$4$};
\draw[-] (4.5,0)--(4.5,1) node[above] {$5$};
\draw[-] (5.5,0)--(5.5,1) node[above] {$6$};
\draw[-] (6.5,0)--(6.5,1) node[above] {$7$};
\node[right] at (7.5,0) {$8$};
\node at (2,-0.5) {$k_1$};
\node at (3,-0.5) {$k_2$};
\node at (4,-0.5) {$k_3$};
\node at (5,-0.5) {$k_4$};
\node at (6,-0.5) {$k_5$};
\end{scope}
\end{tikzpicture}
}
\caption{Eight-point conformal bootstrap equations.}
\label{FigEx8pt}
\end{figure}
As a final example, we consider eight-point correlation functions for which there are four different topologies.  The three independent eight-point conformal bootstrap equations are shown in Figure \ref{FigEx8pt}.  They translate to
\eqna{
&\vev{\vp{i_8}{8}\{\{\vp{i_6}{6}\vp{i_5}{5}\}\vp{i_4}{4}\vp{i_3}{3}\}\vp{i_2}{2}\vp{i_1}{1}|\vp{i_7}{7}}\\
&\qquad=(-1)^{F_{i_1i_2}+F_{i_1i_3}+F_{i_1i_4}+F_{i_1i_5}+F_{i_6i_8}}\vev{\vp{i_6}{6}\vp{i_8}{8}\vp{i_1}{1}\{\vp{i_5}{5}\vp{i_4}{4}\}\vp{i_3}{3}\vp{i_2}{2}|\vp{i_7}{7}}\\
&\qquad=(-1)^{F_{i_1i_7}+F_{i_2i_7}+F_{i_3i_7}+F_{i_4i_7}+F_{i_5i_7}+F_{i_6i_7}+F_{i_1i_8}+F_{i_2i_8}+F_{i_3i_8}+F_{i_4i_8}+F_{i_5i_8}+F_{i_6i_8}+F_{i_7i_8}}\\
&\qquad\phantom{=}\times\vev{\vp{i_7}{7}\vp{i_6}{6}\{\vp{i_5}{5}\vp{i_4}{4}\}\vp{i_3}{3}\vp{i_2}{2}\vp{i_1}{1}|\vp{i_8}{8}}\\
&\qquad=(-1)^{F_{i_1i_7}+F_{i_2i_7}+F_{i_3i_7}+F_{i_4i_7}+F_{i_5i_7}+F_{i_6i_7}+F_{i_1i_8}+F_{i_2i_8}+F_{i_3i_8}+F_{i_4i_8}+F_{i_5i_8}+F_{i_6i_8}+F_{i_7i_8}}\\
&\qquad\phantom{=}\times\vev{\vp{i_7}{7}\vp{i_6}{6}\vp{i_5}{5}\vp{i_4}{4}\vp{i_3}{3}\vp{i_2}{2}\vp{i_1}{1}|\vp{i_8}{8}},
}[EqBoot8]
where fermion crossings are responsible for the overall minus signs that occur in two-dimensional CFTs.  Here, the three independent sets of eight-point conformal bootstrap equations are obtained by equating the first line with the second, the third, and the fourth lines of \eqref{EqBoot8}.  Obviously, they imply the remaining pairings.

For the correlation functions $\vev{\vp{i_8}{8}\{\{\vp{i_6}{6}\vp{i_5}{5}\}\vp{i_4}{4}\vp{i_3}{3}\}\vp{i_2}{2}\vp{i_1}{1}|\vp{i_7}{7}}$ representing the most symmetric eight-point topology, the conformal partial waves are
\eqn{
\begin{gathered}
f_8=(-1)^{-h_{k_1}-h_{k_2}-h_{k_4}}\cOPE{i_2i_1}{k_1}\cOPE{k_3k_1}{k_2}\cOPE{i_4i_3}{k_4}\cOPE{k_5k_4}{k_3}\cOPE{i_6i_5}{k_5}c_{i_8k_2i_7},\\
L_8=z_{13;2}^{h_{i_2}}z_{32;1}^{h_{i_1}}z_{35;4}^{h_{i_4}}z_{54;3}^{h_{i_3}}z_{53;6}^{h_{i_6}}z_{36;5}^{h_{i_5}}z_{17;8}^{h_{i_8}}z_{81;7}^{h_{i_7}},\\
\eta_1^8=\eta_{12;38},\qquad\eta_2^8=\eta_{13;87},\qquad\eta_3^8=\eta_{35;18},\qquad\eta_4^8=\eta_{34;51},\qquad\eta_5^8=\eta_{56;31},
\end{gathered}
}[Eq8pt1]
with
\eqna{
C_8F_8&=\frac{(h_{i_2}-h_{i_1}+h_{k_1})_{n_1}(h_{i_7}-h_{i_8}+h_{k_2})_{n_2}(h_{i_4}-h_{i_3}+h_{k_4})_{n_4}(h_{i_6}-h_{i_5}+h_{k_5})_{n_5}}{(2h_{k_1})_{n_1}(2h_{k_2})_{n_2}(2h_{k_3})_{n_3}(2h_{k_4})_{n_4}(2h_{k_5})_{n_5}}\\
&\phantom{=}\qquad\times(h_{k_3}-h_{k_1}-n_1+h_{k_2})_{n_3+n_2}(-h_{k_3}+h_{k_1}+h_{k_2})_{n_1}\\
&\phantom{=}\qquad\times{}_3F_2\left[\begin{array}{c}-n_3,-n_1,1-2h_{k_1}-n_1\\h_{k_3}-h_{k_1}-n_1+h_{k_2},1+h_{k_3}-h_{k_1}-n_1-h_{k_2}\end{array};1\right]\\
&\phantom{=}\qquad\times(h_{k_5}-h_{k_4}-n_4+h_{k_3})_{n_5+n_3}(-h_{k_5}+h_{k_4}+h_{k_3})_{n_4}\\
&\phantom{=}\qquad\times{}_3F_2\left[\begin{array}{c}-n_5,-n_4,1-2h_{k_4}-n_4\\h_{k_5}-h_{k_4}-n_4+h_{k_3},1+h_{k_5}-h_{k_4}-n_4-h_{k_3}\end{array};1\right].
}

In the case of $\vev{\vp{i_6}{6}\vp{i_8}{8}\vp{i_1}{1}\{\vp{i_5}{5}\vp{i_4}{4}\}\vp{i_3}{3}\vp{i_2}{2}|\vp{i_7}{7}}$, we have instead
\eqn{
\begin{gathered}
f_8=(-1)^{-h_{k_1}-h_{k_2}-h_{k_3}-h_{k_4}}\cOPE{i_3i_2}{k_1}\cOPE{k_5k_1}{k_2}\cOPE{i_1k_2}{k_3}\cOPE{i_5i_4}{k_5}\cOPE{i_8k_3}{k_4}c_{i_6k_4i_7},\\
L_8=z_{24;3}^{h_{i_3}}z_{43;2}^{h_{i_2}}z_{42;5}^{h_{i_5}}z_{25;4}^{h_{i_4}}z_{28;1}^{h_{i_1}}z_{26;8}^{h_{i_8}}z_{27;6}^{h_{i_6}}z_{62;7}^{h_{i_7}},\\
\eta_1^8=\eta_{23;41},\qquad\eta_2^8=\eta_{24;18},\qquad\eta_3^8=\eta_{21;86},\qquad\eta_4^8=\eta_{28;67},\qquad\eta_5^8=\eta_{45;21},
\end{gathered}
}[Eq8pt2]
with
\eqna{
C_8F_8&=\frac{(h_{i_3}-h_{i_2}+h_{k_1})_{n_1}(h_{i_7}-h_{i_6}+h_{k_4})_{n_4}(h_{i_5}-h_{i_4}+h_{k_5})_{n_5}}{(2h_{k_1})_{n_1}(2h_{k_2})_{n_2}(2h_{k_3})_{n_3}(2h_{k_4})_{n_4}(2h_{k_5})_{n_5}}\\
&\phantom{=}\qquad\times(h_{i_1}-h_{k_2}-n_2+h_{k_3})_{n_3}(-h_{i_1}+h_{k_2}+h_{k_3})_{n_2}\\
&\phantom{=}\qquad\times(h_{i_8}-h_{k_3}-n_3+h_{k_4})_{n_4}(-h_{i_8}+h_{k_3}+h_{k_4})_{n_3}\\
&\phantom{=}\qquad\times(h_{k_5}-h_{k_1}-n_1+h_{k_2})_{n_5+n_2}(-h_{k_5}+h_{k_1}+h_{k_2})_{n_1}\\
&\phantom{=}\qquad\times{}_3F_2\left[\begin{array}{c}-n_5,-n_1,1-2h_{k_1}-n_1\\h_{k_5}-h_{k_1}-n_1+h_{k_2},1+h_{k_5}-h_{k_1}-n_1-h_{k_2}\end{array};1\right].
}

For $\vev{\vp{i_7}{7}\vp{i_6}{6}\{\vp{i_5}{5}\vp{i_4}{4}\}\vp{i_3}{3}\vp{i_2}{2}\vp{i_1}{1}|\vp{i_8}{8}}$, the conformal partial waves are
\eqn{
\begin{gathered}
f_8=(-1)^{-h_{k_1}-h_{k_2}-h_{k_3}-h_{k_4}}\cOPE{i_2i_1}{k_1}\cOPE{i_3k_1}{k_2}\cOPE{k_5k_2}{k_3}\cOPE{i_5i_4}{k_5}\cOPE{i_6k_3}{k_4}c_{i_7k_4i_8},\\
L_8=z_{13;2}^{h_{i_2}}z_{32;1}^{h_{i_1}}z_{14;3}^{h_{i_3}}z_{41;5}^{h_{i_5}}z_{15;4}^{h_{i_4}}z_{17;6}^{h_{i_6}}z_{18;7}^{h_{i_7}}z_{71;8}^{h_{i_8}},\\
\eta_1^8=\eta_{12;34},\qquad\eta_2^8=\eta_{13;46},\qquad\eta_3^8=\eta_{14;67},\qquad\eta_4^8=\eta_{16;78},\qquad\eta_5^8=\eta_{45;16},
\end{gathered}
}[Eq8pt3]
with
\eqna{
C_8F_8&=\frac{(h_{i_2}-h_{i_1}+h_{k_1})_{n_1}(h_{i_8}-h_{i_7}+h_{k_4})_{n_4}(h_{i_5}-h_{i_4}+h_{k_5})_{n_5}}{(2h_{k_1})_{n_1}(2h_{k_2})_{n_2}(2h_{k_3})_{n_3}(2h_{k_4})_{n_4}(2h_{k_5})_{n_5}}\\
&\phantom{=}\qquad\times(h_{i_3}-h_{k_1}-n_1+h_{k_2})_{n_2}(-h_{i_3}+h_{k_1}+h_{k_2})_{n_1}\\
&\phantom{=}\qquad\times(h_{i_6}-h_{k_3}-n_3+h_{k_4})_{n_4}(-h_{i_6}+h_{k_3}+h_{k_4})_{n_3}\\
&\phantom{=}\qquad\times(h_{k_5}-h_{k_2}-n_2+h_{k_3})_{n_5+n_3}(-h_{k_5}+h_{k_2}+h_{k_3})_{n_2}\\
&\phantom{=}\qquad\times{}_3F_2\left[\begin{array}{c}-n_5,-n_2,1-2h_{k_2}-n_2\\h_{k_5}-h_{k_2}-n_2+h_{k_3},1+h_{k_5}-h_{k_2}-n_2-h_{k_3}\end{array};1\right].
}

Finally, for $\vev{\vp{i_7}{7}\vp{i_6}{6}\vp{i_5}{5}\vp{i_4}{4}\vp{i_3}{3}\vp{i_2}{2}\vp{i_1}{1}|\vp{i_8}{8}}$, we obtain the conformal partial waves for the comb topology as
\eqn{
\begin{gathered}
f_8=(-1)^{-h_{k_1}-h_{k_2}-h_{k_3}-h_{k_4}-h_{k_5}}\cOPE{i_2i_1}{k_1}\cOPE{i_3k_1}{k_2}\cOPE{i_4k_2}{k_3}\cOPE{i_5k_3}{k_4}\cOPE{i_6k_4}{k_5}c_{i_7k_5i_8},\\
L_8=z_{13;2}^{h_{i_2}}z_{32;1}^{h_{i_1}}z_{14;3}^{h_{i_3}}z_{15;4}^{h_{i_4}}z_{16;5}^{h_{i_5}}z_{17;6}^{h_{i_6}}z_{18;7}^{h_{i_7}}z_{71;8}^{h_{i_8}},\\
\eta_1^8=\eta_{12;34},\qquad\eta_2^8=\eta_{13;45},\qquad\eta_3^8=\eta_{14;56},\qquad\eta_4^8=\eta_{15;67},\qquad\eta_5^8=\eta_{16;78},
\end{gathered}
}[Eq8pt4]
with
\eqna{
C_8F_8&=\frac{(h_{i_2}-h_{i_1}+h_{k_1})_{n_1}(h_{i_8}-h_{i_7}+h_{k_5})_{n_5}}{(2h_{k_1})_{n_1}(2h_{k_2})_{n_2}(2h_{k_3})_{n_3}(2h_{k_4})_{n_4}(2h_{k_5})_{n_5}}\\
&\phantom{=}\qquad\times(h_{i_3}-h_{k_1}-n_1+h_{k_2})_{n_2}(-h_{i_3}+h_{k_1}+h_{k_2})_{n_1}\\
&\phantom{=}\qquad\times(h_{i_4}-h_{k_2}-n_2+h_{k_3})_{n_3}(-h_{i_4}+h_{k_2}+h_{k_3})_{n_2}\\
&\phantom{=}\qquad\times(h_{i_5}-h_{k_3}-n_3+h_{k_4})_{n_4}(-h_{i_5}+h_{k_3}+h_{k_4})_{n_3}\\
&\phantom{=}\qquad\times(h_{i_6}-h_{k_4}-n_4+h_{k_5})_{n_5}(-h_{i_6}+h_{k_4}+h_{k_5})_{n_4},
}
as expected \cite{Rosenhaus:2018zqn}.

Therefore, equating the conformal partial wave decompositions \eqref{EqCPW} of eight-point correlation functions as in \eqref{EqBoot8}, using the conformal partial waves \eqref{Eq8pt1}, \eqref{Eq8pt2}, \eqref{Eq8pt3}, and \eqref{Eq8pt4}, gives rise to the complete eight-point conformal bootstrap.


\section{Discussion and Conclusion}\label{SecConc}

In this paper, we developed and proved a complete set of rules for global one- and two-dimensional higher-point conformal partial waves in arbitrary topology.  We proved the rules based on the known position space operator product expansion by determining its action on products of powers of position space distances.  The methods used to obtain these rules have been known for a long time, but have not been applied to $M$-point functions.  With our results, all quantities appearing in correlation functions that are determined by conformal invariance can be written explicitly.  Hence, with the CFT data, \textit{i.e.}\ the spectrum of quasi-primary operators with their dimensions $h$ and $\hb$ as well as the OPE coefficients, it is straightforward to compute any global $M$-point correlation function.

The rules that we introduced in this paper apply for a fixed choice of OPE limits.  The generalization of the rules to higher-dimensional conformal field theories, including the extra conformal cross-ratios, for scalar conformal blocks with any choice of OPE limits will be presented in a forthcoming publication \cite{Fortin:2020aaa}.

Moreover, now that the global conformal blocks are determined, it would be of interest to investigate if local higher-point conformal blocks could be computed following the usual method used for four-point Virasoro blocks.  Also, from the AdS/CFT-correspondence, higher-point conformal blocks could perhaps be useful in the study of bulk AdS${}_3$.


\ack{
The authors would like to thank Sarah Hoback, Sarthak Parikh, and Valentina Prilepina for useful discussions.  The work of JFF is supported by NSERC.  WJM is supported by the China Scholarship Council and in part by NSERC.  The work of WS is supported in part by DOE HEP grant DE-SC00-17660.
}


\setcounter{section}{0}
\renewcommand{\thesection}{\Alph{section}}

\section{Proof of the Rules}\label{SAppProof}

In this appendix, we provide the proof of the complete set of rules for arbitrary higher-point correlation functions in one- and two-dimensional CFTs.  Due to the factorization property of the OPE in $2d$ CFTs, the proof is presented for $1d$ CFTs without loss of generality.  The proof is constructive: we first build the initial comb structure and then we add extra comb structures following one of the three possible patterns discussed below.  At each step, we verify that the built structure satisfies the rules, completing the proof.

In the proof, we rely on standard hypergeometric identities like
\eqn{
\begin{gathered}
{}_2F_1\left[\begin{array}{c}-n,b\\c\end{array};1\right]=\frac{(c-b)_n}{(c)_n},\\
{}_3F_2\left[\begin{array}{c}-n,b,c\\d,1+b+c-d-n\end{array};1\right]=\frac{(d-b)_n(d-c)_n}{(d)_n(d-b-c)_n},
\end{gathered}
}[EqpFq]
(for $n$ a non-negative integer) to eliminate superfluous sums as well as the binomial identity
\eqn{\left(\frac{z_{jb}}{z_{ja}}\right)^n=\sum_{s\geq0}\frac{(-1)^s(-n)_s}{s!}\left(\frac{z_{ab}}{z_{ja}}\right)^s,}[EqId]
to introduce the proper conformal cross-ratios.

Moreover, to simplify the notation, we always reshuffle the position space coordinates such that the OPE is performed as in \eqref{EqOPE1d}.  We also omit most subscripts and superscripts.


\subsection{Initial Comb}

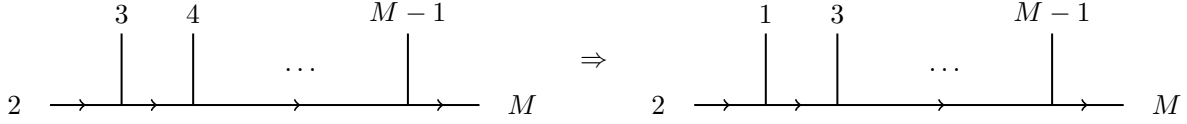
\begin{figure}[t]
\centering
\resizebox{16cm}{!}{%
\begin{tikzpicture}[thick]
\begin{scope}
\node at (0,0) {$2$};
\draw[-,postaction={decorate,decoration={markings,mark=at position 0.5cm with {\arrow{>}},mark=at position 1.5cm with {\arrow{>}},mark=at position 3.5cm with {\arrow{>}},mark=at position 5.5cm with {\arrow{>}}}}] (0.5,0)--(6.5,0);
\node at (7.1,0) {$M$};
\draw[-] (1.5,0)--(1.5,1) node[above] {$3$};
\draw[-] (2.5,0)--(2.5,1) node[above] {$4$};
\node at (4,0.5) {$\ldots$};
\draw[-] (5.5,0)--(5.5,1) node[above] {$M-1$};
\node at (8.1,0.6) {$\Rightarrow$};
\end{scope}
\begin{scope}[xshift=9cm]
\node at (0,0) {$2$};
\draw[-,postaction={decorate,decoration={markings,mark=at position 0.5cm with {\arrow{>}},mark=at position 1.5cm with {\arrow{>}},mark=at position 3.5cm with {\arrow{>}},mark=at position 5.5cm with {\arrow{>}}}}] (0.5,0)--(6.5,0);
\node at (7.1,0) {$M$};
\draw[-] (1.5,0)--(1.5,1) node[above] {$1$};
\draw[-] (2.5,0)--(2.5,1) node[above] {$3$};
\node at (4,0.5) {$\ldots$};
\draw[-] (5.5,0)--(5.5,1) node[above] {$M-1$};
\end{scope}
\end{tikzpicture}
}
\caption{Proof by induction for the initial comb structure with $M-1$ points.  The arrows dictate the flow of position space coordinates in the topologies, \textit{i.e.}\ the choice of OPE limits, following our convention.}
\label{FigComb}
\end{figure}

First, it is straightforward to check that the four-point conformal partial waves satisfy the rules of Section \ref{SecCF}.  Therefore, we assume that the initial $(M-1)$-point comb structure satisfies the rules, then we generate the $M$-point comb structure applying the OPE to finally verify that it also satisfies the rules, as depicted in Figure \ref{FigComb}.  As a consequence of this computation, the comb structure with our choice of OPE vertices satisfies our rules.

Thus, we assume that the $(M-1)$-point conformal partial wave
\eqn{W_{M-1(h_{k_2},\ldots,h_{k_{M-3}})}^{(h_{k_1},h_{i_3},\ldots,h_{i_{M-1}})}=L_{M-1}\left[\prod_{2\leq a\leq M-3}(\eta_a^{M-1})^{h_{k_a}}\right]G_{M-1},}
satisfies our rules, \textit{i.e.}\ with the leg \eqref{EqL} expressed as
\eqn{L_{M-1}=z_{43;2}^{h_{k_1}}z_{M-1,2;M}^{h_{i_M}}\prod_{3\leq a\leq M-1}z_{2,a+1;a}^{h_{i_a}},}
the conformal cross-ratios \eqref{Eqeta} given by
\eqn{\eta_a^{M-1}=\eta_{2,a+1;a+2,a+3}\qquad2\leq a\leq M-3,}
and the conformal block written as
\eqna{
G_{M-1}&=\sum_{\{n_a\}\geq0}\frac{(h_{i_3}-h_{k_1}+h_{k_2})_{n_2}(h_{i_M}-h_{i_{M-1}}+h_{k_{M-3}})_{n_{M-3}}}{(2h_{k_2})_{n_2}}\\
&\phantom{=}\qquad\times\prod_{2\leq a\leq M-4}\frac{(h_{i_{a+2}}-h_{k_a}-n_a+h_{k_{a+1}})_{n_{a+1}}(-h_{i_{a+2}}+h_{k_a}+h_{k_{a+1}})_{n_a}}{(2h_{k_{a+1}})_{n_{a+1}}}\prod_{2\leq a\leq M-3}\frac{(\eta_a^M)^{n_a}}{n_a!},
}
according to the rules of Section \ref{SecCF}.

Acting with the OPE \eqref{EqOPE1d}, we obtain, after extracting $(-1)^{h_{k_1}}$ from the rule for the OPE coefficients and using \eqref{EqIb},
\eqna{
W_{M(h_{k_1},\ldots,h_{k_{M-3}})}^{(h_{i_1},\ldots,h_{i_M})}&=\frac{(-1)^{h_{k_1}}}{z_{12}^{h_{i_1}+h_{i_2}-h_{k_1}}}{}_1F_1(h_{i_1}-h_{i_2}+h_{k_1},2h_{k_1};z_{12}\partial_2)W_{M-1(h_{k_2},\ldots,h_{k_{M-3}})}^{(h_{k_1},h_{i_3},\ldots,h_{i_{M-1}})}\\
&=L_M\left[\prod_{1\leq a\leq M-3}(\eta_a^M)^{h_{k_a}}\right]\left(\frac{z_{23}}{z_{13}}\right)^{-h_{i_1}-h_{k_1}+h_{i_2}}\\
&\phantom{=}\qquad\times\sum_{\{m_a\}\geq0}\frac{(-1)^{\bar{m}}(h_{i_1}-h_{i_2}+h_{k_1})_{\bar{m}}}{(2h_{k_1})_{\bar{m}}z_{12}^{-\bar{m}}}\left[\prod_{3\leq a\leq M}\frac{(p_a)_{m_a}}{m_a!z_{2a}^{m_a}}\right]G_{M-1},
}
where the proper $M$-point leg is
\eqn{L_M=z_{23;1}^{h_{i_1}}z_{31;2}^{h_{i_2}}z_{M-1,2;M}^{h_{i_M}}\prod_{3\leq a\leq M-1}z_{2,a+1;a}^{h_{i_a}},}
the proper $M$-point conformal cross-ratios are
\eqn{\eta_1^M=\eta_{21;34},\qquad\eta_a^M=\eta_{2,a+1;a+2,a+3}\qquad2\leq a\leq M-3,}
and the different powers are
\eqn{
\begin{gathered}
p_3=h_{k_1}+h_{i_3}-h_{k_2}-n_2,\\
p_4=-h_{i_3}+h_{i_4}+h_{k_1}-h_{k_3}-n_3,\\
p_a=-h_{a-1}+h_{i_a}+h_{k_{a-3}}+n_{a-3}-h_{k_{a-1}}-n_{a-1}\qquad5\leq a\leq M-2,\\
p_{M-1}=-h_{i_{M-2}}+h_{i_{M-1}}+h_{k_{M-4}}+n_{M-4}-h_{i_M},\\
p_M=-h_{i_{M-1}}+h_{i_M}+h_{k_{M-3}}+n_{M-3}.
\end{gathered}
}[Eqpcomb]
Thus, isolating the $M$-point conformal block in the $M$-point conformal partial wave above, we simply need to verify that
\eqn{G_M=\left(\frac{z_{23}}{z_{13}}\right)^{-h_{i_1}-h_{k_1}+h_{i_2}}\sum_{\{m_a\}\geq0}\frac{(-1)^{\bar{m}}(h_{i_1}+h_{k_1}-h_{i_2})_{\bar{m}}}{(2h_{k_1})_{\bar{m}}z_{12}^{-\bar{m}}}\left[\prod_{3\leq a\leq M}\frac{(p_a)_{m_a}}{m_a!z_{2a}^{m_a}}\right]G_{M-1},}[EqGcomb]
satisfies the appropriate rules.  We note that in \eqref{EqGcomb} and throughout, the sums over $\{n_a\}$ appearing in $G_{M-1}$ must be understood as being performed in the summation symbol with the $\{m_a\}$.  This is evident from the powers \eqref{Eqpcomb} which depend explicitly on $\{n_a\}$.  This is done only to simplify the notation and should be clear from the context.

To proceed, we first extract $z_{23}$ from the product in \eqref{EqGcomb} as
\eqna{
G_M&=\sum_{\{m_a\}\geq0}\frac{(-1)^{\bar{m}}(h_{i_1}+h_{k_1}-h_{i_2})_{\bar{m}}(p_3)_{\bar{m}-\sum_{4\leq a\leq M}m_a}}{(2h_{k_1})_{\bar{m}}(\bar{m}-\sum_{4\leq a\leq M}m_a)!}\left(\frac{z_{23}}{z_{13}}\right)^{-h_{i_1}-h_{k_1}+h_{i_2}-\bar{m}}\\
&\phantom{=}\qquad\times\left(\frac{z_{12}}{z_{13}}\right)^{\bar{m}}\left[\prod_{4\leq a\leq M}\frac{(p_a)_{m_a}}{m_a!}\frac{z_{23}^{m_a}}{z_{2a}^{m_a}}\right]G_{M-1},
}
and rewrite the power of $z_{23}/z_{13}$ following \eqref{EqId} to reach
\eqn{G_M=\sum_{\{m_a,s\}\geq0}\frac{(-1)^{\bar{m}}(h_{i_1}+h_{k_1}-h_{i_2})_{\bar{m}+s}(p_3)_{\bar{m}-\sum_{4\leq a\leq M}m_a}}{(2h_{k_1})_{\bar{m}}(\bar{m}-\sum_{4\leq a\leq M}m_a)!s!}\left(\frac{z_{12}}{z_{13}}\right)^{\bar{m}+s}\left[\prod_{4\leq a\leq M}\frac{(p_a)_{m_a}}{m_a!}\frac{z_{23}^{m_a}}{z_{2a}^{m_a}}\right]G_{M-1}.}
We then rename $s=n_1-\bar{m}$ and re-sum over $m_3$ with the help of the first identity in \eqref{EqpFq} to get
\eqna{
G_M&=\sum_{\{n_1,m_a\}\geq0}\frac{(-1)^{\sum_{4\leq a\leq M}m_a}(h_{i_1}+h_{k_1}-h_{i_2})_{n_1}(2h_{k_1}-p_3+\sum_{4\leq a\leq M}m_a)_{n_1-\sum_{4\leq a\leq M}m_a}}{(2h_{k_1})_{n_1}(n_1-\sum_{4\leq a\leq M}m_a)!}\\
&\phantom{=}\qquad\times\left(\frac{z_{12}}{z_{13}}\right)^{n_1}\left[\prod_{4\leq a\leq M}\frac{(p_a)_{m_a}}{m_a!}\frac{z_{23}^{m_a}}{z_{2a}^{m_a}}\right]G_{M-1}.
}
Using \eqref{EqId} for all ratios of conformal cross-ratios appearing in the product, we obtain
\eqna{
G_M&=\sum_{\{n_1,m_a,s_a\}\geq0}\frac{(-1)^{\sum_{4\leq a\leq M}m_a}(h_{i_1}+h_{k_1}-h_{i_2})_{n_1}(2h_{k_1}-p_3+\sum_{4\leq a\leq M}m_a)_{n_1-\sum_{4\leq a\leq M}m_a}}{(2h_{k_1})_{n_1}(n_1-\sum_{4\leq a\leq M}m_a)!}\\
&\phantom{=}\qquad\times\left(\frac{z_{12}}{z_{13}}\right)^{n_1}\left[\prod_{4\leq a\leq M}\frac{(p_a)_{m_a}}{(m_a-s_a)!s_a!}\frac{z_{a3}^{s_a}}{z_{2a}^{s_a}}\right]G_{M-1},
}
where we now change summation indices from $m_a$ to $m_a+s_a$ to evaluate all the sums over $m_a$ [again using the first identity in \eqref{EqpFq}], leading to
\eqna{
G_M&=\sum_{\{n_1,s_a\}\geq0}\frac{(-1)^{\sum_{4\leq a\leq M}s_a}(h_{i_1}+h_{k_1}-h_{i_2})_{n_1}(2h_{k_1}-\sum_{3\leq a\leq M}p_a)_{n_1-\sum_{4\leq a\leq M}s_a}}{(2h_{k_1})_{n_1}(n_1-\sum_{4\leq a\leq M}s_a)!}\\
&\phantom{=}\qquad\times\left(\frac{z_{12}}{z_{13}}\right)^{n_1}\left[\prod_{4\leq a\leq M}\frac{(p_a)_{s_a}}{s_a!}\frac{z_{a3}^{s_a}}{z_{2a}^{s_a}}\right]G_{M-1}.
}
From the definitions of the powers \eqref{Eqpcomb}, we see that
\eqn{\sum_{3\leq a\leq M}p_a=2h_{k_1},}
hence the Pochhammer symbol $(0)_{n_1-\sum_{4\leq a\leq M}s_a}$ forces $\sum_{4\leq a\leq M}s_a=n_1$ and we can fix $s_4=n_1-\sum_{5\leq a\leq M}s_a$ to reach
\eqna{
G_M&=\sum_{\{n_1,s_a\}\geq0}\frac{(-1)^{n_1}(h_{i_1}+h_{k_1}-h_{i_2})_{n_1}}{(2h_{k_1})_{n_1}}\left(\frac{z_{12}}{z_{13}}\right)^{n_1}\left[\prod_{4\leq a\leq M}\frac{(p_a)_{s_a}}{s_a!}\frac{z_{a3}^{s_a}}{z_{2a}^{s_a}}\right]G_{M-1}\\
&=\sum_{\{n_1,s_a\}\geq0}\frac{(h_{i_1}+h_{k_1}-h_{i_2})_{n_1}(p_4)_{n_1-\sum_{5\leq a\leq M}s_a}}{(2h_{k_1})_{n_1}(n_1-\sum_{5\leq a\leq M}s_a)!}\\
&\phantom{=}\qquad\times\left(\frac{z_{12}z_{43}}{z_{13}z_{42}}\right)^{n_1}\left[\prod_{5\leq a\leq M}\frac{(p_a)_{s_a}}{s_a!}\left(\frac{z_{24}z_{3a}}{z_{2a}z_{34}}\right)^{s_a}\right]G_{M-1}.
}
At this stage, we observe the appearance of the new conformal cross-ratio $\eta_1^M=\eta_{21;34}$.

To generate the remaining conformal cross-ratios, we use the fact that
\eqn{\frac{z_{24}z_{3a}}{z_{2a}z_{34}}=\eta_{24,3a}=1+\sum_{5\leq b\leq a}(-1)^b\prod_{2\leq c\leq b-3}\eta_c^M,}
and write
\eqna{
G_M&=\sum_{\{n_1,s_a\}\geq0}\frac{(h_{i_1}+h_{k_1}-h_{i_2})_{n_1}(p_4)_{n_1-\sum_{5\leq a\leq M}s_a}}{(2h_{k_1})_{n_1}(n_1-\sum_{5\leq a\leq M}s_a)!}(\eta_1^M)^{n_1}G_{M-1}\\
&\phantom{=}\qquad\times\prod_{5\leq a\leq M}\frac{(p_a)_{s_a}}{s_a!}\left[1+\sum_{5\leq b\leq a}(-1)^b\prod_{2\leq c\leq b-3}\eta_c^M\right]^{s_a}\\
&=\sum_{\{n_1,t_{ab}\}\geq0}\frac{(h_{i_1}+h_{k_1}-h_{i_2})_{n_1}(p_4)_{n_1-\sum_{5\leq a\leq M}t_{a0}}}{(2h_{k_1})_{n_1}(n_1-\sum_{5\leq a\leq M}t_{a0})!}(\eta_1^M)^{n_1}\\
&\phantom{=}\qquad\times\left[\prod_{5\leq a\leq M}(-1)^{t_{a0}}(p_a)_{t_{a0}}(\eta_{a-3}^M)^{\sum_{a\leq b\leq M}t_{b,a-4}}\prod_{0\leq b\leq a-4}\frac{(-1)^{t_{ab}}}{(t_{ab}-t_{a,b+1})!}\right]G_{M-1},
}
where $t_{a0}=s_a$ and $t_{a,a-3}=0$.  In the last equality, we simply expanded using the binomial theorem repetitively, introducing in the process several sums with indices of summation $t_{ab}$.

We now evaluate the sums over $t_{a0}$ using the ${}_2F_1$ identity \eqref{EqpFq} after performing the change of variables $t_{a0}\to t_{a0}+t_{a1}$, leading to
\eqna{
G_M&=\sum_{\{n_1,t_{ab}\}\geq0}\frac{(h_{i_1}+h_{k_1}-h_{i_2})_{n_1}(h_{k_1}-h_{i_3}+h_{k_2}+n_2+\sum_{5\leq a\leq M}t_{a1})_{n_1-\sum_{5\leq a\leq M}t_{a1}}}{(2h_{k_1})_{n_1}(n_1-\sum_{5\leq a\leq M}t_{a1})!}(\eta_1^M)^{n_1}\\
&\phantom{=}\qquad\times\left[\prod_{5\leq a\leq M}(p_a)_{t_{a1}}(\eta_{a-3}^M)^{\sum_{a\leq b\leq M}t_{b,a-4}}\prod_{1\leq b\leq a-4}\frac{(-1)^{t_{ab}}}{(t_{ab}-t_{a,b+1})!}\right]G_{M-1},
}
where we used
\eqn{\sum_{4\leq a\leq M}p_a=h_{k_1}-h_{i_3}+h_{k_2}+n_2,}
from the definitions \eqref{Eqpcomb}.

By defining $t_{51}=t_2-\sum_{6\leq a\leq M}t_{a1}$, we get
\eqna{
G_M&=\sum_{\{n_1,t_2,t_{ab}\}\geq0}\frac{(h_{i_1}+h_{k_1}-h_{i_2})_{n_1}(h_{k_1}-h_{i_3}+h_{k_2}+n_2+t_2)_{n_1-t_2}}{(2h_{k_1})_{n_1}(n_1-t_2)!}\frac{(-1)^{t_2}(p_5)_{t_2-\sum_{6\leq a\leq M}t_{a1}}}{(t_2-\sum_{6\leq a\leq M}t_{a1})!}\\
&\phantom{=}\qquad\times(\eta_1^M)^{n_1}(\eta_2^M)^{t_2}\left[\prod_{6\leq a\leq M}\frac{(p_a)_{t_{a1}}(\eta_{a-3}^M)^{\sum_{a\leq b\leq M}t_{b,a-4}}}{(t_{a1}-t_{a2})!}\prod_{2\leq b\leq a-4}\frac{(-1)^{t_{ab}}}{(t_{ab}-t_{a,b+1})!}\right]G_{M-1},
}
which allows us to evaluate the sums over $t_{a1}$ after completing the change of variables $t_{a1}\to t_{a1}+t_{a2}$ [with the first identity in \eqref{EqpFq}], implying
\eqna{
G_M&=\sum_{\{n_1,t_2,t_{ab}\}\geq0}\frac{(h_{i_1}+h_{k_1}-h_{i_2})_{n_1}(h_{k_1}-h_{i_3}+h_{k_2}+n_2+t_2)_{n_1-t_2}}{(2h_{k_1})_{n_1}(n_1-t_2)!}\\
&\phantom{=}\qquad\times\frac{(-1)^{t_2}(h_{k_2}-h_{i_4}+h_{k_3}+n_2+n_3+\sum_{6\leq a\leq M}t_{a2})_{t_2-\sum_{6\leq a\leq M}t_{a2}}}{(t_2-\sum_{6\leq a\leq M}t_{a2})!}\\
&\phantom{=}\qquad\times(\eta_1^M)^{n_1}(\eta_2^M)^{t_2}\left[\prod_{6\leq a\leq M}(p_a)_{t_{a2}}(\eta_{a-3}^M)^{\sum_{a\leq b\leq M}t_{b,a-4}}\prod_{2\leq b\leq a-4}\frac{(-1)^{t_{ab}}}{(t_{ab}-t_{a,b+1})!}\right]G_{M-1},
}
where we replaced
\eqn{\sum_{5\leq a\leq M}p_a=h_{k_2}-h_{i_4}+h_{k_3}+n_2+n_3,}
using the definitions of the powers \eqref{Eqpcomb}.

Defining $t_a=\sum_{a+3\leq b\leq M}t_{b,a-1}$ and repeating the previous procedure, it is straightforward to evaluate all the remaining sums over $t_{ab}$ apart from $t_{M,M-4}=t_{M-3}$, leading to
\eqna{
G_M&=\sum_{\{n_1,t_a\}\geq0}\frac{(h_{i_1}+h_{k_1}-h_{i_2})_{n_1}(h_{k_1}-h_{i_3}+h_{k_2}+n_2+t_2)_{n_1-t_2}}{(2h_{k_1})_{n_1}(n_1-t_2)!}(\eta_1^M)^{n_1}\\
&\phantom{=}\qquad\times\left[\prod_{2\leq a\leq M-4}\frac{(-1)^{t_a}(h_{k_a}-h_{i_{a+2}}+h_{k_{a+1}}+n_a+n_{a+1}+t_{a+1})_{t_a-t_{a+1}}}{(t_a-t_{a+1})!}(\eta_a^M)^{t_a}\right]\\
&\phantom{=}\qquad\times\frac{(-1)^{t_{M-3}}(h_{i_M}+h_{k_{M-3}}-h_{i_{M-1}}+n_{M-3})_{t_{M-3}}}{t_{M-3}!}(\eta_{M-3}^M)^{t_{M-3}}G_{M-1}.
}
Using the explicit definition of $G_{M-1}$, we obtain
\eqna{
G_M&=\sum_{\{n_a,t_a\}\geq0}\frac{(h_{i_1}+h_{k_1}-h_{i_2})_{n_1}(h_{k_1}-h_{i_3}+h_{k_2}+n_2+t_2)_{n_1-t_2}(h_{i_3}-h_{k_1}+h_{k_2})_{n_2}}{(2h_{k_1})_{n_1}(2h_{k_2})_{n_2}(n_1-t_2)!}\\
&\phantom{=}\qquad\times\frac{(-1)^{t_{M-3}}(h_{i_M}+h_{k_{M-3}}-h_{i_{M-1}})_{n_{M-3}+t_{M-3}}}{t_{M-3}!}(\eta_1^M)^{n_1}\\
&\phantom{=}\qquad\times\prod_{2\leq a\leq M-4}\frac{(-1)^{t_a}(h_{k_a}-h_{i_{a+2}}+h_{k_{a+1}}+n_a+n_{a+1}+t_{a+1})_{t_a-t_{a+1}}}{(t_a-t_{a+1})!}\\
&\phantom{=}\qquad\times\prod_{2\leq a\leq M-4}\frac{(h_{i_{a+2}}-h_{k_a}-n_a+h_{k_{a+1}})_{n_{a+1}}(-h_{i_{a+2}}+h_{k_a}+h_{k_{a+1}})_{n_a}}{(2h_{k_{a+1}})_{n_{a+1}}}\prod_{2\leq a\leq M-3}\frac{(\eta_a^M)^{n_a+t_a}}{n_a!},
}
where we can transform $n_a\to n_a-t_a$ for $a\geq2$ and evaluate the sums over $t_a$ starting from $t_{M-3}$, using this time the second identity in \eqref{EqpFq}, to reach
\eqna{
G_M&=\sum_{\{n_a\}\geq0}\frac{(h_{i_1}-h_{i_2}+h_{k_1})_{n_1}(h_{i_M}-h_{i_{M-1}}+h_{k_{M-3}})_{n_{M-3}}}{(2h_{k_1})_{n_1}}\\
&\phantom{=}\qquad\times\prod_{1\leq a\leq M-4}\frac{(h_{i_{a+2}}-h_{k_a}-n_a+h_{k_{a+1}})_{n_{a+1}}(-h_{i_{a+2}}+h_{k_a}+h_{k_{a+1}})_{n_a}}{(2h_{k_{a+1}})_{n_{a+1}}}\prod_{1\leq a\leq M-3}\frac{(\eta_a^M)^{n_a}}{n_a!},
}
which verifies the rules.

Hence, the comb structure satisfies the rules introduced in Section \ref{SecCF}.  It will now serve as the initial comb structure on which we will append extra comb structures to construct the full topology.

Before proceeding, we note first that one can provide a simpler proof of the rules for the comb topology by starting from the result of \cite{Rosenhaus:2018zqn}, which can also be proven easily by recurrence from the OPE, by changing variables to the conformal cross-ratios used here,=and by re-summing the additional sums.


\subsection{Extra Combs}

At this point---now that the initial comb structure has been shown to satisfy the rules for any number of points---we need to prove that the rules are correct when extra comb structures are added to the initial comb.  To do so, we assume that the rules are satisfied for some arbitrary topology and add one OPE vertex as boundary condition for the extra comb structure.  After the rules are shown to be correct for the extra comb structure with only one OPE vertex, we once again assume that the rules are valid for an extra comb structure with $q-1$ OPE vertices attached to the arbitrary topology and use the OPE to generate an additional OPE vertex to the extra comb structure.  We finally verify that the rules are consistent for the arbitrary topology to which an extra comb structure with $q$ OPE vertices is glued.  This procedure thus proves the rules for any topology by induction.

To properly add an extra comb structure to an arbitrary topology, it is necessary to separate the possible topologies into three different types.  The types, illustrated in Figure \ref{FigType}, change according to where the extra comb structure is attached, with type $n$ implying the extra comb is glued to a $nI$ OPE vertex.
\begin{figure}[!t]
\centering
\resizebox{11cm}{!}{%
\begin{tikzpicture}[thick]
\begin{scope}
\draw[-,postaction={decorate,decoration={markings,mark=at position 0.5 with {\arrow{<}}}}] (0,0)++(90:1)--+(30:1);
\draw[-] (0,0)++(90:1)--+(150:1) node[above left]{$2$};
\draw[-,postaction={decorate,decoration={markings,mark=at position 0.5 with {\arrow{<}}}}] (0,0)--+(90:1);
\draw[-] (0,0)--+(-150:0.5) node[circle,draw,below left,minimum size=24pt]{$\beta_0$};
\draw[-,postaction={decorate,decoration={markings,mark=at position 0.65 with {\arrow{>}}}}] (0,0)--+(-30:0.5) node[circle,draw,below right,minimum size=24pt]{$\alpha_0$};
\draw[-,postaction={decorate,decoration={markings,mark=at position 0.5 with {\arrow{<}}}}] (0,0)++(90:1)++(30:1)--+(90:0.5) node[above]{$\alpha_1$};
\draw[-] (0,0)++(90:1)++(30:1)--+(-30:0.5) node[below right]{$\alpha_2$};
\node at (3.5,1) {$\Rightarrow$};
\node at (3.5,-2) {Type 1};
\end{scope}
\begin{scope}[xshift=7cm]
\draw[-,postaction={decorate,decoration={markings,mark=at position 0.5 with {\arrow{<}}}}] (0,0)++(90:1)--+(30:1);
\draw[-,postaction={decorate,decoration={markings,mark=at position 0.32cm with {\arrow{<<}},mark=at position 1.32cm with {\arrow{<<}},mark=at position 2.32cm with {\arrow{<<}}markings,mark=at position 2.82cm with {\arrow{<<}}}}] (0,0)++(90:1)--+(150:3) node[above left]{$2$};
\draw[-] (0,0)++(90:1)++(150:2.5)--+(60:0.5) node[above right]{$1$};
\draw[-] (0,0)++(90:1)++(150:2.0)--+(60:0.5) node[above right]{$\gamma_3$};
\node[rotate=-30] at (-0.9,1.9) {$\cdots$};
\draw[-] (0,0)++(90:1)++(150:0.5)--+(60:0.5) node[above right]{$\gamma_{q-1}$};
\draw[-,postaction={decorate,decoration={markings,mark=at position 0.5 with {\arrow{<}}}}] (0,0)--+(90:1);
\draw[-] (0,0)--+(-150:0.5) node[circle,draw,below left,minimum size=24pt]{$\beta_0$};
\draw[-,postaction={decorate,decoration={markings,mark=at position 0.65 with {\arrow{>}}}}] (0,0)--+(-30:0.5) node[circle,draw,below right,minimum size=24pt]{$\alpha_0$};
\draw[-,postaction={decorate,decoration={markings,mark=at position 0.5 with {\arrow{<}}}}] (0,0)++(90:1)++(30:1)--+(90:0.5) node[above]{$\alpha_1$};
\draw[-] (0,0)++(90:1)++(30:1)--+(-30:0.5) node[below right]{$\alpha_2$};
\end{scope}
\begin{scope}[yshift=-7cm]
\draw[-,postaction={decorate,decoration={markings,mark=at position 0.5 with {\arrow{<}}}}] (0,0)++(90:1)--+(30:1);
\draw[-] (0,0)++(90:1)--+(150:1) node[above left]{$2$};
\draw[-,postaction={decorate,decoration={markings,mark=at position 0.5 with {\arrow{<}}}}] (0,0)--+(90:1);
\draw[-] (0,0)--+(-150:0.5) node[circle,draw,below left,minimum size=24pt]{$\beta_0$};
\draw[-,postaction={decorate,decoration={markings,mark=at position 0.65 with {\arrow{>}}}}] (0,0)--+(-30:0.5) node[circle,draw,below right,minimum size=24pt]{$\alpha_0$};
\draw[-,postaction={decorate,decoration={markings,mark=at position 0.5 with {\arrow{<}}}}] (0,0)++(90:1)++(30:1)--+(90:1);
\draw[-,postaction={decorate,decoration={markings,mark=at position 0.5 with {\arrow{<}}}}] (0,0)++(90:1)++(30:1)++(90:1)--+(150:0.5) node[circle,draw,above left,minimum size=24pt]{$\alpha_1$};
\draw[-] (0,0)++(90:1)++(30:1)++(90:1)--+(30:0.5) node[circle,draw,above right,minimum size=24pt]{$\beta_1$};
\draw[-] (0,0)++(90:1)++(30:1)--+(-30:0.5) node[below right]{$\alpha_2$};
\node at (3.5,1) {$\Rightarrow$};
\node at (3.5,-2) {Type 2};
\end{scope}
\begin{scope}[xshift=7cm,yshift=-7cm]
\draw[-,postaction={decorate,decoration={markings,mark=at position 0.5 with {\arrow{<}}}}] (0,0)++(90:1)--+(30:1);
\draw[-,postaction={decorate,decoration={markings,mark=at position 0.32cm with {\arrow{<<}},mark=at position 1.32cm with {\arrow{<<}},mark=at position 2.32cm with {\arrow{<<}}markings,mark=at position 2.82cm with {\arrow{<<}}}}] (0,0)++(90:1)--+(150:3) node[above left]{$2$};
\draw[-] (0,0)++(90:1)++(150:2.5)--+(60:0.5) node[above right]{$1$};
\draw[-] (0,0)++(90:1)++(150:2.0)--+(60:0.5) node[above right]{$\gamma_3$};
\node[rotate=-30] at (-0.9,1.9) {$\cdots$};
\draw[-] (0,0)++(90:1)++(150:0.5)--+(60:0.5) node[above right]{$\gamma_{q-1}$};
\draw[-,postaction={decorate,decoration={markings,mark=at position 0.5 with {\arrow{<}}}}] (0,0)--+(90:1);
\draw[-] (0,0)--+(-150:0.5) node[circle,draw,below left,minimum size=24pt]{$\beta_0$};
\draw[-,postaction={decorate,decoration={markings,mark=at position 0.65 with {\arrow{>}}}}] (0,0)--+(-30:0.5) node[circle,draw,below right,minimum size=24pt]{$\alpha_0$};
\draw[-,postaction={decorate,decoration={markings,mark=at position 0.5 with {\arrow{<}}}}] (0,0)++(90:1)++(30:1)--+(90:1);
\draw[-,postaction={decorate,decoration={markings,mark=at position 0.5 with {\arrow{<}}}}] (0,0)++(90:1)++(30:1)++(90:1)--+(150:0.5) node[circle,draw,above left,minimum size=24pt]{$\alpha_1$};
\draw[-] (0,0)++(90:1)++(30:1)++(90:1)--+(30:0.5) node[circle,draw,above right,minimum size=24pt]{$\beta_1$};
\draw[-] (0,0)++(90:1)++(30:1)--+(-30:0.5) node[below right]{$\alpha_2$};
\end{scope}
\begin{scope}[yshift=-14cm]
\draw[-,postaction={decorate,decoration={markings,mark=at position 0.5 with {\arrow{<}}}}] (0,0)++(90:1)--+(30:1);
\draw[-] (0,0)++(90:1)--+(150:1) node[above left]{$2$};
\draw[-,postaction={decorate,decoration={markings,mark=at position 0.5 with {\arrow{<}}}}] (0,0)--+(90:1);
\draw[-] (0,0)--+(-150:0.5) node[circle,draw,below left,minimum size=24pt]{$\beta_0$};
\draw[-,postaction={decorate,decoration={markings,mark=at position 0.65 with {\arrow{>}}}}] (0,0)--+(-30:0.5) node[circle,draw,below right,minimum size=24pt]{$\alpha_0$};
\draw[-,postaction={decorate,decoration={markings,mark=at position 0.5 with {\arrow{<}}}}] (0,0)++(90:1)++(30:1)--+(90:1);
\draw[-,postaction={decorate,decoration={markings,mark=at position 0.5 with {\arrow{<}}}}] (0,0)++(90:1)++(30:1)++(90:1)--+(150:0.5) node[circle,draw,above left,minimum size=24pt]{$\alpha_1$};
\draw[-] (0,0)++(90:1)++(30:1)++(90:1)--+(30:0.5) node[circle,draw,above right,minimum size=24pt]{$\beta_1$};
\draw[-] (0,0)++(90:1)++(30:1)--+(-30:1);
\draw[-] (0,0)++(90:1)++(30:1)++(-30:1)--+(30:0.5) node[circle,draw,above right,minimum size=24pt]{$\alpha_2$};
\draw[-] (0,0)++(90:1)++(30:1)++(-30:1)--+(-90:0.5) node[circle,draw,below,minimum size=24pt]{$\beta_2$};
\node at (3.5,1) {$\Rightarrow$};
\node at (3.5,-2) {Type 3};
\end{scope}
\begin{scope}[xshift=7cm,yshift=-14cm]
\draw[-,postaction={decorate,decoration={markings,mark=at position 0.5 with {\arrow{<}}}}] (0,0)++(90:1)--+(30:1);
\draw[-,postaction={decorate,decoration={markings,mark=at position 0.32cm with {\arrow{<<}},mark=at position 1.32cm with {\arrow{<<}},mark=at position 2.32cm with {\arrow{<<}}markings,mark=at position 2.82cm with {\arrow{<<}}}}] (0,0)++(90:1)--+(150:3) node[above left]{$2$};
\draw[-] (0,0)++(90:1)++(150:2.5)--+(60:0.5) node[above right]{$1$};
\draw[-] (0,0)++(90:1)++(150:2.0)--+(60:0.5) node[above right]{$\gamma_3$};
\node[rotate=-30] at (-0.9,1.9) {$\cdots$};
\draw[-] (0,0)++(90:1)++(150:0.5)--+(60:0.5) node[above right]{$\gamma_{q-1}$};
\draw[-,postaction={decorate,decoration={markings,mark=at position 0.5 with {\arrow{<}}}}] (0,0)--+(90:1);
\draw[-] (0,0)--+(-150:0.5) node[circle,draw,below left,minimum size=24pt]{$\beta_0$};
\draw[-,postaction={decorate,decoration={markings,mark=at position 0.65 with {\arrow{>}}}}] (0,0)--+(-30:0.5) node[circle,draw,below right,minimum size=24pt]{$\alpha_0$};
\draw[-,postaction={decorate,decoration={markings,mark=at position 0.5 with {\arrow{<}}}}] (0,0)++(90:1)++(30:1)--+(90:1);
\draw[-,postaction={decorate,decoration={markings,mark=at position 0.5 with {\arrow{<}}}}] (0,0)++(90:1)++(30:1)++(90:1)--+(150:0.5) node[circle,draw,above left,minimum size=24pt]{$\alpha_1$};
\draw[-] (0,0)++(90:1)++(30:1)++(90:1)--+(30:0.5) node[circle,draw,above right,minimum size=24pt]{$\beta_1$};
\draw[-] (0,0)++(90:1)++(30:1)--+(-30:1);
\draw[-] (0,0)++(90:1)++(30:1)++(-30:1)--+(30:0.5) node[circle,draw,above right,minimum size=24pt]{$\alpha_2$};
\draw[-] (0,0)++(90:1)++(30:1)++(-30:1)--+(-90:0.5) node[circle,draw,below,minimum size=24pt]{$\beta_2$};
\end{scope}
\end{tikzpicture}
}
\caption{Types of arbitrary topologies on which an extra comb structure can be glued.  The blobs represent arbitrary substructures while the arrows dictate the flow of position space coordinates in the topologies, \textit{i.e.}\ the choice of OPE limits, following our convention.}
\label{FigType}
\end{figure}
We note that the blobs represent any substructures in the initial arbitrary topology (with the parameters representing position space coordinates) while the arrows show the comb structure of interest (\textit{i.e.}\ the OPE limits) in the arbitrary topology to which the extra comb structure is glued.  This particularity allows us to determine the leg factor and the conformal cross-ratios that carry the position space coordinate (chosen without loss of generality to be $z_2$) relevant to the OPE differential operator.


\subsubsection{Type 1: Boundary Condition}

We first assume that the $(M-1)$-point conformal partial wave
\eqn{W_{M-1(h_{k_2},\ldots,h_{k_{M-3}})}^{(h_{k_1},h_{i_3},\ldots,h_{i_M})}=L_{M-1}\left[\prod_{2\leq a\leq M-3}(\eta_a^{M-1})^{h_{k_a}}\right]G_{M-1},}
satisfies the rules.  Therefore, the conformal block is given by \eqref{EqG} and is of the form
\eqn{G_{M-1}=\sum_{\{n_a\}\geq0}C_{M-1}F_{M-1}\frac{(\eta_a^{M-1})^{n_a}}{n_a!},}
with the proper factors (originating from the rules) associated to the arbitrary topology of Figure \ref{FigType}.  With our convention for the OPE limits, the only $z_2$-dependent quantities in the conformal partial wave are the leg factors and conformal cross-ratios
\eqn{
\begin{gathered}
L_{M-1}=z_{\alpha_1\beta_0;2}^{h_{k_1}}z_{2\alpha_2;\alpha_1}^{h_{i_{\alpha_1}}}z_{\alpha_12;\alpha_2}^{h_{i_{\alpha_2}}}\bar{L}_{M-1},\\
\eta_2^{M-1}=\eta_{\alpha_1\alpha_2;2\beta_0},\qquad\eta_3^{M-1}=\eta_{\alpha_12;\beta_0\alpha_0},
\end{gathered}
}
where $\bar{L}_{M-1}$ represents the remaining leg contributions.  Hence it is straightforward to act with the OPE once \eqref{EqOPE1d} using \eqref{EqIb} to generate
\eqna{
W_{M(h_{k_1},\ldots,h_{k_{M-3}})}^{(h_{i_1},\ldots,h_{i_M})}&=\frac{1}{z_{12}^{h_{i_1}+h_{i_2}-h_{k_1}}}{}_1F_1(h_{i_1}+h_{k_1}-h_{i_2},2h_{k_1};z_{12}\partial_2)W_{M-1(h_{k_2},\ldots,h_{k_{M-3}})}^{(h_{k_1},h_{i_3},\ldots,h_{i_{M-1}})}\\
&=L_M\left[\prod_{1\leq a\leq M-3}(\eta_a^{M})^{h_{k_a}}\right]\left(\frac{z_{2\alpha_1}}{z_{1\alpha_1}}\right)^{-h_{i_1}-h_{k_1}+h_{i_2}}\\
&\phantom{=}\qquad\times\sum_{\{n_a\},n,m_0,m_1\geq0}C_{M-1}F_{M-1}\left[\prod_{2\leq a\leq M-3}\frac{(\eta_a^M)^{n_a}}{n_a!}\right]\frac{z_{12}^n}{z_{2\alpha_1}^{n-m_0-m_1}z_{2\alpha_2}^{m_1}z_{2\beta_0}^{m_0}}\\
&\phantom{=}\qquad\times\frac{(-1)^n(h_{i_1}+h_{k_1}-h_{i_2})_n(h_{k_1}+h_{i_{\alpha_1}}-h_{i_{\alpha_2}}-h_{k_3}-n_3)_{n-m_0-m_1}}{(2h_{k_1})_n(n-m_0-m_1)!}\\
&\phantom{=}\qquad\times\frac{(h_{i_{\alpha_2}}-h_{i_{\alpha_1}}+h_{k_2}+n_2)_{m_1}(h_{k_1}-h_{k_2}+h_{k_3}-n_2+n_3)_{m_0}}{m_1!m_0!},
}
where the proper leg and cross-ratios are
\eqn{
\begin{gathered}
L_M=z_{2\alpha_1;1}^{h_{i_1}}z_{\alpha_11;2}^{h_{i_2}}z_{2\alpha_2;\alpha_1}^{h_{i_{\alpha_1}}}z_{\alpha_12;\alpha_2}^{h_{i_{\alpha_2}}}\bar{L}_{M-1},\\
\eta_1^M=\eta_{21;\alpha_1\beta_0},\qquad\eta_a^M=\eta_a^{M-1}\qquad2\leq a\leq M-3.
\end{gathered}
}

Therefore, the $M$-point conformal block is given by
\eqna{
G_M&=\left(\frac{z_{2\alpha_1}}{z_{1\alpha_1}}\right)^{-h_{i_1}-h_{k_1}+h_{i_2}}\sum_{\{n_a\},n,m_0,m_1\geq0}C_{M-1}F_{M-1}\left[\prod_{2\leq a\leq M-3}\frac{(\eta_a^M)^{n_a}}{n_a!}\right]\frac{z_{12}^n}{z_{2\alpha_1}^{n-m_0-m_1}z_{2\alpha_2}^{m_1}z_{2\beta_0}^{m_0}}\\
&\phantom{=}\qquad\times\frac{(-1)^n(h_{i_1}+h_{k_1}-h_{i_2})_n(h_{k_1}+h_{i_{\alpha_1}}-h_{i_{\alpha_2}}-h_{k_3}-n_3)_{n-m_0-m_1}}{(2h_{k_1})_n(n-m_0-m_1)!}\\
&\phantom{=}\qquad\times\frac{(h_{i_{\alpha_2}}-h_{i_{\alpha_1}}+h_{k_2}+n_2)_{m_1}(h_{k_1}-h_{k_2}+h_{k_3}-n_2+n_3)_{m_0}}{m_1!m_0!}\\
&=\sum_{\{n_a\},n,n_1,m_0,m_1\geq0}C_{M-1}F_{M-1}\left[\prod_{2\leq a\leq M-3}\frac{(\eta_a^M)^{n_a}}{n_a!}\right]\left(\frac{z_{12}}{z_{1\alpha_1}}\right)^{n_1}\frac{z_{2\alpha_1}^{m_0+m_1}}{z_{2\alpha_2}^{m_1}z_{2\beta_0}^{m_0}}\\
&\phantom{=}\qquad\times\frac{(-1)^n(h_{i_1}+h_{k_1}-h_{i_2})_{n_1}(h_{k_1}+h_{i_{\alpha_1}}-h_{i_{\alpha_2}}-h_{k_3}-n_3)_{n-m_0-m_1}}{(2h_{k_1})_n(n_1-n)!(n-m_0-m_1)!}\\
&\phantom{=}\qquad\times\frac{(h_{i_{\alpha_2}}-h_{i_{\alpha_1}}+h_{k_2}+n_2)_{m_1}(h_{k_1}-h_{k_2}+h_{k_3}-n_2+n_3)_{m_0}}{m_1!m_0!},
}[EqGtype1]
where we used \eqref{EqId} for $(z_{2\alpha_1}/z_{1\alpha_1})^{-h_{i_1}-h_{k_1}+h_{i_2}-n}$ and we shifted the new index of summation in the second equality.  We must now prove that the $M$-point conformal block \eqref{EqGtype1} satisfies our rules by evaluating all superfluous sums.

We first redefine $n\to n+m_0+m_1$ and sum over $n$ using the ${}_2F_1$ identity \eqref{EqpFq} to reach
\eqna{
G_M&=\sum_{\{n_a\},m_0,m_1\geq0}C_{M-1}F_{M-1}\left[\prod_{2\leq a\leq M-3}\frac{(\eta_a^M)^{n_a}}{n_a!}\right]\left(\frac{z_{12}}{z_{1\alpha_1}}\right)^{n_1}\left(\frac{z_{2\alpha_1}}{z_{2\alpha_2}}\right)^{m_1}\left(\frac{z_{2\alpha_1}}{z_{2\beta_0}}\right)^{m_0}\\
&\phantom{=}\qquad\times\frac{(-1)^{m_0+m_1}(h_{i_1}+h_{k_1}-h_{i_2})_{n_1}(h_{k_1}-h_{i_{\alpha_1}}+h_{i_{\alpha_2}}+h_{k_3}+n_3+m_0+m_1)_{n_1-m_0-m_1}}{(2h_{k_1})_{n_1}(n_1-m_0-m_1)!}\\
&\phantom{=}\qquad\times\frac{(h_{i_{\alpha_2}}-h_{i_{\alpha_1}}+h_{k_2}+n_2)_{m_1}(h_{k_1}-h_{k_2}+h_{k_3}-n_2+n_3)_{m_0}}{m_1!m_0!}.
}
Using \eqref{EqId} for $(z_{2\alpha_1}/z_{2\alpha_2})^{m_1}$ and $(z_{2\alpha_1}/z_{2\beta_0})^{m_0}$, we find that
\eqna{
G_M&=\sum_{\{n_a\},m_0,m_1,s_0,s_1\geq0}C_{M-1}F_{M-1}\left[\prod_{2\leq a\leq M-3}\frac{(\eta_a^M)^{n_a}}{n_a!}\right]\left(\frac{z_{12}}{z_{1\alpha_1}}\right)^{n_1}\left(\frac{z_{\alpha_2\alpha_1}}{z_{2\alpha_2}}\right)^{s_1}\left(\frac{z_{\beta_0\alpha_1}}{z_{2\beta_0}}\right)^{s_0}\\
&\phantom{=}\qquad\times\frac{(-1)^{m_0+m_1}(h_{i_1}+h_{k_1}-h_{i_2})_{n_1}(h_{k_1}-h_{i_{\alpha_1}}+h_{i_{\alpha_2}}+h_{k_3}+n_3+m_0+m_1)_{n_1-m_0-m_1}}{(2h_{k_1})_{n_1}(n_1-m_0-m_1)!}\\
&\phantom{=}\qquad\times\frac{(h_{i_{\alpha_2}}-h_{i_{\alpha_1}}+h_{k_2}+n_2)_{m_1}(h_{k_1}-h_{k_2}+h_{k_3}-n_2+n_3)_{m_0}}{(m_1-s_1)!s_1!(m_0-s_0)!s_0!}.
}
We can now rename the indices of summation $m_0\to m_0+s_0$ and $m_1\to m_1+s_1$ and perform the sums over $m_0$ and $m_1$ with the help of the first identity in \eqref{EqpFq}, leading to
\eqna{
G_M&=\sum_{\{n_a\},s_0,s_1\geq0}C_{M-1}F_{M-1}\left[\prod_{2\leq a\leq M-3}\frac{(\eta_a^M)^{n_a}}{n_a!}\right]\left(\frac{z_{12}}{z_{1\alpha_1}}\right)^{n_1}\left(\frac{z_{\alpha_2\alpha_1}}{z_{2\alpha_2}}\right)^{s_1}\left(\frac{z_{\beta_0\alpha_1}}{z_{2\beta_0}}\right)^{s_0}\\
&\phantom{=}\qquad\times\frac{(-1)^{s_0+s_1}(h_{i_1}+h_{k_1}-h_{i_2})_{n_1}(0)_{n_1-s_0-s_1}}{(2h_{k_1})_{n_1}(n_1-s_0-s_1)!}\\
&\phantom{=}\qquad\times\frac{(h_{i_{\alpha_2}}-h_{i_{\alpha_1}}+h_{k_2}+n_2)_{s_1}(h_{k_1}-h_{k_2}+h_{k_3}-n_2+n_3)_{n_1-s_1}}{s_1!s_0!}.
}
The Pochhammer symbol with vanishing argument forces $s_0=n_1-s_1$ which allows us to simplify the $M$-point conformal block \eqref{EqGtype1} to
\eqna{
G_M&=\sum_{\{n_a\},s_1\geq0}C_{M-1}F_{M-1}\left[\prod_{2\leq a\leq M-3}\frac{(\eta_a^M)^{n_a}}{n_a!}\right]\left(\frac{z_{12}z_{\beta_0\alpha_1}}{z_{1\alpha_1}z_{\beta_02}}\right)^{n_1}\left(\frac{z_{2\beta_0}z_{\alpha_1\alpha_2}}{z_{2\alpha_2}z_{\alpha_1\beta_0}}\right)^{s_1}\\
&\phantom{=}\qquad\times\frac{(h_{i_1}+h_{k_1}-h_{i_2})_{n_1}(h_{i_{\alpha_2}}-h_{i_{\alpha_1}}+h_{k_2}+n_2)_{s_1}(h_{k_1}-h_{k_2}+h_{k_3}-n_2+n_3)_{n_1-s_1}}{(2h_{k_1})_{n_1}s_1!(n_1-s_1)!}.
}
We finally see the conformal cross-ratios $\eta_1^M=\eta_{21;\alpha_1\beta_0}$ and $\eta_2^M=\eta_{2\beta_0;\alpha_1\alpha_2}$ appear.

Extracting the known part of the $(M-1)$-point conformal block of type $1$ following our rule, we have
\eqn{C_{M-1}=\frac{(h_{k_1}-h_{k_2}-n_2+h_{k_3})_{n_3}(-h_{k_1}+h_{k_2}+h_{k_3})_{n_2}(h_{i_{\alpha_2}}-h_{i_{\alpha_1}}+h_{k_2})_{n_2}}{(2h_{k_2})_{n_2}}\bar{C}_{M-1},}
where $\bar{C}_{M-1}$ is undetermined (it is defined by the arbitrary topology) and most importantly independent of $n_2$.  Hence, we can rewrite the $M$-point conformal block as
\eqna{
G_M&=\sum_{\{n_a\},s_1\geq0}\bar{C}_{M-1}F_{M-1}\left[\prod_{3\leq a\leq M-3}\frac{(\eta_a^M)^{n_a}}{n_a!}\right](\eta_1^M)^{n_1}\frac{(\eta_2^M)^{n_2+s_1}}{n_2!}\frac{(-h_{k_1}+h_{k_2}+h_{k_3})_{n_2}}{(2h_{k_2})_{n_2}}\\
&\phantom{=}\qquad\times\frac{(h_{i_1}+h_{k_1}-h_{i_2})_{n_1}(h_{i_{\alpha_2}}-h_{i_{\alpha_1}}+h_{k_2})_{n_2+s_1}(h_{k_1}-h_{k_2}-n_2+h_{k_3})_{n_1+n_3-s_1}}{(2h_{k_1})_{n_1}s_1!(n_1-s_1)!},
}
which is easy to re-sum after changing variables as $n_2\to n_2-s_1$, leading to
\eqna{
G_M&=\sum_{\{n_a\}\geq0}\bar{C}_{M-1}F_{M-1}\left[\prod_{1\leq a\leq M-3}\frac{(\eta_a^M)^{n_a}}{n_a!}\right]\frac{(h_{i_1}-h_{i_2}+h_{k_1})_{n_1}(h_{i_{\alpha_2}}-h_{i_{\alpha_1}}+h_{k_2})_{n_2}}{(2h_{k_1})_{n_1}}\\
&\phantom{=}\qquad\times\frac{(h_{k_1}-h_{k_2}-n_2+h_{k_3})_{n_1+n_3}(-h_{k_1}+h_{k_2}+h_{k_3})_{n_2}}{(2h_{k_2})_{n_2}}\\
&\phantom{=}\qquad\times{}_3F_2\left[\begin{array}{c}-n_1,-n_2,1-2h_{k_2}-n_2\\h_{k_1}-h_{k_2}-n_2+h_{k_3},1+h_{k_1}-h_{k_2}-n_2-h_{k_3}\end{array};1\right],
}
where the ${}_3F_2$ originates from the sum over $s_1$.  Comparing with Section \ref{SecCF}, we see that the boundary condition for the gluing of an extra comb structure for type $1$ topologies satisfies our rules.


\subsubsection{Type 1: Full Extra Comb}

Now that the boundary condition for an extra comb structure glued to an arbitrary topology of type $1$ has been verified to follow the rules, we are ready to generate a full comb structure.  Again, we proceed by induction, assuming that the $(q-1)$-point extra comb structure satisfies our rules, using the OPE to generate the $q$-point extra comb structure, and verifying that the resulting conformal block satisfies the rules of Section \ref{SecCF}.

From the rules, the only $z_2$-dependent quantities in the leg and conformal cross-ratios are
\eqn{L_{M-1}=z_{\gamma_4\gamma_3;2}^{h_{k_1}}\left[\prod_{3\leq a\leq q-1}z_{2\gamma_{a+1};\gamma_a}^{h_{i_{\gamma_a}}}\right]z_{2\alpha_2;\alpha_1}^{h_{i_{\alpha_1}}}z_{\alpha_12;\alpha_2}^{h_{i_{\alpha_2}}}\bar{L}_{M-1},}
and
\eqn{
\begin{gathered}
\eta_a^{M-1}=\eta_{2\gamma_{a+1};\gamma_{a+2}\gamma_{a+3}}\qquad2\leq a\leq q-3,\\
\eta_{q-2}^{M-1}=\eta_{2\gamma_{q-1};\alpha_1\beta_0},\qquad\eta_{q-1}^{M-1}=\eta_{\alpha_1\alpha_2;2\beta_0},\qquad\eta_q^{M-1}=\eta_{\alpha_12;\beta_0\alpha_0},
\end{gathered}
}
where $\bar{L}_{M-1}$ is fixed by the topology and we define $\gamma_q=\alpha_1$ for convenience.  Moreover, extracting once again the $n_2$-dependent part of the $(M-1)$-point conformal block, we have
\eqna{
G_{M-1}&=\sum_{\{n_a\}\geq0}\bar{C}_{M-1}\bar{F}_{M-1}\left[\prod_{2\leq a\leq M-3}\frac{(\eta_a^{M-1})^{n_a}}{n_a!}\right]\frac{(h_{i_{\gamma_3}}-h_{k_1}+h_{k_2})_{n_2}}{(2h_{k_2})_{n_2}}\frac{(h_{i_{\alpha_2}}-h_{i_{\alpha_1}}+h_{k_{q-1}})_{n_{q-1}}}{(2h_{k_{q-1}})_{n_{q-1}}}\\
&\phantom{=}\qquad\times\prod_{2\leq a\leq q-3}\frac{(h_{i_{\gamma_{a+2}}}-h_{k_a}-n_a+h_{k_{a+1}})_{n_{a+1}}(-h_{i_{\gamma_{a+2}}}+h_{k_a}+h_{k_{a+1}})_{n_a}}{(2h_{k_{a+1}})_{n_{a+1}}}\\
&\phantom{=}\qquad\times\frac{(h_{k_{q-2}}-h_{k_{q-1}}-n_{q-1}+h_{k_q})_{n_{q-2}+n_q}(-h_{k_{q-2}}+h_{k_{q-1}}+h_{k_q})_{n_{q-1}}}{(2h_{k_q})_{n_q}}\\
&\phantom{=}\qquad\times{}_3F_2\left[\begin{array}{c}-n_{q-2},-n_{q-1},1-2h_{k_{q-1}}-n_{q-1}\\h_{k_{q-2}}-h_{k_{q-1}}-n_{q-1}+h_{k_q},1+h_{k_{q-2}}-h_{k_{q-1}}-n_{q-1}-h_{k_q}\end{array};1\right],
}
following the same notation than in the previous section, with $\bar{C}_{M-1}$ and $\bar{F}_{M-1}$ having no dependence in $n_2$.

Acting with the OPE \eqref{EqOPE1d} using \eqref{EqIb}, we find that the $M$-point conformal block is
\eqna{
G_M&=\left(\frac{z_{2\gamma_3}}{z_{1\gamma_3}}\right)^{-h_{i_1}-h_{k_1}+h_{i_2}}\sum_{\{n_a,m_a\},n\geq0}\frac{(-1)^n(h_{i_1}+h_{k_1}-h_{i_2})_n(p_3)_{n-m_0-m_1-\sum_{4\leq a\leq q}m_a}}{(2h_{k_1})_n(n-m_0-m_1-\sum_{4\leq a\leq q}m_a)!}\\
&\phantom{=}\qquad\times\frac{z_{12}^n}{z_{2\gamma_3}^{n-m_0-m_1-\sum_{4\leq a\leq q}m_a}z_{2\beta_0}^{m_0}z_{2\alpha_2}^{m_1}}\frac{(p_0)_{m_0}(p_1)_{m_1}}{m_0!m_1!}\prod_{4\leq a\leq q}\frac{(p_a)_{m_a}}{m_a!z_{2\gamma_a}^{m_a}}\\
&\phantom{=}\qquad\times\bar{C}_{M-1}\bar{F}_{M-1}\left[\prod_{2\leq a\leq M-3}\frac{(\eta_a^{M-1})^{n_a}}{n_a!}\right]\frac{(h_{i_{\gamma_3}}-h_{k_1}+h_{k_2})_{n_2}}{(2h_{k_2})_{n_2}}\frac{(h_{i_{\alpha_2}}-h_{i_{\alpha_1}}+h_{k_{q-1}})_{n_{q-1}}}{(2h_{k_{q-1}})_{n_{q-1}}}\\
&\phantom{=}\qquad\times\prod_{2\leq a\leq q-3}\frac{(h_{i_{\gamma_{a+2}}}-h_{k_a}-n_a+h_{k_{a+1}})_{n_{a+1}}(-h_{i_{\gamma_{a+2}}}+h_{k_a}+h_{k_{a+1}})_{n_a}}{(2h_{k_{a+1}})_{n_{a+1}}}\\
&\phantom{=}\qquad\times\frac{(h_{k_{q-2}}-h_{k_{q-1}}-n_{q-1}+h_{k_q})_{n_{q-2}+n_q}(-h_{k_{q-2}}+h_{k_{q-1}}+h_{k_q})_{n_{q-1}}}{(2h_{k_q})_{n_q}}\\
&\phantom{=}\qquad\times{}_3F_2\left[\begin{array}{c}-n_{q-2},-n_{q-1},1-2h_{k_{q-1}}-n_{q-1}\\h_{k_{q-2}}-h_{k_{q-1}}-n_{q-1}+h_{k_q},1+h_{k_{q-2}}-h_{k_{q-1}}-n_{q-1}-h_{k_q}\end{array};1\right],
}
where
\eqn{
\begin{gathered}
p_0=-h_{k_{q-1}}+h_{k_{q-2}}+h_{k_q}-n_{q-1}+n_{q-2}+n_q,\\
p_1=-h_{i_{\alpha_1}}+h_{i_{\alpha_2}}+h_{k_{q-1}}+n_{q-1},\\
p_3=h_{k_1}+h_{i_{\gamma_3}}-h_{k_2}-n_2,\\
p_4=-h_{i_{\gamma_3}}+h_{i_{\gamma_4}}+h_{k_1}-h_{k_3}-n_3,\\
p_a=-h_{i_{\gamma_{a-1}}}+h_{i_{\gamma_a}}+h_{k_{a-3}}+n_{a-3}-h_{k_{a-1}}-n_{a-1}\qquad5\leq a\leq q-2,\\
p_{q-1}=-h_{i_{\gamma_{q-2}}}+h_{i_{\gamma_{q-1}}}+h_{k_{q-4}}+n_{q-4}-h_{k_{q-2}}-n_{q-2},\\
p_q=h_{i_{\alpha_1}}-h_{i_{\alpha_2}}-h_{i_{\gamma_{q-1}}}+h_{k_{q-3}}-h_{k_q}+n_{q-3}-n_q.
\end{gathered}
}[Eqptype1]
As a consequence of \eqref{EqId} for $(z_{2\gamma_3}/z_{12})^{-h_{i_1}-h_{k_1}+h_{i_2}-n}$, we obtain
\eqna{
G_M&=\sum_{\{n_a,m_a\},n\geq0}\frac{(-1)^n(h_{i_1}+h_{k_1}-h_{i_2})_{n_1}(p_3)_{n-m_0-m_1-\sum_{4\leq a\leq q}m_a}}{(2h_{k_1})_n(n-m_0-m_1-\sum_{4\leq a\leq q}m_a)!(n_1-n)!}\\
&\phantom{=}\qquad\times\left(\frac{z_{12}}{z_{1\gamma_3}}\right)^{n_1}\left(\frac{z_{2\gamma_3}}{z_{2\beta_0}}\right)^{m_0}\left(\frac{z_{2\gamma_3}}{z_{2\alpha_2}}\right)^{m_1}\frac{(p_0)_{m_0}(p_1)_{m_1}}{m_0!m_1!}\prod_{4\leq a\leq q}\frac{(p_a)_{m_a}}{m_a!}\frac{z_{2\gamma_3}^{m_a}}{z_{2\gamma_a}^{m_a}}\\
&\phantom{=}\qquad\times\bar{C}_{M-1}\bar{F}_{M-1}\left[\prod_{2\leq a\leq M-3}\frac{(\eta_a^{M-1})^{n_a}}{n_a!}\right]\frac{(h_{i_{\gamma_3}}-h_{k_1}+h_{k_2})_{n_2}}{(2h_{k_2})_{n_2}}\frac{(h_{i_{\alpha_2}}-h_{i_{\alpha_1}}+h_{k_{q-1}})_{n_{q-1}}}{(2h_{k_{q-1}})_{n_{q-1}}}\\
&\phantom{=}\qquad\times\prod_{2\leq a\leq q-3}\frac{(h_{i_{\gamma_{a+2}}}-h_{k_a}-n_a+h_{k_{a+1}})_{n_{a+1}}(-h_{i_{\gamma_{a+2}}}+h_{k_a}+h_{k_{a+1}})_{n_a}}{(2h_{k_{a+1}})_{n_{a+1}}}\\
&\phantom{=}\qquad\times\frac{(h_{k_{q-2}}-h_{k_{q-1}}-n_{q-1}+h_{k_q})_{n_{q-2}+n_q}(-h_{k_{q-2}}+h_{k_{q-1}}+h_{k_q})_{n_{q-1}}}{(2h_{k_q})_{n_q}}\\
&\phantom{=}\qquad\times{}_3F_2\left[\begin{array}{c}-n_{q-2},-n_{q-1},1-2h_{k_{q-1}}-n_{q-1}\\h_{k_{q-2}}-h_{k_{q-1}}-n_{q-1}+h_{k_q},1+h_{k_{q-2}}-h_{k_{q-1}}-n_{q-1}-h_{k_q}\end{array};1\right],
}[EqGtype1comb]
after shifting the new index of summation.  Once again, we simply need to evaluate the superfluous sums to verify that the $M$-point conformal block \eqref{EqGtype1comb} satisfies the rules.  To simplify the notation, in the following the last four lines of \eqref{EqGtype1comb} will be denoted by $H_{M-1}$ and $\sum_{4\leq a\leq q}m_a=\bar{m}$.

We first implement the change of summation index $n\to n+m_0+m_1+\bar{m}$ and evaluate the sum over $n$ with the help of the first identity in \eqref{EqpFq} to reach
\eqna{
G_M&=\sum_{\{n_a,m_a\}\geq0}\frac{(-1)^{m_0+m_1+\bar{m}}(h_{i_1}+h_{k_1}-h_{i_2})_{n_1}(h_{k_1}-h_{i_{\gamma_3}}+h_{k_2}+n_2+m_0+m_1+\bar{m})_{n_1-m_0-m_1-\bar{m}}}{(2h_{k_1})_{n_1}(n_1-m_0-m_1-\bar{m})!}\\
&\phantom{=}\qquad\times\left(\frac{z_{12}}{z_{1\gamma_3}}\right)^{n_1}\left(\frac{z_{2\gamma_3}}{z_{2\beta_0}}\right)^{m_0}\left(\frac{z_{2\gamma_3}}{z_{2\alpha_2}}\right)^{m_1}\frac{(p_0)_{m_0}(p_1)_{m_1}}{m_0!m_1!}\left[\prod_{4\leq a\leq q}\frac{(p_a)_{m_a}}{m_a!}\frac{z_{2\gamma_3}^{m_a}}{z_{2\gamma_a}^{m_a}}\right]H_{M-1}.
}
We then use \eqref{EqId} for all factors of $(z_{2\gamma_3}/z_{2x_a})^{m_a}$ with $x\in\{\alpha,\beta,\gamma\}$ to eliminate all factors of $z_{2\gamma_3}$ and obtain
\eqna{
G_M&=\sum_{\{n_a,m_a,s_a\}\geq0}\frac{(-1)^{m_0+m_1+\bar{m}}(h_{i_1}+h_{k_1}-h_{i_2})_{n_1}(h_{k_1}-h_{i_{\gamma_3}}+h_{k_2}+n_2+m_0+m_1+\bar{m})_{n_1-m_0-m_1-\bar{m}}}{(2h_{k_1})_{n_1}(n_1-m_0-m_1-\bar{m})!}\\
&\phantom{=}\qquad\times\left(\frac{z_{12}}{z_{1\gamma_3}}\right)^{n_1}\left(\frac{z_{\beta_0\gamma_3}}{z_{2\beta_0}}\right)^{s_0}\left(\frac{z_{\alpha_2\gamma_3}}{z_{2\alpha_2}}\right)^{s_1}\frac{(p_0)_{m_0}(p_1)_{m_1}}{(m_0-s_0)!s_0!(m_1-s_1)!s_1!}\left[\prod_{4\leq a\leq q}\frac{(p_a)_{m_a}}{(m_a-s_a)!s_a!}\frac{z_{\gamma_a\gamma_3}^{s_a}}{z_{2\gamma_a}^{s_a}}\right]H_{M-1}.
}
Shifting $m_a\to m_a+s_a$, we can evaluate the sums over $m_a$ using the first identity in \eqref{EqpFq} repetitively and rewrite the $M$-point conformal block \eqref{EqGtype1comb} as
\eqna{
G_M&=\sum_{\{n_a,s_a\}\geq0}\frac{(-1)^{s_0+s_1+\bar{s}}(h_{i_1}+h_{k_1}-h_{i_2})_{n_1}(0)_{n_1-s_0-s_1-\bar{s}}}{(2h_{k_1})_{n_1}(n_1-s_0-s_1-\bar{s})!}\\
&\phantom{=}\qquad\times\left(\frac{z_{12}}{z_{1\gamma_3}}\right)^{n_1}\left(\frac{z_{\beta_0\gamma_3}}{z_{2\beta_0}}\right)^{s_0}\left(\frac{z_{\alpha_2\gamma_3}}{z_{2\alpha_2}}\right)^{s_1}\frac{(p_0)_{s_0}(p_1)_{s_1}}{s_0!s_1!}\left[\prod_{4\leq a\leq q}\frac{(p_a)_{s_a}}{s_a!}\frac{z_{\gamma_a\gamma_3}^{s_a}}{z_{2\gamma_a}^{s_a}}\right]H_{M-1},
}
since
\eqn{p_0+p_1+\sum_{4\leq a\leq q}p_a=h_{k_1}-h_{i_{\gamma_3}}+h_{k_2}+n_2,}
from \eqref{Eqptype1}.

Setting $s_4=n_1-s_0-s_1-\sum_{5\leq a\leq q}s_a$ from the Pochhammer symbol with vanishing argument, we have
\eqna{
G_M&=\sum_{\{n_a,s_a\}\geq0}\frac{(h_{i_1}+h_{k_1}-h_{i_2})_{n_1}(p_4)_{n_1-s_0-s_1-\sum_{5\leq a\leq q}s_a}}{(2h_{k_1})_{n_1}(n_1-s_0-s_1-\sum_{5\leq a\leq q}s_a)!}(\eta_1^M)^{n_1}\\
&\phantom{=}\qquad\times\left(\frac{z_{2\gamma_4}z_{\gamma_3\beta_0}}{z_{2\beta_0}z_{\gamma_3\gamma_4}}\right)^{s_0}\left(\frac{z_{2\gamma_4}z_{\gamma_3\alpha_2}}{z_{2\alpha_2}z_{\gamma_3\gamma_4}}\right)^{s_1}\frac{(p_0)_{s_0}(p_1)_{s_1}}{s_0!s_1!}\left[\prod_{5\leq a\leq q}\frac{(p_a)_{s_a}}{s_a!}\left(\frac{z_{2\gamma_4}z_{\gamma_3\gamma_a}}{z_{2\gamma_a}z_{\gamma_3\gamma_4}}\right)^{s_a}\right]H_{M-1},
}
where we see the extra conformal cross-ratio $\eta_1^M=\eta_{21;\gamma_3\gamma_4}$ appear.  Since for $a\geq2$ the remaining conformal cross-ratios satisfy $\eta_a^M=\eta_a^{M-1}$, we have
\eqn{
\begin{gathered}
\frac{z_{2\gamma_4}z_{\gamma_3\beta_0}}{z_{2\beta_0}z_{\gamma_3\gamma_4}}=\eta_{2\gamma_4;\gamma_3\beta_0}=1+\sum_{5\leq b\leq q+1}(-1)^b\prod_{2\leq c\leq b-3}\eta_c^M,\\
\frac{z_{2\gamma_4}z_{\gamma_3\alpha_2}}{z_{2\alpha_2}z_{\gamma_3\gamma_4}}=\eta_{2\gamma_4;\gamma_3\alpha_2}=1+\sum_{5\leq b\leq q}(-1)^b\prod_{2\leq a\leq b-3}\eta_c^M+(-1)^{q+1}\prod_{2\leq c\leq q-1}\eta_c^M,\\
\frac{z_{2\gamma_4}z_{\gamma_3\gamma_a}}{z_{2\gamma_a}z_{\gamma_3\gamma_4}}=\eta_{2\gamma_4;\gamma_3\gamma_a}=1+\sum_{5\leq b\leq a}(-1)^b\prod_{2\leq c\leq b-3}\eta_c^M\qquad5\leq a\leq q,
\end{gathered}
}
and after applying the binomial theorem several times, we obtain
\eqna{
G_M&=\sum_{\{n_a,t_{ab}\}\geq0}\frac{(h_{i_1}+h_{k_1}-h_{i_2})_{n_1}(p_4)_{n_1-t_{00}-t_{10}-\sum_{5\leq a\leq q}t_{a0}}}{(2h_{k_1})_{n_1}(n_1-t_{00}-t_{10}-\sum_{5\leq a\leq q}t_{a0})!}(\eta_1^M)^{n_1}\\
&\phantom{=}\qquad\times\frac{(p_0)_{t_{00}}(-1)^{t_{0,q-3}}(\eta_{q-2}^M)^{t_{0,q-3}}}{(t_{0,q-4}-t_{0,q-3})!t_{0,q-3}!}\frac{(p_1)_{t_{10}}(-1)^{t_{1,q-3}}(\eta_{q-2}^M)^{t_{1,q-3}}(\eta_{q-1}^M)^{t_{1,q-3}}}{(t_{1,q-4}-t_{1,q-3})!t_{1,q-3}!}\\
&\phantom{=}\qquad\times\left[\prod_{5\leq a\leq q}\frac{(-1)^{t_{a0}+t_{0,a-4}+t_{1,a-4}}(p_a)_{t_{a0}}}{(t_{0,a-5}-t_{0,a-4})!(t_{1,a-5}-t_{1,a-4})!}(\eta_{a-3}^M)^{t_{0,a-4}+t_{1,a-4}+\sum_{a\leq b\leq q}t_{b,a-4}}\right.\\
&\phantom{=}\qquad\times\left.\prod_{0\leq b\leq a-4}\frac{(-1)^{t_{ab}}}{(t_{ab}-t_{a,b+1})!}\right]H_{M-1},
}
which is the $M$-point conformal block in terms of the proper conformal cross-ratios.  Here we defined $t_{a0}=s_a$ for all $a$ as well as $t_{0,q-2}=t_{1,q-2}=t_{a,a-3}=0$ for $a\geq5$.

At this point, we shift $t_{a0}\to t_{a0}+t_{a1}$ and evaluate the sums over $t_{a0}$ using the ${}_2F_1$ identity \eqref{EqpFq}, giving us
\eqna{
G_M&=\sum_{\{n_a,t_{ab}\}\geq0}\frac{(h_{i_1}+h_{k_1}-h_{i_2})_{n_1}(h_{k_1}+h_{k_2}-h_{i_{\gamma_3}}+n_2+t_{01}+t_{11}+\sum_{5\leq a\leq q}t_{a1})_{n_1-t_{01}-t_{11}-\sum_{5\leq a\leq q}t_{a1}}}{(2h_{k_1})_{n_1}(n_1-t_{01}-t_{11}-\sum_{5\leq a\leq q}t_{a1})!}\\
&\phantom{=}\qquad\times(\eta_1^M)^{n_1}\frac{(p_0)_{t_{01}}(-1)^{t_{0,q-3}}(\eta_{q-2}^M)^{t_{0,q-3}}}{\prod_{1\leq a\leq q-3}(t_{0a}-t_{0,a+1})!}\frac{(p_1)_{t_{11}}(-1)^{t_{1,q-3}}(\eta_{q-2}^M)^{t_{1,q-3}}(\eta_{q-1}^M)^{t_{1,q-3}}}{\prod_{1\leq a\leq q-3}(t_{1a}-t_{1,a+1})!}\\
&\phantom{=}\qquad\times\left[\prod_{5\leq a\leq q}(-1)^{t_{0,a-4}+t_{1,a-4}}(p_a)_{t_{a1}}(\eta_{a-3}^M)^{t_{0,a-4}+t_{1,a-4}+\sum_{a\leq b\leq q}t_{b,a-4}}\prod_{1\leq b\leq a-4}\frac{(-1)^{t_{ab}}}{(t_{ab}-t_{a,b+1})!}\right]H_{M-1},
}
with
\eqn{p_0+p_1+\sum_{4\leq a\leq q}p_a=h_{k_1}+h_{k_2}-h_{i_{\gamma_3}}+n_2,}
from \eqref{Eqptype1}.

With the re-definitions $t_{a,a-4}=t_{a-3}-t_{0,a-4}-t_{1,a-4}-\sum_{a+1\leq b\leq q}t_{b,a-4}$ for $5\leq a\leq q$, we can perform the sums over $t_{ab}$ after shifting $t_{ab}\to t_{ab}+t_{a,b+1}$ (starting from the smallest value for $a$, \textit{i.e.}\ summing over $t_{01},t_{11},t_{61},t_{71},\ldots$ followed by $t_{02},t_{12},t_{62},t_{72},\ldots$, \textit{etc.}) following the ${}_2F_1$ identity \eqref{EqpFq} which leads to
\eqna{
G_M&=\sum_{\{n_a,t_a\}\geq0}\frac{(h_{i_1}+h_{k_1}-h_{i_2})_{n_1}(h_{k_1}+h_{k_2}-h_{i_{\gamma_3}}+n_2+t_2)_{n_1-t_2}}{(2h_{k_1})_{n_1}(n_1-t_2)!}\\
&\phantom{=}\qquad\times\frac{(-1)^{t_{q-2}}(p_0)_{t_{q-2}-t_{q-1}}(p_1)_{t_{q-1}}}{(t_{q-2}-t_{q-1})!t_{q-1}!}(\eta_1^M)^{n_1}(\eta_{q-2}^M)^{t_{q-2}}(\eta_{q-1}^M)^{t_{q-1}}\\
&\phantom{=}\qquad\times\left[\prod_{2\leq a\leq q-3}\frac{(-1)^{t_a}(h_{k_a}+h_{k_{a+1}}-h_{i_{\gamma_{a+2}}}+n_a+n_{a+1}+t_{a+1})_{t_a-t_{a+1}}}{(t_a-t_{a+1})!}(\eta_a^M)^{t_a}\right]H_{M-1},
}
where we defined $t_{0,q-3}=t_{q-2}-t_{q-1}$, $t_{1,q-3}=t_{q-1}$ and we used \eqref{Eqptype1}.

To proceed, we re-introduce $H_{M-1}$ and get
\eqna{
G_M&=\sum_{\{n_a,t_a\}\geq0}\frac{(h_{i_1}+h_{k_1}-h_{i_2})_{n_1}(h_{k_1}+h_{k_2}-h_{i_{\gamma_3}}+n_2+t_2)_{n_1-t_2}}{(2h_{k_1})_{n_1}(n_1-t_2)!}\\
&\phantom{=}\qquad\times\frac{(-1)^{t_{q-2}}(p_0)_{t_{q-2}-t_{q-1}}(p_1)_{t_{q-1}}}{(t_{q-2}-t_{q-1})!t_{q-1}!}(\eta_1^M)^{n_1}(\eta_{q-2}^M)^{t_{q-2}}(\eta_{q-1}^M)^{t_{q-1}}\\
&\phantom{=}\qquad\times\prod_{2\leq a\leq q-3}\frac{(-1)^{t_a}(h_{k_a}+h_{k_{a+1}}-h_{i_{\gamma_{a+2}}}+n_a+n_{a+1}+t_{a+1})_{t_a-t_{a+1}}}{(t_a-t_{a+1})!}(\eta_a^M)^{t_a}\\
&\phantom{=}\qquad\times\bar{C}_{M-1}\bar{F}_{M-1}\left[\prod_{2\leq a\leq M-3}\frac{(\eta_a^{M-1})^{n_a}}{n_a!}\right]\frac{(h_{i_{\gamma_3}}-h_{k_1}+h_{k_2})_{n_2}}{(2h_{k_2})_{n_2}}\frac{(h_{i_{\alpha_2}}-h_{i_{\alpha_1}}+h_{k_{q-1}})_{n_{q-1}}}{(2h_{k_{q-1}})_{n_{q-1}}}\\
&\phantom{=}\qquad\times\prod_{2\leq a\leq q-3}\frac{(h_{i_{\gamma_{a+2}}}-h_{k_a}-n_a+h_{k_{a+1}})_{n_{a+1}}(-h_{i_{\gamma_{a+2}}}+h_{k_a}+h_{k_{a+1}})_{n_a}}{(2h_{k_{a+1}})_{n_{a+1}}}\\
&\phantom{=}\qquad\times\frac{(h_{k_{q-2}}-h_{k_{q-1}}-n_{q-1}+h_{k_q})_{n_{q-2}+n_q}(-h_{k_{q-2}}+h_{k_{q-1}}+h_{k_q})_{n_{q-1}}}{(2h_{k_q})_{n_q}}\\
&\phantom{=}\qquad\times{}_3F_2\left[\begin{array}{c}-n_{q-2},-n_{q-1},1-2h_{k_{q-1}}-n_{q-1}\\h_{k_{q-2}}-h_{k_{q-1}}-n_{q-1}+h_{k_q},1+h_{k_{q-2}}-h_{k_{q-1}}-n_{q-1}-h_{k_q}\end{array};1\right].
}
We then expand the ${}_3F_2$ with index of summation $s$, we shift $n_{q-1}\to n_{q-1}-t_{q-1}$, we rename $t_{q-1}=t-s$ to perform the sum over $s$ using the first identity of \eqref{EqpFq}, and finally we express the sum over $t$ in terms of a ${}_3F_2$ to reach
\eqna{
G_M&=\sum_{\{n_a,t_a\}\geq0}\frac{(h_{i_1}+h_{k_1}-h_{i_2})_{n_1}(h_{k_1}+h_{k_2}-h_{i_{\gamma_3}}+n_2+t_2)_{n_1-t_2}}{(2h_{k_1})_{n_1}(n_1-t_2)!}(\eta_1^M)^{n_1}\frac{(-\eta_{q-2}^M)^{t_{q-2}}}{t_{q-2}!}\\
&\phantom{=}\qquad\times\prod_{2\leq a\leq q-3}\frac{(-1)^{t_a}(h_{k_a}+h_{k_{a+1}}-h_{i_{\gamma_{a+2}}}+n_a+n_{a+1}+t_{a+1})_{t_a-t_{a+1}}}{(t_a-t_{a+1})!}(\eta_a^M)^{t_a}\\
&\phantom{=}\qquad\times\bar{C}_{M-1}\bar{F}_{M-1}\left[\prod_{2\leq a\leq M-3}\frac{(\eta_a^M)^{n_a}}{n_a!}\right]\frac{(h_{i_{\gamma_3}}-h_{k_1}+h_{k_2})_{n_2}}{(2h_{k_2})_{n_2}}\frac{(h_{i_{\alpha_2}}-h_{i_{\alpha_1}}+h_{k_{q-1}})_{n_{q-1}}}{(2h_{k_{q-1}})_{n_{q-1}}}\\
&\phantom{=}\qquad\times\prod_{2\leq a\leq q-3}\frac{(h_{i_{\gamma_{a+2}}}-h_{k_a}-n_a+h_{k_{a+1}})_{n_{a+1}}(-h_{i_{\gamma_{a+2}}}+h_{k_a}+h_{k_{a+1}})_{n_a}}{(2h_{k_{a+1}})_{n_{a+1}}}\\
&\phantom{=}\qquad\times\frac{(h_{k_{q-2}}-h_{k_{q-1}}-n_{q-1}+h_{k_q})_{n_{q-2}+n_q+t_{q-2}}(-h_{k_{q-2}}+h_{k_{q-1}}+h_{k_q})_{n_{q-1}}}{(2h_{k_q})_{n_q}}\\
&\phantom{=}\qquad\times{}_3F_2\left[\begin{array}{c}-n_{q-2}-t_{q-2},-n_{q-1},1-2h_{k_{q-1}}-n_{q-1}\\h_{k_{q-2}}-h_{k_{q-1}}-n_{q-1}+h_{k_q},1+h_{k_{q-2}}-h_{k_{q-1}}-n_{q-1}-h_{k_q}\end{array};1\right],
}
using again the definitions \eqref{Eqptype1}.

To complete the proof, we shift $n_a\to n_a-t_a$ for $2\leq a\leq q-2$ and compute the sums over $t_a$ using the ${}_3F_2$ identity in \eqref{EqpFq} to reach
\eqna{
G_M&=\sum_{\{n_a\}\geq0}\frac{(h_{i_1}-h_{i_2}+h_{k_1})_{n_1}}{(2h_{k_1})_{n_1}}\frac{(h_{i_{\alpha_2}}-h_{i_{\alpha_1}}+h_{k_{q-1}})_{n_{q-1}}}{(2h_{k_{q-1}})_{n_{q-1}}}\bar{C}_{M-1}\bar{F}_{M-1}\prod_{1\leq a\leq M-3}\frac{(\eta_a^M)^{n_a}}{n_a!}\\
&\phantom{=}\qquad\times\prod_{1\leq a\leq q-3}\frac{(h_{i_{\gamma_{a+2}}}-h_{k_a}-n_a+h_{k_{a+1}})_{n_{a+1}}(-h_{i_{\gamma_{a+2}}}+h_{k_a}+h_{k_{a+1}})_{n_a}}{(2h_{k_{a+1}})_{n_{a+1}}}\\
&\phantom{=}\qquad\times\frac{(h_{k_{q-2}}-h_{k_{q-1}}-n_{q-1}+h_{k_q})_{n_{q-2}+n_q}(-h_{k_{q-2}}+h_{k_{q-1}}+h_{k_q})_{n_{q-1}}}{(2h_{k_q})_{n_q}}\\
&\phantom{=}\qquad\times{}_3F_2\left[\begin{array}{c}-n_{q-2},-n_{q-1},1-2h_{k_{q-1}}-n_{q-1}\\h_{k_{q-2}}-h_{k_{q-1}}-n_{q-1}+h_{k_q},1+h_{k_{q-2}}-h_{k_{q-1}}-n_{q-1}-h_{k_q}\end{array};1\right],
}
which satisfies the rules.  We note that the proof for the type $1$ full extra comb is reminiscent of the proof for the initial comb.  As seen in the next subsections, the same is true for types $2$ and $3$, simplifying their proofs.

Therefore, gluing an extra comb structure unto an arbitrary topology of the first type following our convention for the OPE limits demonstrates that our rules are valid in that case.


\subsubsection{Type 2: Boundary Condition}

Following the same steps than for the boundary condition of type $1$, we find that the only $z_2$-dependent quantities in the $(M-1)$-point conformal partial wave are the leg factors and conformal cross-ratios
\eqn{
\begin{gathered}
L_{M-1}=z_{\alpha_1\beta_0;2}^{h_{k_1}}z_{\alpha_12;\alpha_2}^{h_{i_{\alpha_2}}}\bar{L}_{M-1},\\
\eta_2^{M-1}=\eta_{\alpha_1\alpha_2;2\beta_0},\qquad\eta_3^{M-1}=\eta_{\alpha_12;\beta_0\alpha_0},\qquad\eta_4^{M-1}=\eta_{\alpha_1\beta_1;\alpha_22}.
\end{gathered}
}
As a consequence, the $M$-point conformal partial wave resulting from the action of the OPE \eqref{EqOPE1d} is
\eqna{
W_{M(h_{k_1},\ldots,h_{k_{M-3}})}^{(h_{i_1},\ldots,h_{i_M})}&=\frac{1}{z_{12}^{h_{i_1}+h_{i_2}-h_{k_1}}}{}_1F_1(h_{i_1}+h_{k_1}-h_{i_2},2h_{k_1};z_{12}\partial_2)W_{M-1(h_{k_2},\ldots,h_{k_{M-3}})}^{(h_{k_1},h_{i_3},\ldots,h_{i_{M-1}})}\\
&=L_M\left[\prod_{1\leq a\leq M-3}(\eta_a^{M})^{h_{k_a}}\right]\left(\frac{z_{2\alpha_1}}{z_{1\alpha_1}}\right)^{-h_{i_1}-h_{k_1}+h_{i_2}}\\
&\phantom{=}\qquad\times\sum_{\{n_a\},n,m_0,m_1\geq0}C_{M-1}F_{M-1}\left[\prod_{2\leq a\leq M-3}\frac{(\eta_a^M)^{n_a}}{n_a!}\right]\frac{z_{12}^n}{z_{2\alpha_1}^{n-m_0-m_1}z_{2\alpha_2}^{m_1}z_{2\beta_0}^{m_0}}\\
&\phantom{=}\qquad\times\frac{(-1)^n(h_{i_1}+h_{k_1}-h_{i_2})_n(h_{k_1}+h_{k_4}+n_4-h_{i_{\alpha_2}}-h_{k_3}-n_3)_{n-m_0-m_1}}{(2h_{k_1})_n(n-m_0-m_1)!}\\
&\phantom{=}\qquad\times\frac{(h_{i_{\alpha_2}}-h_{k_4}-n_4+h_{k_2}+n_2)_{m_1}(h_{k_1}-h_{k_2}+h_{k_3}-n_2+n_3)_{m_0}}{m_1!m_0!},
}
where we used \eqref{EqIb}.  Here the proper leg and cross-ratios are
\eqn{
\begin{gathered}
L_M=z_{2\alpha_1;1}^{h_{i_1}}z_{\alpha_11;2}^{h_{i_2}}z_{\alpha_12;\alpha_2}^{h_{i_{\alpha_2}}}\bar{L}_{M-1},\\
\eta_1^M=\eta_{21;\alpha_1\beta_0},\qquad\eta_a^M=\eta_a^{M-1}\qquad2\leq a\leq M-3,
\end{gathered}
}
which imply that the $M$-point conformal block is given by
\eqna{
G_M&=\left(\frac{z_{2\alpha_1}}{z_{1\alpha_1}}\right)^{-h_{i_1}-h_{k_1}+h_{i_2}}\sum_{\{n_a\},n,m_0,m_1\geq0}C_{M-1}F_{M-1}\left[\prod_{2\leq a\leq M-3}\frac{(\eta_a^M)^{n_a}}{n_a!}\right]\frac{z_{12}^n}{z_{2\alpha_1}^{n-m_0-m_1}z_{2\alpha_2}^{m_1}z_{2\beta_0}^{m_0}}\\
&\phantom{=}\qquad\times\frac{(-1)^n(h_{i_1}+h_{k_1}-h_{i_2})_n(h_{k_1}+h_{k_4}+n_4-h_{i_{\alpha_2}}-h_{k_3}-n_3)_{n-m_0-m_1}}{(2h_{k_1})_n(n-m_0-m_1)!}\\
&\phantom{=}\qquad\times\frac{(h_{i_{\alpha_2}}-h_{k_4}-n_4+h_{k_2}+n_2)_{m_1}(h_{k_1}-h_{k_2}+h_{k_3}-n_2+n_3)_{m_0}}{m_1!m_0!}\\
&=\sum_{\{n_a\},s_1\geq0}C_{M-1}F_{M-1}\left[\prod_{2\leq a\leq M-3}\frac{(\eta_a^M)^{n_a}}{n_a!}\right](\eta_1^M)^{n_1}(\eta_2^M)^{s_1}\\
&\phantom{=}\qquad\times\frac{(h_{i_1}+h_{k_1}-h_{i_2})_{n_1}(h_{i_{\alpha_2}}-h_{k_4}-n_4+h_{k_2}+n_2)_{s_1}(h_{k_1}-h_{k_2}+h_{k_3}-n_2+n_3)_{n_1-s_1}}{(2h_{k_1})_{n_1}s_1!(n_1-s_1)!},
}[EqGtype2]
since \eqref{EqGtype2} corresponds to \eqref{EqGtype1} with the replacement $h_{i_{\alpha_1}}\to h_{k_4}+n_4$.

From the known part of the $(M-1)$-point conformal block of type $2$, assuming our rules, we can write
\eqn{C_{M-1}=\frac{(h_{k_1}-h_{k_2}-n_2+h_{k_3})_{n_3}(-h_{k_1}+h_{k_2}+h_{k_3})_{n_2}(h_{i_{\alpha_2}}-h_{k_4}-n_4+h_{k_2})_{n_2}}{(2h_{k_2})_{n_2}}\bar{C}_{M-1},}
where $\bar{C}_{M-1}$ is determined by the arbitrary topology and does not depend on $n_2$.  This is again equivalent to the type $1$ boundary condition case with $h_{i_{\alpha_1}}\to h_{k_4}+n_4$, hence we can express the $M$-point conformal block \eqref{EqGtype2} as
\eqna{
G_M&=\sum_{\{n_a\}\geq0}\bar{C}_{M-1}F_{M-1}\left[\prod_{1\leq a\leq M-3}\frac{(\eta_a^M)^{n_a}}{n_a!}\right]\frac{(h_{i_1}-h_{i_2}+h_{k_1})_{n_1}(h_{i_{\alpha_2}}-h_{k_4}-n_4+h_{k_2})_{n_2}}{(2h_{k_1})_{n_1}}\\
&\phantom{=}\qquad\times\frac{(h_{k_1}-h_{k_2}-n_2+h_{k_3})_{n_1+n_3}(-h_{k_1}+h_{k_2}+h_{k_3})_{n_2}}{(2h_{k_2})_{n_2}}\\
&\phantom{=}\qquad\times{}_3F_2\left[\begin{array}{c}-n_1,-n_2,1-2h_{k_2}-n_2\\h_{k_1}-h_{k_2}-n_2+h_{k_3},1+h_{k_1}-h_{k_2}-n_2-h_{k_3}\end{array};1\right],
}
which satisfies our rules as dictated in Section \ref{SecCF}.  In conclusion, our rules are valid for the boundary condition when gluing an extra comb structure for type $2$ topologies.


\subsubsection{Type 2: Full Extra Comb}

Since the type $2$ boundary condition is valid, we can follow the same path than for the type $1$ extra comb structure and proceed by induction to verify that the addition of an extra comb structure on an arbitrary topology of type $2$ is consistent with our rules.  Hence, we assume that the rules are verified for a $(q-1)$-point extra comb structure and compute the $q$-point extra comb structure using the OPE to show that the resulting conformal block matches our expectation.

As seen from Figure \ref{FigType}, the rules imply that all $z_2$-dependence can be found in the leg
\eqn{L_{M-1}=z_{\gamma_4\gamma_3;2}^{h_{k_1}}\left[\prod_{3\leq a\leq q-1}z_{2\gamma_{a+1};\gamma_a}^{h_{i_{\gamma_a}}}\right]z_{\alpha_12;\alpha_2}^{h_{i_{\alpha_2}}}\bar{L}_{M-1},}
and conformal cross-ratios
\eqn{
\begin{gathered}
\eta_a^{M-1}=\eta_{2\gamma_{a+1};\gamma_{a+2}\gamma_{a+3}}\qquad2\leq a\leq q-3,\\
\eta_{q-2}^{M-1}=\eta_{2\gamma_{q-1};\alpha_1\beta_0},\qquad\eta_{q-1}^{M-1}=\eta_{\alpha_1\alpha_2;2\beta_0},\qquad\eta_q^{M-1}=\eta_{\alpha_12;\beta_0\alpha_0},\qquad\eta_{q+1}^{M-1}=\eta_{\alpha_1\beta_1;\alpha_22},
\end{gathered}
}
where $\bar{L}_{M-1}$ is $z_2$-independent and $\gamma_q=\alpha_1$.  With $\bar{C}_{M-1}$ and $\bar{F}_{M-1}$ being $n_2$-independent, the $n_2$-dependent part of the $(M-1)$-point conformal block is
\eqna{
G_{M-1}&=\sum_{\{n_a\}\geq0}\bar{C}_{M-1}\bar{F}_{M-1}\left[\prod_{2\leq a\leq M-3}\frac{(\eta_a^{M-1})^{n_a}}{n_a!}\right]\frac{(h_{i_{\gamma_3}}-h_{k_1}+h_{k_2})_{n_2}}{(2h_{k_2})_{n_2}}\frac{(h_{i_{\alpha_2}}-h_{k_{q+1}}-n_{q+1}+h_{k_{q-1}})_{n_{q-1}}}{(2h_{k_{q-1}})_{n_{q-1}}}\\
&\phantom{=}\qquad\times\prod_{2\leq a\leq q-3}\frac{(h_{i_{\gamma_{a+2}}}-h_{k_a}-n_a+h_{k_{a+1}})_{n_{a+1}}(-h_{i_{\gamma_{a+2}}}+h_{k_a}+h_{k_{a+1}})_{n_a}}{(2h_{k_{a+1}})_{n_{a+1}}}\\
&\phantom{=}\qquad\times\frac{(h_{k_{q-2}}-h_{k_{q-1}}-n_{q-1}+h_{k_q})_{n_{q-2}+n_q}(-h_{k_{q-2}}+h_{k_{q-1}}+h_{k_q})_{n_{q-1}}}{(2h_{k_q})_{n_q}}\\
&\phantom{=}\qquad\times{}_3F_2\left[\begin{array}{c}-n_{q-2},-n_{q-1},1-2h_{k_{q-1}}-n_{q-1}\\h_{k_{q-2}}-h_{k_{q-1}}-n_{q-1}+h_{k_q},1+h_{k_{q-2}}-h_{k_{q-1}}-n_{q-1}-h_{k_q}\end{array};1\right],
}
following Figure \ref{FigType} and the rules.

With the help of \eqref{EqIb}, it is trivial to find that the OPE \eqref{EqOPE1d} leads to the $M$-point conformal block
\eqna{
G_M&=\left(\frac{z_{2\gamma_3}}{z_{1\gamma_3}}\right)^{-h_{i_1}-h_{k_1}+h_{i_2}}\sum_{\{n_a,m_a\},n\geq0}\frac{(-1)^n(h_{i_1}+h_{k_1}-h_{i_2})_n(p_3)_{n-m_0-m_1-\sum_{4\leq a\leq q}m_a}}{(2h_{k_1})_n(n-m_0-m_1-\sum_{4\leq a\leq q}m_a)!}\\
&\phantom{=}\qquad\times\frac{z_{12}^n}{z_{2\gamma_3}^{n-m_0-m_1-\sum_{4\leq a\leq q}m_a}z_{2\beta_0}^{m_0}z_{2\alpha_2}^{m_1}}\frac{(p_0)_{m_0}(p_1)_{m_1}}{m_0!m_1!}\prod_{4\leq a\leq q}\frac{(p_a)_{m_a}}{m_a!z_{2\gamma_a}^{m_a}}\\
&\phantom{=}\qquad\times\bar{C}_{M-1}\bar{F}_{M-1}\left[\prod_{2\leq a\leq M-3}\frac{(\eta_a^{M-1})^{n_a}}{n_a!}\right]\frac{(h_{i_{\gamma_3}}-h_{k_1}+h_{k_2})_{n_2}}{(2h_{k_2})_{n_2}}\frac{(h_{i_{\alpha_2}}-h_{k_{q+1}}-n_{q+1}+h_{k_{q-1}})_{n_{q-1}}}{(2h_{k_{q-1}})_{n_{q-1}}}\\
&\phantom{=}\qquad\times\prod_{2\leq a\leq q-3}\frac{(h_{i_{\gamma_{a+2}}}-h_{k_a}-n_a+h_{k_{a+1}})_{n_{a+1}}(-h_{i_{\gamma_{a+2}}}+h_{k_a}+h_{k_{a+1}})_{n_a}}{(2h_{k_{a+1}})_{n_{a+1}}}\\
&\phantom{=}\qquad\times\frac{(h_{k_{q-2}}-h_{k_{q-1}}-n_{q-1}+h_{k_q})_{n_{q-2}+n_q}(-h_{k_{q-2}}+h_{k_{q-1}}+h_{k_q})_{n_{q-1}}}{(2h_{k_q})_{n_q}}\\
&\phantom{=}\qquad\times{}_3F_2\left[\begin{array}{c}-n_{q-2},-n_{q-1},1-2h_{k_{q-1}}-n_{q-1}\\h_{k_{q-2}}-h_{k_{q-1}}-n_{q-1}+h_{k_q},1+h_{k_{q-2}}-h_{k_{q-1}}-n_{q-1}-h_{k_q}\end{array};1\right],
}[EqGtype2comb]
where
\eqn{
\begin{gathered}
p_0=-h_{k_{q-1}}+h_{k_{q-2}}+h_{k_q}-n_{q-1}+n_{q-2}+n_q,\\
p_1=-h_{k_{q+1}}-n_{q+1}+h_{i_{\alpha_2}}+h_{k_{q-1}}+n_{q-1},\\
p_3=h_{k_1}+h_{i_{\gamma_3}}-h_{k_2}-n_2,\\
p_4=-h_{i_{\gamma_3}}+h_{i_{\gamma_4}}+h_{k_1}-h_{k_3}-n_3,\\
p_a=-h_{i_{\gamma_{a-1}}}+h_{i_{\gamma_a}}+h_{k_{a-3}}+n_{a-3}-h_{k_{a-1}}-n_{a-1}\qquad5\leq a\leq q-2,\\
p_{q-1}=-h_{i_{\gamma_{q-2}}}+h_{i_{\gamma_{q-1}}}+h_{k_{q-4}}+n_{q-4}-h_{k_{q-2}}-n_{q-2},\\
p_q=h_{k_{q+1}}+n_{q+1}-h_{i_{\alpha_2}}-h_{i_{\gamma_{q-1}}}+h_{k_{q-3}}-h_{k_q}+n_{q-3}-n_q.
\end{gathered}
}[Eqptype2]
Comparing \eqref{EqGtype2comb} and \eqref{Eqptype2} with \eqref{EqGtype1comb} and \eqref{Eqptype1}, respectively, we note that \eqref{EqGtype2comb} is nothing but \eqref{EqGtype1comb} with the change $h_{i_{\alpha_1}}\to h_{k_{q+1}}+n_{q+1}$.  As a result, we thus have (with $\eta_1^M=z_{21;\gamma_3\gamma_4}$ as before)
\eqna{
G_M&=\sum_{\{n_a\}\geq0}\frac{(h_{i_1}-h_{i_2}+h_{k_1})_{n_1}}{(2h_{k_1})_{n_1}}\frac{(h_{i_{\alpha_2}}-h_{k_{q+1}}-n_{q+1}+h_{k_{q-1}})_{n_{q-1}}}{(2h_{k_{q-1}})_{n_{q-1}}}\bar{C}_{M-1}\bar{F}_{M-1}\prod_{1\leq a\leq M-3}\frac{(\eta_a^M)^{n_a}}{n_a!}\\
&\phantom{=}\qquad\times\prod_{1\leq a\leq q-3}\frac{(h_{i_{\gamma_{a+2}}}-h_{k_a}-n_a+h_{k_{a+1}})_{n_{a+1}}(-h_{i_{\gamma_{a+2}}}+h_{k_a}+h_{k_{a+1}})_{n_a}}{(2h_{k_{a+1}})_{n_{a+1}}}\\
&\phantom{=}\qquad\times\frac{(h_{k_{q-2}}-h_{k_{q-1}}-n_{q-1}+h_{k_q})_{n_{q-2}+n_q}(-h_{k_{q-2}}+h_{k_{q-1}}+h_{k_q})_{n_{q-1}}}{(2h_{k_q})_{n_q}}\\
&\phantom{=}\qquad\times{}_3F_2\left[\begin{array}{c}-n_{q-2},-n_{q-1},1-2h_{k_{q-1}}-n_{q-1}\\h_{k_{q-2}}-h_{k_{q-1}}-n_{q-1}+h_{k_q},1+h_{k_{q-2}}-h_{k_{q-1}}-n_{q-1}-h_{k_q}\end{array};1\right],
}
which is in agreement with our rules.

Following our convention for the OPE limits, we conclude that the rules of Section \ref{SecCF} are correct for the addition of an extra comb structure unto an arbitrary topology of the second type.


\subsubsection{Type 3: Boundary Condition}

We once again adapt the procedure from the boundary condition of type $2$ to type $3$.  First, we observe that the $z_2$-dependence of the $(M-1)$-point conformal partial wave is located in the leg factors and conformal cross-ratios
\eqn{
\begin{gathered}
L_{M-1}=z_{\alpha_1\beta_0;2}^{h_{k_1}}\bar{L}_{M-1},\\
\eta_2^{M-1}=\eta_{\alpha_1\alpha_2;2\beta_0},\qquad\eta_3^{M-1}=\eta_{\alpha_12;\beta_0\alpha_0},\qquad\eta_4^{M-1}=\eta_{\alpha_1\beta_1;\alpha_22},\qquad\eta_5^{M-1}=\eta_{\alpha_2\beta_2;\alpha_12},
\end{gathered}
}
where $\bar{L}_{M-1}$ does not depend on $z_2$ and is fixed by the topology.  From the OPE \eqref{EqOPE1d} and the identity \eqref{EqIb}, the $M$-point conformal partial wave is
\eqna{
W_{M(h_{k_1},\ldots,h_{k_{M-3}})}^{(h_{i_1},\ldots,h_{i_M})}&=\frac{1}{z_{12}^{h_{i_1}+h_{i_2}-h_{k_1}}}{}_1F_1(h_{i_1}+h_{k_1}-h_{i_2},2h_{k_1};z_{12}\partial_2)W_{M-1(h_{k_2},\ldots,h_{k_{M-3}})}^{(h_{k_1},h_{i_3},\ldots,h_{i_{M-1}})}\\
&=L_M\left[\prod_{1\leq a\leq M-3}(\eta_a^{M})^{h_{k_a}}\right]\left(\frac{z_{2\alpha_1}}{z_{1\alpha_1}}\right)^{-h_{i_1}-h_{k_1}+h_{i_2}}\\
&\phantom{=}\qquad\times\sum_{\{n_a\},n,m_0,m_1\geq0}C_{M-1}F_{M-1}\left[\prod_{2\leq a\leq M-3}\frac{(\eta_a^M)^{n_a}}{n_a!}\right]\frac{z_{12}^n}{z_{2\alpha_1}^{n-m_0-m_1}z_{2\alpha_2}^{m_1}z_{2\beta_0}^{m_0}}\\
&\phantom{=}\qquad\times\frac{(-1)^n(h_{i_1}+h_{k_1}-h_{i_2})_n(h_{k_1}+h_{k_4}+n_4-h_{k_5}-n_5-h_{k_3}-n_3)_{n-m_0-m_1}}{(2h_{k_1})_n(n-m_0-m_1)!}\\
&\phantom{=}\qquad\times\frac{(h_{k_5}+n_5-h_{k_4}-n_4+h_{k_2}+n_2)_{m_1}(h_{k_1}-h_{k_2}+h_{k_3}-n_2+n_3)_{m_0}}{m_1!m_0!},
}
with the following leg and conformal cross-ratios
\eqn{
\begin{gathered}
L_M=z_{2\alpha_1;1}^{h_{i_1}}z_{\alpha_11;2}^{h_{i_2}}\bar{L}_{M-1},\\
\eta_1^M=\eta_{21;\alpha_1\beta_0},\qquad\eta_a^M=\eta_a^{M-1}\qquad2\leq a\leq M-3,
\end{gathered}
}
as expected from our rules.

Isolating the $M$-point conformal block, we have
\eqna{
G_M&=\left(\frac{z_{2\alpha_1}}{z_{1\alpha_1}}\right)^{-h_{i_1}-h_{k_1}+h_{i_2}}\sum_{\{n_a\},n,m_0,m_1\geq0}C_{M-1}F_{M-1}\left[\prod_{2\leq a\leq M-3}\frac{(\eta_a^M)^{n_a}}{n_a!}\right]\frac{z_{12}^n}{z_{2\alpha_1}^{n-m_0-m_1}z_{2\alpha_2}^{m_1}z_{2\beta_0}^{m_0}}\\
&\phantom{=}\qquad\times\frac{(-1)^n(h_{i_1}+h_{k_1}-h_{i_2})_n(h_{k_1}+h_{k_4}+n_4-h_{k_5}-n_5-h_{k_3}-n_3)_{n-m_0-m_1}}{(2h_{k_1})_n(n-m_0-m_1)!}\\
&\phantom{=}\qquad\times\frac{(h_{k_5}-n_5-h_{k_4}-n_4+h_{k_2}+n_2)_{m_1}(h_{k_1}-h_{k_2}+h_{k_3}-n_2+n_3)_{m_0}}{m_1!m_0!}\\
&=\sum_{\{n_a\},s_1\geq0}C_{M-1}F_{M-1}\left[\prod_{2\leq a\leq M-3}\frac{(\eta_a^M)^{n_a}}{n_a!}\right](\eta_1^M)^{n_1}(\eta_2^M)^{s_1}\\
&\phantom{=}\qquad\times\frac{(h_{i_1}+h_{k_1}-h_{i_2})_{n_1}(h_{k_5}-n_5-h_{k_4}-n_4+h_{k_2}+n_2)_{s_1}(h_{k_1}-h_{k_2}+h_{k_3}-n_2+n_3)_{n_1-s_1}}{(2h_{k_1})_{n_1}s_1!(n_1-s_1)!},
}[EqGtype3]
where in the last equality we used the fact that \eqref{EqGtype3} is analog to \eqref{EqGtype2} but with $h_{i_{\alpha_2}}\to h_{k_5}+n_5$.

From the rules of Section \ref{SecCF}, extracting the known part of the $(M-1)$-point conformal block of type $3$ leads to
\eqna{
C_{M-1}&=\frac{(h_{k_1}-h_{k_2}-n_2+h_{k_3})_{n_3}(-h_{k_1}+h_{k_2}+h_{k_3})_{n_2}(h_{k_5}-h_{k_4}-n_4+h_{k_2})_{n_5+n_2}}{(2h_{k_2})_{n_2}}\bar{C}_{M-1}\\
&=(h_{k_5}-h_{k_4}-n_4+h_{k_2})_{n_5}\\
&\phantom{=}\qquad\times\frac{(h_{k_1}-h_{k_2}-n_2+h_{k_3})_{n_3}(-h_{k_1}+h_{k_2}+h_{k_3})_{n_2}(h_{k_5}+n_5-h_{k_4}-n_4+h_{k_2})_{n_2}}{(2h_{k_2})_{n_2}}\bar{C}_{M-1},
}
where $\bar{C}_{M-1}$ is independent of $n_2$ (it is undetermined, it is only fixed when the arbitrary topology is chosen).  Up to the factor $(h_{k_5}+h_{k_2}-h_{k_4}-n_4)_{n_5}$ which does not play a role in the remaining re-summations, this result is equivalent to the type $2$ boundary condition result with $h_{i_{\alpha_2}}\to h_{k_5}+n_5$.  Consequently, we derive the $M$-point conformal block \eqref{EqGtype3} as
\eqna{
G_M&=\sum_{\{n_a\}\geq0}\bar{C}_{M-1}F_{M-1}\left[\prod_{1\leq a\leq M-3}\frac{(\eta_a^M)^{n_a}}{n_a!}\right]\frac{(h_{i_1}-h_{i_2}+h_{k_1})_{n_1}(h_{k_5}+n_5-h_{k_4}-n_4+h_{k_2})_{n_2}}{(2h_{k_1})_{n_1}}\\
&\phantom{=}\qquad\times(h_{k_5}-h_{k_4}-n_4+h_{k_2})_{n_5}\frac{(h_{k_1}-h_{k_2}-n_2+h_{k_3})_{n_1+n_3}(-h_{k_1}+h_{k_2}+h_{k_3})_{n_2}}{(2h_{k_2})_{n_2}}\\
&\phantom{=}\qquad\times{}_3F_2\left[\begin{array}{c}-n_1,-n_2,1-2h_{k_2}-n_2\\h_{k_1}-h_{k_2}-n_2+h_{k_3},1+h_{k_1}-h_{k_2}-n_2-h_{k_3}\end{array};1\right]\\
&=\sum_{\{n_a\}\geq0}\bar{C}_{M-1}F_{M-1}\left[\prod_{1\leq a\leq M-3}\frac{(\eta_a^M)^{n_a}}{n_a!}\right]\frac{(h_{i_1}-h_{i_2}+h_{k_1})_{n_1}(h_{k_5}-h_{k_4}-n_4+h_{k_2})_{n_5+n_2}}{(2h_{k_1})_{n_1}}\\
&\phantom{=}\qquad\times\frac{(h_{k_1}-h_{k_2}-n_2+h_{k_3})_{n_1+n_3}(-h_{k_1}+h_{k_2}+h_{k_3})_{n_2}}{(2h_{k_2})_{n_2}}\\
&\phantom{=}\qquad\times{}_3F_2\left[\begin{array}{c}-n_1,-n_2,1-2h_{k_2}-n_2\\h_{k_1}-h_{k_2}-n_2+h_{k_3},1+h_{k_1}-h_{k_2}-n_2-h_{k_3}\end{array};1\right],
}
which satisfies the rules discussed in Section \ref{SecCF}.  We conclude that the rules are valid for the type $3$ boundary condition.


\subsubsection{Type 3: Full Extra Comb}

With the appropriate boundary condition, we are once again ready to verify by induction the rules of Section \ref{SecCF} when an extra comb structure is glued to an arbitrary topology of the third type.

From the rules and Figure \ref{FigType}, we deduce that the $z_2$-dependence is located in the leg
\eqn{L_{M-1}=z_{\gamma_4\gamma_3;2}^{h_{k_1}}\left[\prod_{3\leq a\leq q-1}z_{2\gamma_{a+1};\gamma_a}^{h_{i_{\gamma_a}}}\right]\bar{L}_{M-1},}
and the conformal cross-ratios
\eqn{
\begin{gathered}
\eta_a^{M-1}=\eta_{2\gamma_{a+1};\gamma_{a+2}\gamma_{a+3}}\qquad2\leq a\leq q-3,\\
\eta_{q-2}^{M-1}=\eta_{2\gamma_{q-1};\alpha_1\beta_0},\qquad\eta_{q-1}^{M-1}=\eta_{\alpha_1\alpha_2;2\beta_0},\qquad\eta_q^{M-1}=\eta_{\alpha_12;\beta_0\alpha_0},\\
\eta_{q+1}^{M-1}=\eta_{\alpha_1\beta_1;\alpha_22},\qquad\eta_{q+2}^{M-1}=\eta_{\alpha_2\beta_2;\alpha_12},
\end{gathered}
}
with $\bar{L}_{M-1}$ independent of $z_2$ and $\gamma_q=\alpha_1$.  Denoting by $\bar{C}_{M-1}$ and $\bar{F}_{M-1}$ the $n_2$-independent of the $(M-1)$-point conformal blocs, we have
\eqna{
G_{M-1}&=\sum_{\{n_a\}\geq0}\bar{C}_{M-1}\bar{F}_{M-1}\left[\prod_{2\leq a\leq M-3}\frac{(\eta_a^{M-1})^{n_a}}{n_a!}\right]\frac{(h_{i_{\gamma_3}}-h_{k_1}+h_{k_2})_{n_2}}{(2h_{k_2})_{n_2}}\frac{(h_{k_{q+2}}+n_{q+2}-h_{k_{q+1}}-n_{q+1}+h_{k_{q-1}})_{n_{q-1}}}{(2h_{k_{q-1}})_{n_{q-1}}}\\
&\phantom{=}\qquad\times\prod_{2\leq a\leq q-3}\frac{(h_{i_{\gamma_{a+2}}}-h_{k_a}-n_a+h_{k_{a+1}})_{n_{a+1}}(-h_{i_{\gamma_{a+2}}}+h_{k_a}+h_{k_{a+1}})_{n_a}}{(2h_{k_{a+1}})_{n_{a+1}}}\\
&\phantom{=}\qquad\times\frac{(h_{k_{q-2}}-h_{k_{q-1}}-n_{q-1}+h_{k_q})_{n_{q-2}+n_q}(-h_{k_{q-2}}+h_{k_{q-1}}+h_{k_q})_{n_{q-1}}}{(2h_{k_q})_{n_q}}\\
&\phantom{=}\qquad\times{}_3F_2\left[\begin{array}{c}-n_{q-2},-n_{q-1},1-2h_{k_{q-1}}-n_{q-1}\\h_{k_{q-2}}-h_{k_{q-1}}-n_{q-1}+h_{k_q},1+h_{k_{q-2}}-h_{k_{q-1}}-n_{q-1}-h_{k_q}\end{array};1\right]\\
&\phantom{=}\qquad\times(h_{k_{q+2}}-h_{k_{q+1}}-n_{q+1}+h_{k_{q-1}})_{n_{q+2}},
}
by direct application of the rules.  Note the re-writing of one Pochhammer symbol for future convenience.

The action of the OPE \eqref{EqOPE1d} using the identity \eqref{EqIb} thus implies that the $M$-point conformal block is
\eqna{
G_M&=\left(\frac{z_{2\gamma_3}}{z_{1\gamma_3}}\right)^{-h_{i_1}-h_{k_1}+h_{i_2}}\sum_{\{n_a,m_a\},n\geq0}\frac{(-1)^n(h_{i_1}+h_{k_1}-h_{i_2})_n(p_3)_{n-m_0-m_1-\sum_{4\leq a\leq q}m_a}}{(2h_{k_1})_n(n-m_0-m_1-\sum_{4\leq a\leq q}m_a)!}\\
&\phantom{=}\qquad\times\frac{z_{12}^n}{z_{2\gamma_3}^{n-m_0-m_1-\sum_{4\leq a\leq q}m_a}z_{2\beta_0}^{m_0}z_{2\alpha_2}^{m_1}}\frac{(p_0)_{m_0}(p_1)_{m_1}}{m_0!m_1!}\prod_{4\leq a\leq q}\frac{(p_a)_{m_a}}{m_a!z_{2\gamma_a}^{m_a}}\\
&\phantom{=}\qquad\times\bar{C}_{M-1}\bar{F}_{M-1}\left[\prod_{2\leq a\leq M-3}\frac{(\eta_a^{M-1})^{n_a}}{n_a!}\right]\frac{(h_{i_{\gamma_3}}-h_{k_1}+h_{k_2})_{n_2}}{(2h_{k_2})_{n_2}}\frac{(h_{k_{q+2}}+n_{q+2}-h_{k_{q+1}}-n_{q+1}+h_{k_{q-1}})_{n_{q-1}}}{(2h_{k_{q-1}})_{n_{q-1}}}\\
&\phantom{=}\qquad\times\prod_{2\leq a\leq q-3}\frac{(h_{i_{\gamma_{a+2}}}-h_{k_a}-n_a+h_{k_{a+1}})_{n_{a+1}}(-h_{i_{\gamma_{a+2}}}+h_{k_a}+h_{k_{a+1}})_{n_a}}{(2h_{k_{a+1}})_{n_{a+1}}}\\
&\phantom{=}\qquad\times\frac{(h_{k_{q-2}}-h_{k_{q-1}}-n_{q-1}+h_{k_q})_{n_{q-2}+n_q}(-h_{k_{q-2}}+h_{k_{q-1}}+h_{k_q})_{n_{q-1}}}{(2h_{k_q})_{n_q}}\\
&\phantom{=}\qquad\times{}_3F_2\left[\begin{array}{c}-n_{q-2},-n_{q-1},1-2h_{k_{q-1}}-n_{q-1}\\h_{k_{q-2}}-h_{k_{q-1}}-n_{q-1}+h_{k_q},1+h_{k_{q-2}}-h_{k_{q-1}}-n_{q-1}-h_{k_q}\end{array};1\right]\\
&\phantom{=}\qquad\times(h_{k_{q+2}}-h_{k_{q+1}}-n_{q+1}+h_{k_{q-1}})_{n_{q+2}},
}[EqGtype3comb]
where
\eqn{
\begin{gathered}
p_0=-h_{k_{q-1}}+h_{k_{q-2}}+h_{k_q}-n_{q-1}+n_{q-2}+n_q,\\
p_1=-h_{k_{q+1}}-n_{q+1}+h_{k_{q+2}}+n_{q+2}+h_{k_{q-1}}+n_{q-1},\\
p_3=h_{k_1}+h_{i_{\gamma_3}}-h_{k_2}-n_2,\\
p_4=-h_{i_{\gamma_3}}+h_{i_{\gamma_4}}+h_{k_1}-h_{k_3}-n_3,\\
p_a=-h_{i_{\gamma_{a-1}}}+h_{i_{\gamma_a}}+h_{k_{a-3}}+n_{a-3}-h_{k_{a-1}}-n_{a-1}\qquad5\leq a\leq q-2,\\
p_{q-1}=-h_{i_{\gamma_{q-2}}}+h_{i_{\gamma_{q-1}}}+h_{k_{q-4}}+n_{q-4}-h_{k_{q-2}}-n_{q-2},\\
p_q=h_{k_{q+1}}+n_{q+1}-h_{k_{q+2}}-n_{q+2}-h_{i_{\gamma_{q-1}}}+h_{k_{q-3}}-h_{k_q}+n_{q-3}-n_q,
\end{gathered}
}[Eqptype3]
with again the new conformal cross-ratio given by $\eta_1^M=z_{21;\gamma_3\gamma_4}$.  A direct comparison between \eqref{EqGtype3comb} and \eqref{Eqptype3} on one side and \eqref{EqGtype2comb} and \eqref{Eqptype2} on the other side shows that \eqref{EqGtype3comb} corresponds to \eqref{EqGtype2comb} where $h_{i_{\alpha_2}}\to h_{k_{q+2}}+n_{q+2}$ up to the factor $(h_{k_{q+2}}-h_{k_{q+1}}-n_{q+1}+h_{k_{q-1}})_{n_{q+2}}$.  Since this factor is inconsequential in the re-summations, we reach the result
\eqna{
G_M&=\sum_{\{n_a\}\geq0}\frac{(h_{i_1}-h_{i_2}+h_{k_1})_{n_1}}{(2h_{k_1})_{n_1}}\frac{(h_{k_{q+2}}+n_{q+2}-h_{k_{q+1}}-n_{q+1}+h_{k_{q-1}})_{n_{q-1}}}{(2h_{k_{q-1}})_{n_{q-1}}}\bar{C}_{M-1}\bar{F}_{M-1}\\
&\phantom{=}\qquad\times\prod_{1\leq a\leq q-3}\frac{(h_{i_{\gamma_{a+2}}}-h_{k_a}-n_a+h_{k_{a+1}})_{n_{a+1}}(-h_{i_{\gamma_{a+2}}}+h_{k_a}+h_{k_{a+1}})_{n_a}}{(2h_{k_{a+1}})_{n_{a+1}}}\prod_{1\leq a\leq M-3}\frac{(\eta_a^M)^{n_a}}{n_a!}\\
&\phantom{=}\qquad\times\frac{(h_{k_{q-2}}-h_{k_{q-1}}-n_{q-1}+h_{k_q})_{n_{q-2}+n_q}(-h_{k_{q-2}}+h_{k_{q-1}}+h_{k_q})_{n_{q-1}}}{(2h_{k_q})_{n_q}}\\
&\phantom{=}\qquad\times{}_3F_2\left[\begin{array}{c}-n_{q-2},-n_{q-1},1-2h_{k_{q-1}}-n_{q-1}\\h_{k_{q-2}}-h_{k_{q-1}}-n_{q-1}+h_{k_q},1+h_{k_{q-2}}-h_{k_{q-1}}-n_{q-1}-h_{k_q}\end{array};1\right]\\
&\phantom{=}\qquad\times(h_{k_{q+2}}-h_{k_{q+1}}-n_{q+1}+h_{k_{q-1}})_{n_{q+2}}\\
&=\sum_{\{n_a\}\geq0}\frac{(h_{i_1}-h_{i_2}+h_{k_1})_{n_1}}{(2h_{k_1})_{n_1}}\frac{(h_{k_{q+2}}-h_{k_{q+1}}-n_{q+1}+h_{k_{q-1}})_{n_{q+2}+n_{q-1}}}{(2h_{k_{q-1}})_{n_{q-1}}}\bar{C}_{M-1}\bar{F}_{M-1}\\
&\phantom{=}\qquad\times\prod_{1\leq a\leq q-3}\frac{(h_{i_{\gamma_{a+2}}}-h_{k_a}-n_a+h_{k_{a+1}})_{n_{a+1}}(-h_{i_{\gamma_{a+2}}}+h_{k_a}+h_{k_{a+1}})_{n_a}}{(2h_{k_{a+1}})_{n_{a+1}}}\prod_{1\leq a\leq M-3}\frac{(\eta_a^M)^{n_a}}{n_a!}\\
&\phantom{=}\qquad\times\frac{(h_{k_{q-2}}-h_{k_{q-1}}-n_{q-1}+h_{k_q})_{n_{q-2}+n_q}(-h_{k_{q-2}}+h_{k_{q-1}}+h_{k_q})_{n_{q-1}}}{(2h_{k_q})_{n_q}}\\
&\phantom{=}\qquad\times{}_3F_2\left[\begin{array}{c}-n_{q-2},-n_{q-1},1-2h_{k_{q-1}}-n_{q-1}\\h_{k_{q-2}}-h_{k_{q-1}}-n_{q-1}+h_{k_q},1+h_{k_{q-2}}-h_{k_{q-1}}-n_{q-1}-h_{k_q}\end{array};1\right],
}
which matches with the rules applied to Figure \ref{FigType}.

As a consequence, the rules of Section \ref{SecCF} are consistent when an extra comb structure is glued unto an arbitrary topology of the third type.  This thus completes the proof of the rules in all cases.


\bibliography{CFT1d2d}

\end{document}